\documentclass[longauth]{aa} 
\usepackage{graphicx}
\usepackage{txfonts}
\begin{document}
   \title{First simultaneous optical/near-infrared imaging \\ of an X-ray selected, high-redshift cluster of galaxies with GROND}

   \subtitle{The galaxy population of XMMU\,J0338.7$+$0030 at $z = 1.1$}

   \author{D. Pierini
          \inst{1} \fnmsep\thanks{Currently visiting astronomer at the MPE},
           R. \u{S}uhada
          \inst{2,1},
           R. Fassbender
          \inst{1},
           A. Nastasi
          \inst{1},
           H. B\"{o}hringer
          \inst{1},
           M. Salvato
          \inst{3,4,1},
           G. W. Pratt
          \inst{5},
           M. Lerchster
          \inst{1,*},
           P. Rosati
          \inst{6},
           J. S. Santos
          \inst{7},
           A. de Hoon
          \inst{8},
           J. Kohnert
          \inst{8},
           G. Lamer
          \inst{8},
           J. J. Mohr
          \inst{9,1,10},
           M. M\"{u}hlegger
          \inst{1,*},
           H. Quintana
          \inst{11},
           A. Schwope
          \inst{8},
           V. Biffi
          \inst{1},
           G. Chon
          \inst{1},
           S. Giodini
          \inst{12},
           J. Koppenhoefer
          \inst{1,2},
           M. Verdugo
          \inst{1},
           F. Ziparo
          \inst{1},
           P. M. J. Afonso
          \inst{13,1},
           C. Clemens
          \inst{1,*},
           J. Greiner
          \inst{1},
           T. Kr\"{u}hler
          \inst{14,10,1},
           A. K\"{u}pc\"{u} Yolda\c{s}
          \inst{15},
           F. Olivares E.
          \inst{1},
           A. Rossi
          \inst{16},
          \and
           A. Yolda\c{s}
          \inst{15}
          }

   \institute{Max-Planck-Institut f\"ur extraterrestrische Physik (MPE),
              Giessenbachstrasse, D-85748 Garching, Germany
	 \and
              University Observatory Munich,
              Scheinerstrasse 1, D-81679 Munich, Germany
	 \and
              Max-Planck-Institut f\"{u}r Plasmaphysik,
              Boltzmannstrasse 2, D-85748 Garching, Germany
	 \and
	      Caltech,
              1200 East California Blvd, PMA 249-17, Pasadena, CA 91125, USA
         \and
              Laboratoire AIM, IRFU/Service d'Astrophysique - CEA/DSM - CNRS - Universit\'e Paris Diderot,
              B\^at. 709, CEA-Saclay, F-91191 Gif-sur-Yvette Cedex, France
         \and
              European Southern Observatory (ESO),
              Karl-Schwarzschild-Strasse 2, D-85748 Garching, Germany
         \and
              European Space Astronomy Centre (ESAC), P.O. Box 78,
              28691 Villanueva de la Ca\~{n}ada, Madrid, Spain
         \and
              Astrophysikalisches Institut Potsdam,
              An der Sternwarte 16, D-14482 Potsdam, Germany
         \and
              Department of Physics, Ludwig-Maximilians-Universit\"{a}t,
              Scheinerstrasse 1, D-81679 Munich, Germany
         \and
              Excellence Cluster Universe,
              Boltzmannstrasse 2, D-85748 Garching, Germany
         \and
              Departamento de Astronom\'{\i}a y Astrof\'{\i}sica, Pontificia Universidad Cat\'olica de Chile,
              Casilla 306, Santiago, 22, Chile
         \and
              Leiden Observatory,
              P.O. Box 9513, NL-2300 RA Leiden, The Netherlands 
         \and
              American River College, Physics and Astronomy Dept.,
              4700 College Oak Drive, Sacramento, CA 95841
         \and
              Dark Cosmology Centre, Niels Bohr Institute,
              University of Copenhagen,
              Juliane Maries Vej 30, 2100 C{\o}benhavn \O, Denmark
         \and
	      Institute of Astronomy, University of Cambridge,
              Madingley Road, Cambridge CB3 0HA, United Kingdom
         \and
              Th\"{u}ringer Landessternwarte Tautenburg,
              Sternwarte 5, D-07778 Tautenburg, Germany
             }

   \date{Received ... ..., ...; accepted ... ..., ...}

 
  \abstract
   {The XMM-{\em Newton} Distant Cluster Project
   is a serendipitous survey for clusters of galaxies at redshifts $z \ge 0.8$
   based on deep archival XMM-{\em Newton} observations.
   X-ray sources identified as extended
   are screened against existing optical all-sky surveys for galaxies,
   in case of candidate high-$z$ clusters followed up
   with imaging at 4m-class telescopes
   and, ultimately, multi-object spectroscopy at 8m-class telescopes.
   Low-significance candidate high-$z$ clusters
   are followed up with the seven-channel imager GROND
   ({{\em Gamma-Ray Burst Optical and Near-Infrared Detector}})
   that is mounted at a 2m-class telescope.
   Its unique capability of simultaneous imaging in the
   $\mathrm{g^{\prime}, r^{\prime}, i^{\prime}, z^{\prime}, J, H, Ks}$ bands
   enables the use of the photometric redshift technique.}
   {Observing strategy, data reduction and analysis, depth and accuracy
   of the simultaneous multi-wavelength photometry are discussed
   with the goal of establishing GROND as a useful instrument
   to confirm X-ray selected (high-$z$) clusters.}
   {The test case is XMMU\,J0338.7$+$0030,
   suggested to be at $z \sim 1.45 \pm 0.15$ ($1 \mathrm{\sigma}$)
   from the analysis of the $z - H$ vs $H$ colour--magnitude diagram
   obtained from the follow-up imaging.
   Later VLT-FORS2 spectroscopy enabled us to identify four members,
   which set this cluster at $z = 1.097 \pm 0.002$ ($1 \mathrm{\sigma}$).   
   To reach a better knowledge of its galaxy population,
   we observed XMMU\,J0338.7$+$0030 with GROND for about $6~\mathrm{hr}$.
   The publicly available photo-$z$ code {{\em le Phare}} was used.}
   {The Ks-band number counts of the non-stellar sources
   out of the 832 detected down to $z^{\prime} \sim 26~\mathrm{AB~mag}$
   ($1 \mathrm{\sigma}$) in the $3.9 \times 4.3~\mathrm{arcmin}^2$ region
   of XMMU\,J0338.7$+$0030 imaged at all GROND bands clearly exceed
   those computed in deep fields/survey areas
   at $\sim 20.5$--$22.5~\mathrm{AB~mag}$.
   The photo-$z$'s of the three imaged spectroscopic members yield
   $z = 1.12 \pm 0.09$ ($1 \mathrm{\sigma}$).
   The spatial distribution and the properties of the GROND sources
   with a photo-$z$ in the range 1.01--1.23 confirm the correspondence
   of the X-ray source with a galaxy over-density at a significance
   of at least $4.3~\mathrm{\sigma}$.
   Candidate members that are spectro-photometrically classified
   as elliptical galaxies define a red locus
   in the $i^{\prime} - z^{\prime}$ vs $z^{\prime}$ colour--magnitude diagram
   that is consistent with the red sequence of the cluster RDCS\,J0910$+$5422
   at $z = 1.106$.
   XMMU\,J0338.7$+$0030 hosts also a population of bluer late-type spirals
   and irregulars.
   The starbursts among the photometric members populate both loci,
   consistently with previous results.}
   {The analysis of the available data set indicates that XMMU\,J0338.7$+$0030
   is a low-mass cluster ($M_{200} \sim 10^{14}~\mathrm{M}_{\sun}$)
   at $z = 1.1$.
   With the photometric accuracy yielded by the present unoptimized
   multi-band observations with GROND, we not only confirm the spectroscopic
   redshift of this cluster but also show that it hosts a galaxy population
   that can still undergo significant bursts of star-formation activity.}


   \authorrunning{D. Pierini et al.}
   \titlerunning{I. The galaxy population of XMMU\,J0338.7$+$0030 at $z=1.1$}
   \maketitle
%

\section{Introduction}

   A process of self-similar gravitational clustering in an expanding universe
   leads to the build-up of structures in a hierarchical fashion across time
   (White \& Rees 1978; White et al. 1987).
   Ensembles of galaxies and hot ($T \ge 10^6~\mathrm{K}$) plasma
   within gravitationally bound haloes dominated by dark matter
   define groups and clusters, with total masses
   from $\sim 10^{13}$ to $\sim 10^{15}~\mathrm{M}_{\sun}$, respectively.
   These objects offer a biased view of the evolution
   of the large-scale structure (Kaiser 1984)
   that is easy to model theoretically from ab-initio principles.

   The identification of a group/cluster requires observations
   at rest-frame optical/near-infrared (IR) wavelengths
   (where the photospheric stellar emission dominates),
   in X-rays (where the hot plasma\footnote{Whether the hot X-ray emitting plasma belongs to a group or a cluster, we will refer to it as the intracluster medium (ICM).} emits), or at sub-millimeter wavelengths (owing to the Sunyaev-Zel'dovich - SZ - effect, Sunyaev \& Zel'dovich 1972).
   In particular, the X-ray or SZ-effect detection ensures that a galaxy system
   is truly bound and associated with a deep gravitational potential well.
   X-ray observations of groups/clusters can now be pushed
   towards the distant universe (i.e., $z \ge 0.8$)
   thanks to the {\em Chandra} and XMM-{\em Newton} space-borne observatories
   (e.g., Stanford et al. 2001, 2002; Rosati et al. 2004; Mullis et al. 2005;
   Finoguenov et al. 2007; \u{S}uhada et al. 2010;
   however, see Rosati et al. 1998).
   With the SZ-effect this has become possible very recently
   but only for massive clusters (Staniszewski et al. 2009;
   Vanderlinde et al. 2010; Marriage et al. 2011).

   Compiling a sample of distant galaxy clusters has a twofold importance.
   Firstly, it offers a way to trace structure growth in the far universe
   that enables the study of the effects of cold dark matter (CDM)
   and dark energy (DE) on the cosmic structure evolution
   (e.g., Albrecht et al. 2006; Vikhlinin et al. 2009; Mantz et al. 2010a).
   Secondly, it fosters the investigation of the evolution
   of the galaxy population, the baryon mass component of groups/clusters,
   the thermodynamics and chemical abundance of the ICM
   (e.g., Postman et al. 2005; Mei et al. 2006a,b; Balestra et al. 2007;
   Gobat et al. 2008; Maughan et al. 2008; Giodini et al. 2009;
   Mantz et al. 2010b; Santos et al. 2010; Strazzullo et al. 2010).
   About twenty clusters at $z = 0.8$--2.1 have been discovered
   and studied in some detail (e.g., Stanford et al. 2001, 2002, 2006;
   Blakeslee et al. 2003; Mullis et al. 2005; Demarco et al. 2007;
   Hilton et al. 2007, 2009; Eisenhardt et al. 2008; Lamer et al. 2008;
   Menci et al. 2008; Wilson et al. 2009; Kurk et al. 2009;
   Tanaka, Finoguenov \& Ueda 2010; Papovich et al. 2010; Henry et al. 2010;
   Gobat et al. 2011; Fassbender et al. 2011a).
   However, a better sampling in redshift and total mass is necessary
   to build a representative sample of high-$z$ groups and clusters.
   Existing X-ray surveys allow clusters in the redshift range 0.8--1.5
   to be detected, with about 1.5 such clusters per square degree
   down to a flux level of $\sim 6 \times 10^{-15}~\mathrm{erg~s^{-1}~cm^{-2}}$
   (\u{S}uhada et al. 2012).

   The XMM-{\em Newton} Distant Cluster Project (XDCP)
   is a serendipitous survey for galaxy clusters at $z \ge 0.8$
   based on deep archival XMM-{\em Newton} observations:
   it now covers an area of $80~\mathrm{deg}^2$ (B\"ohringer et al. 2005;
   Fassbender 2007; Fassbender et al. 2011c).
   XDCP is concerned with the build-up of a representative and complete sample
   of distant clusters and the accurate determination of the total mass
   of each system.
   Both aspects are crucial for a test of the current understanding
   of structure formation and evolution in $\mathrm{\Lambda}$CDM universes
   (e.g., Hoyle, Jimenez \& Verde 2011).
   In XDCP, X-ray sources identified as extended
   are screened against existing optical all-sky surveys
   and eventually followed up with imaging in the z, H or I, J bands
   at 4m-class telescopes (occasionally in the r, Z bands
   at 8m-class telescopes).
   If the distribution of the galaxies associated with an X-ray source
   exhibits evidence of a so-called ``red sequence''
   (e.g. Bower, Lucey \& Ellis 1992; Kodama et al. 1998; Gladders \& Yee 2000;
   see Arimoto \& Yoshii 1987 for its physical interpretation)
   in the ensuing colour--magnitude diagram,
   a candidate high-$z$ cluster is robustly selected
   and proposed for spectroscopic confirmation at 8m-class telescopes.
   This strategy has already yielded more than 30 spectroscopically confirmed
   distant clusters since the discovery of XMMU\,J2235.3$-$2557
   at $z = 1.393$ (Mullis et al. 2005).
   They include XMMU\,J0338.8$+$0021 at $z = 1.49$ (Nastasi et al. 2011),
   XMMU\,J1007.4+1237 at $z = 1.555$ (Fassbender et al. 2011b)
   and XMMU\,J0044.0$-$2033 at $z = 1.579$ (Santos et al. 2011).

   Among the main results from XDCP we quote the multi-wavelength studies
   of XMMU\,J1229$+$0151 at $z = 0.975$ (Santos et al. 2009)
   and XMMU\,J2235.3$-$2557 (Rosati et al. 2009; Strazzullo et al. 2010);
   the weak-lensing estimate of the total mass of XMMU\,J2235.3$-$2557
   (Jee et al. 2009); the total mass estimate of XMMU\,J100750.5$+$125818
   ($z = 1.082$) from strong lensing, optical spectroscopy and X-ray imaging
   (Schwope et al. 2010); the pan-chromatic study of XMMU\,J1230.3$+$1339
   at $z = 0.975$ (Fassbender et al. 2011a) and its total mass estimate
   from weak-lensing analysis of ground-based imaging (Lerchster et al. 2011);
   the evidence of starburst activity in the core of XMMU\,J1007.4+1237
   (Fassbender et al. 2011b).

   In order to maximize the numerical throughput from XDCP,   
   low-significance candidate high-$z$ clusters
   are followed up with the seven-channel imager GROND
   ({\em Gamma-Ray Burst Optical and Near-Infrared Detector};
   Greiner et al. 2008), that is mounted at a 2m-class telescope,
   before being proposed for multi-object spectroscopy (MOS)
   at 8m-class telescopes.
   The optical confirmation of these distant clusters lies in the discovery
   of a significant feature in the distribution of the photometric redshifts
   of galaxies selected in a region corresponding to
   the extended X-ray emission.
   Otherwise, it relies upon the fact that colour- or photo-$z$-selected
   galaxies within a given projected distance from the centroid
   of the extended X-ray emission define an over-density
   with respect to the surrounding region and/or a red sequence
   in a given colour--magnitude diagram.

   Here we report on the feasibility of this approach by illustrating results
   obtained from the multi-wavelength observations of XMMU\,J0338.7$+$0030.
   Follow-up $\mathrm{z, H}$ photometry was originally obtained
   with the near-IR, wide-field camera OMEGA2000 mounted at the prime focus
   of the 3.5m telescope at Calar Alto, Spain.
   A redshift $z \sim 1.45 \pm 0.15$ was inferred
   from comparison between the average $z - H$ colour of the reddest galaxies
   within $45^{\prime \prime}$ from the X-ray position of XMMU\,J0338.7$+$0030
   and the synthetic $z - H$ colours expected for a large suite
   of simple stellar population (SSP) models (see Fassbender et al. 2011c).
   The synthetic photometry was computed using the spectro-photometric
   evolutionary synthesis model {\em P\'EGASE} v2
   (Fioc \& Rocca-Volmerange 1997) in analogy with Pierini et al. (2005)
   and Wilman et al. (2008).
   A calibration of the best-performing SSP model was originally based
   on observed red-sequence galaxies in 10 X-ray selected clusters
   with spectroscopic redshifts up to 1.5 (Fassbender 2007).
   This fiducial model corresponds to a Salpeter (1955)
   stellar initial mass function (IMF) between 0.1 and $120~\mathrm{M}_{\sun}$
   (as all the others), a solar metallicity and a formation redshift
   equal to 5.

   In addition to its being potentially one of the few clusters at $z \ge 1.4$
   known so far, XMMU\,J0338.7$+$0030 represented a test case
   for XDCP sources at the detection limit of the survey.
   It was an ideal target for GROND imaging because its region is covered
   by the {\em Sloan} Digital Sky Survey (SDSS; York et al. 2000)
   and the 2 Micron All Sky Survey (2MASS; Skrutskie et al. 2006),
   which enables a robust photometric calibration for all seven bands.
   For the same region, sparse redshift information was later available
   from observations with the spectrograph FORS2
   mounted at the ESO {\em Very Large Telescope} (VLT) at Paranal, Chile.

   The XMM-{\em Newton}, OMEGA2000, GROND and FORS2 observations
   of XMMU\,J0338.7$+$0030, plus the reduction and analysis
   of the ensuing data, are discussed in Sect.~2.
   Estimates of the X-ray bolometric luminosity, temperature
   and total mass of this cluster, plus the spectroscopic determination
   of its redshift, are given there.
   Section 3 contains results obtained from the GROND data, i.e.
   galaxy number counts, the photo-$z$ distribution of the detected sources,
   the two-dimensional (2-D) spatial distribution of photo-$z$-selected
   galaxies and their distribution in specific colour--magnitude diagrams.
   Discussion and conclusions appear in Sect. 4 and 5, respectively.

   Throughout we adopt a $\mathrm{\Lambda}$CDM cosmological model
   ($\mathrm{\Omega_m} = 0.3$, $\mathrm{\Omega_\Lambda} = 0.7$)
   with $\mathrm{H}_0 = 70~\mathrm{km~s}^{-1}~\mathrm{Mpc}^{-1}$,
   that is broadly consistent with the main results
   from the $5~\mathrm{yr}$ operations
   of the {\em Wilkinson Microwave Anisotropy Probe}
   (WMAP5, Dunkley et al. 2009; Komatsu et al. 2009).
   Accordingly, at the spectroscopic redshift of XMMU\,J0338.7$+$0030
   ($z = 1.097$; Sect.~2.4.2), the age of the Universe is $5.4~\mathrm{Gyr}$,
   the cosmic evolution factor $E(z) = H(z) / H_0$ is equal to 1.86,
   the luminosity distance is $7409.2~\mathrm{Mpc}$,
   the angular scale is $8.168~\mathrm{kpc}/ ^{\prime \prime}$.
   Hence, an angular distance of $1^{\prime}$ corresponds
   to $\sim 490~\mathrm{kpc}$ at the cluster distance.


\section{Observations, data reduction and analysis}

\subsection{X-ray imaging with XMM-Newton}

\subsubsection{Initial X-ray selection}

   \begin{figure*}
   \centering
   \vskip 0.35truecm
   \includegraphics[width=9cm]{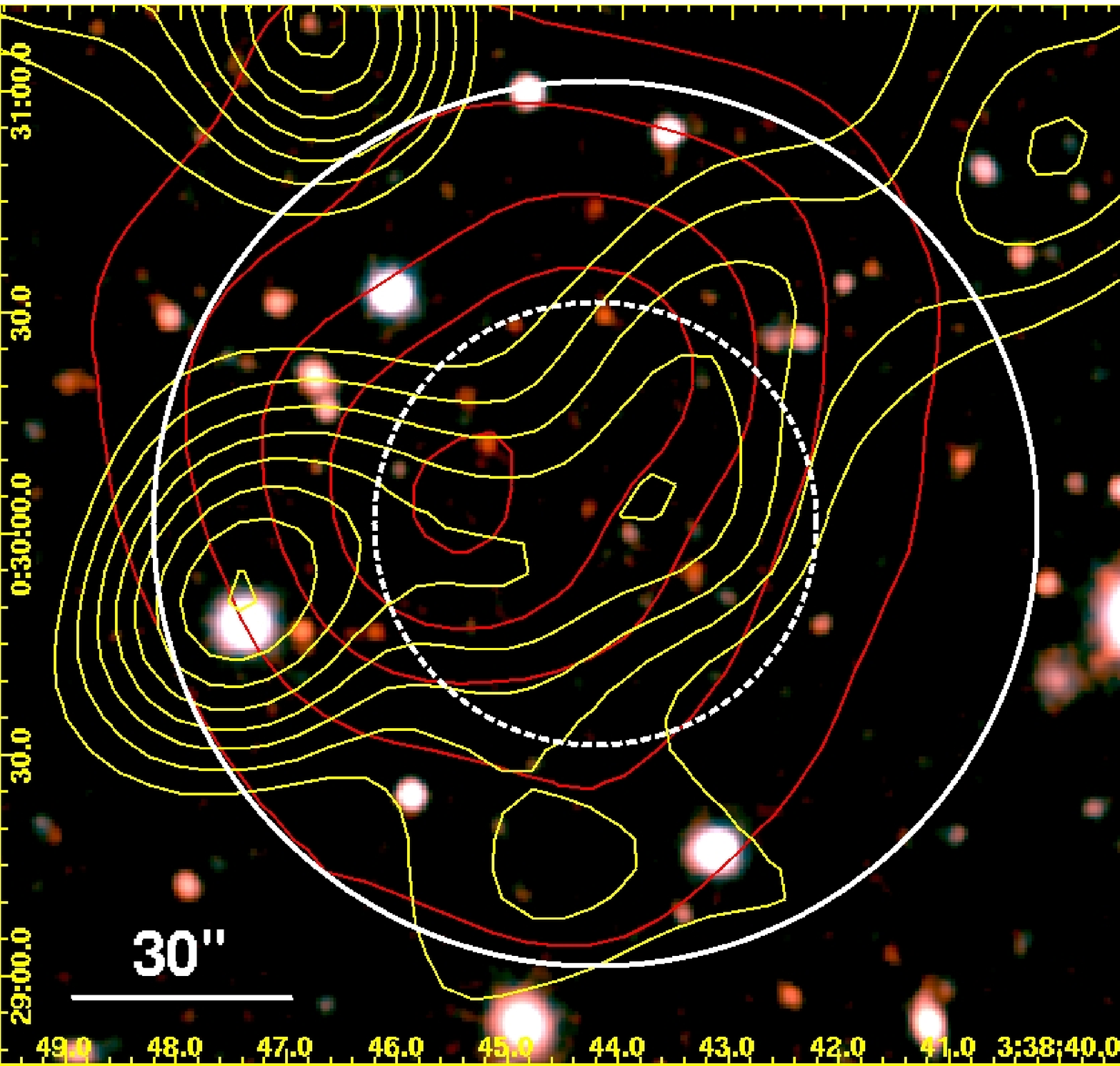}
   \vskip 0.25truecm
      \caption{A z$+$H colour composite image ($4.5^{\prime}$ on the side)
               of the environment of XMMU\,J0338.7$+$0030
               with an overlaid contour map of the X-ray emission
               detected by XMM-{\em Newton} (in yellow).
               These log-spaced X-ray contours correspond
               to significance levels of 2--$16 \mathrm{\sigma}$;
               they were derived from the adaptively smoothed, combined images
               of the cluster environment.
               Red contours highlight the over-density of galaxies
               with very red $z - H$ colours
               ($3.0 \le z - H \le 4.8~\mathrm{Vega~mag}$),
               as obtained from the OMEGA2000 imaging
               (see Sect.~2.2.2).
               White dashed and solid lines encircle regions
               within $30^{\prime \prime}$ and $60^{\prime \prime}$
               from the X-ray centroid of the cluster
               (R.A.(J2000.0): 03h38m44.2s, Dec(J2000.0): $+$00d30m01.8s),
               respectively; they guide the eye.
               The red contour closest to the X-ray position of the cluster
               corresponds to a significance of $4.6 \mathrm{\sigma}$
               above the mean background.
              }
         \label{FigXrayzHimage}
   \end{figure*}

   A total of 470 XMM-{\em Newton} (Jansen et al. 2001) archival fields
   with nominal exposure times longer than $10~\mathrm{ks}$
   have so far been processed and analyzed in XDCP as extensively described
   in Fassbender (2007) and Fassbender et al. (2011c).
   The serendipitous, weak X-ray source identified as XMMU\,J0338.7$+$0030
   was detected in the field with observation identification number
   (OBSID) 0036540101.
   The target was SDSS\,033829.31$+$002156.3
   (R.A.(J2000.0): 03h38m38.5s, Dec(J2000.0): $+$00d39m28s),
    a quasar at $z=5.07091$ (P\'eroux et al. 2001) imaged on February 22, 2002
   with XMM-{\em Newton} for a nominal exposure time of $22.9~\mathrm{ks}$.

   This serendipitous X-ray source was found at an off-axis angle
   of $9.6^{\prime}$ (R.A.(J2000.0): 03h38m44.2s, Dec(J2000.0): $+$00d30m01.8s)
   and a significance of about $5 \mathrm{\sigma}$
   during the early XDCP source detection run performed with \texttt{SAS} v6.1.
   The tasks \texttt{eboxdetect} was used for a sliding box detection,
   followed by a subsequent maximum likelihood analysis
   with \texttt{emldetect}.
   This yielded an extent of 3.48 pixels (i.e., 13.92 arcsec),
   a source extent likelyhood of about $2 \mathrm{\sigma}$ and a flux of
   $\sim 8 \times 10^{-15}~\mathrm{erg~s^{-1}~cm^{-2}}$
   for $\sim 32$ PN counts, close to the detection threshold.
   Hence, XMMU\,J0338.7$+$0030 appeared as a marginally extended X-ray source
   at the XMM-{\em Newton} detection threshold for extended sources.
   An X-ray contour map is overlaid on the z$+$H colour composite image
   obtained from the medium--deep imaging with OMEGA2000 (Sect. 2.2.1)
   in Fig.~1.

   The circular area centred on the X-ray emission centroid
   of XMMU\,J0338.7$+$0030 and delimited by a radius of $1^{\prime}$
   is assumed as the ``bona fide'' cluster region throughout.

\subsubsection{The upgraded X-ray picture of XMMU\,J0338.7$+$0030}

   The XMM-{\em Newton} field containing XMMU\,J0338.7$+$0030 was reprocessed
   with \texttt{SAS}\,v10.0.0 following \u{S}uhada et al. (2011).
   These observations are strongly flared and an automatic flare removal
   is impossible; hence, the contaminated periods were excised manually
   and the automatic two-step flare cleaning process
   was run on the remaining good part of the data
   for the removal of high background periods.
   The loss of information is severe: clean net exposure times on axis
   amount to $9~\mathrm{ks}$ for either MOS camera
   and $7.8~\mathrm{ks}$ for the PN instrument,
   whereas the effective (vignetted) clean net exposure times on source
   amount to 4.5--$5.5~\mathrm{ks}$ for all detectors.
   Even after cleaning a slight residual quiescent contamination
   (about 10\% above normal) remains; large part of it should be captured
   by the background model, so this anomalous contamination
   likely has a negligible effect.

   With \texttt{SAS}\,v10.0.0 XMMU\,J0338.7$+$0030 is detected
   as a very weak source without a significant extent,
   located about $10^{\prime \prime}$ to the east
   with respect to the X-ray position originally estimated (Sect.~2.1.1).
   A Kolmogorov--Smirnov (KS) test with the null hypothesis that
   the extracted X-ray profile corresponds to that of a point-like source
   gives $-\mathrm{log}~P_{\mathrm{KS}} = 2.08$
   and $-\mathrm{log}~P_{\mathrm{KS}} = 1.57$ for the PN and MOS data,
   respectively.
   In other words, there is a chance of 0.8\% or 2.7\%
   that XMMU\,J0338.7$+$0030 is a point-like source.
   The result based on the PN instrument is likely unrealistic,
   because PN has a chip gap very close to the centre.
   Furthermore, the performed test uses the on-axis point-spread-function
   (PSF) of XMM-{\em Newton}, so the output probabilities
   are slightly underestimated for the off-axis source
   identified as XMMU\,J0338.7$+$0030.
   We conclude that this source is weak and marginally extended.

   This is supported by the regular behaviour of the growth curves,
   where the growth curve analysis (GCA) method of B\"ohringer et al. (2000)
   was applied to measure the flux in the soft 0.5--$2~\mathrm{keV}$ band.
   The two growth curves obtained with the MOS cameras
   or the PN instrument agree (see Fig.~2).
   However, they are not independent: the latter was slightly corrected
   to obtain a rough agreement, because PN has a chip gap very close
   to the centre as already mentioned.
   The double component background model was used in the first instance;
   the alternative use of a spline fitting led to the same results.
   As a result, we measured a flux
   $f_{\mathrm{X,500}} \simeq 7.1 (\pm 2.3) \times 10^{-15}~\mathrm{erg~s^{-1}~cm^{-2}}$
   within a circular aperture of radius equal to $R_{500}$\footnote{$R_{\mathrm{\Delta}}$ ($\mathrm{\Delta} = 500$) is the radius within which the total mass density of a group/cluster is equal to $\mathrm{\Delta}$ times the critical density of the universe ($\rho_{\mathrm{c}}$). Correspondingly, $M_{\mathrm{\Delta}} = \mathrm{\Delta}~\rho_{\mathrm{c}} (z)~(4 \pi / 3)~R_{\mathrm{\Delta}}^3$ is the total mass within $R_{\mathrm{\Delta}}$.} (corresponding to $51.2^{\prime \prime}$).
   At the cluster redshift, this translates into an X-ray luminosity
   of $L_{\mathrm{X,500}} \simeq 4.5 (\pm 1.4) \times 10^{43}~\mathrm{erg~s^{-1}}$
   in the soft band or $L^{\mathrm{bol}}_{\mathrm{X,500}} \simeq 1.1 (\pm 0.3) \times 10^{44}~\mathrm{erg~s^{-1}}$ for the bolometric luminosity.
   The latter quantity was inferred as described in \u{S}uhada et al. (2010).
   In brief, the growth curve was used to iteratively obtain
   a self-consistent set of parameters using the X-ray luminosity--temperature
   and X-ray luminosity--total mass relations from Pratt et al. (2009).

   Owing to the faintness of the source with less than 50 net counts
   in the soft band, the determination of additional structural 
   or spectral parameters is currently not feasible.
   The potential flux contribution of point sources can presently
   not be accounted for either, which implies that the stated luminosities
   have to be taken as upper limits.
   However, based on the luminosity scaling relation in Pratt et al. (2009),
   first rough estimates on the expected temperature
   of $T_{\mathrm{X}} \sim 2.4~\mathrm{keV}$ and the total mass
   of $M_{500} \sim 7 \times 10^{13}~\mathrm{M_{\sun}}$
   (i.e., $M_{200} \sim 10^{14}~\mathrm{M_{\sun}}$) can be obtained
   from the previous values.
   Hence, XMMU\,J0338.7$+$0030 appears to be a system
   at the boundary between X-ray selected groups and clusters,
   as XMMU\,J1532.2$-$0836 at $z = 1.358$ (\u{S}uhada et al. 2011)
   and XMMU\,J0338.8$+$0021 at $z = 1.49$ (Nastasi et al. 2011).

   An exposure-corrected count image of the innermost region
   of the cluster is produced after masking point-like sources (cf. Fig.~1)
   and is then adaptively smoothed with a circular top hat filter.
   The ensuing X-ray contours are reproduced in the right panel
   of Fig.~4 (Sect.~2.3.1) using a squared scale.

   \begin{figure}
   \centering
   \vskip -0.15truecm
      \includegraphics[width=9cm]{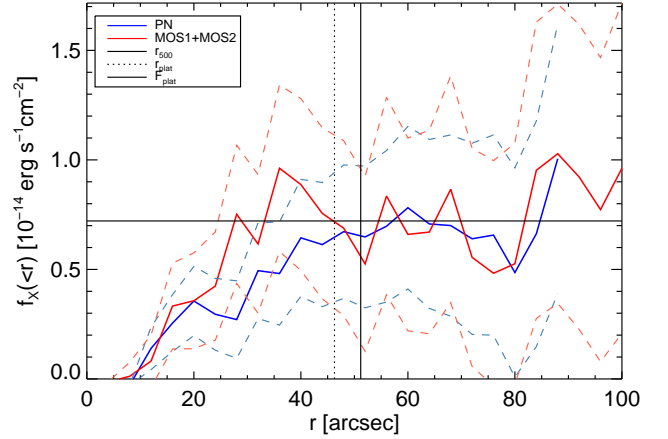}
      \caption{Results from the application of the GCA method
               of B\"ohringer et al. (2000) in the soft 0.5--$2~\mathrm{keV}$
               band to the MOS and PN observations of XMMU\,J0338.7$+$0030.
               The two growth curves are well behaved
               and agree with each other.}
         \label{FigXraygc}
   \end{figure}

\subsection{Near-infrared follow-up imaging with OMEGA2000}

\subsubsection{Observations and data reduction}

   XMMU\,J0338.7$+$0030 was observed with the near-IR camera OMEGA2000
   (Bailer-Jones, Bizenberger \& Storz 2000)
   that is mounted at the 3.5m telescope at the Centro Astron\'omico
   Hispano Alem\'an (CAHA) at Calar Alto, Spain.
   OMEGA2000 provides a field-of-view (FoV)
   of $15.4^{\prime} \times 15.4^{\prime}$.
   In particular, medium--deep imaging of the XMMU\,J0338.7$+$0030 region
   in the H-band ($50~\mathrm{min}$) and z-band ($53~\mathrm{min}$)
   was obtained on January 5, 2006 under good but non-photometric conditions
   of the sky and with a seeing (or PSF)
   that varied between $1.0^{\prime \prime}$ and $1.5^{\prime \prime}$ (z-band)
   or $1.6^{\prime \prime}$ (H-band) in full-width-at-half-maximum (FWHM).
   The same region was reobserved in z-band ($5~\mathrm{min}$)
   on October 30, 2006 in photometric conditions of the sky.
   Observations of a designated SDSS standard star (Smith et al. 2002)
   were performed then to enable a photometric calibration
   of the science frames.

   These data were reduced with the OMEGA2000 near-IR pipeline
   (Fassbender 2007), which performs the basic frame calibrations
   (flat-fielding, bad pixel correction), a two-pass background subtraction
   with masked objects, frame alignment and optimally weighted co-addition
   of the individual $40~\mathrm{s}$ ($60~\mathrm{s}$) exposures in H-band
   (z-band).
   For each band, the individual frames were stacked without PSF-matching,
   which gave a measured seeing of $1.3^{\prime \prime}$ (z-band)
   or $1.4^{\prime \prime}$ (H-band) in the final image.

   For the CAHA data, source detection, extraction of photometry
   and star/galaxy separation are described in detail
   in Fassbdender et al. (2007).
   The {\em SExtractor} (Bertin \& Arnouts 1996) photometry was calibrated
   to the Vega magnitude system using 2MASS point sources (Cutri et al. 2003)
   in H-band and designated SDSS standard star observations in z-band,
   cross-checked with SDSS photometry in the science frame.
   In particular, total magnitudes were measured
   in a standard Kron (\cite{kron80}) aperture of factor 2.5
   and minimum radius of 3.5.
   The limiting (Vega) magnitudes\footnote{These limiting magnitudes correspond to a 50\% completeness, as estimated from the number counts obtained from the OMEGA2000 imaging only, as in Fassbender et al. (2011a).} were determined
   to be $H_\mathrm{lim} \sim 21.2$ and $z_\mathrm{lim} \sim 23.1$.
   They correspond to the expected apparent magnitudes of an old,
   passively evolving galaxy at the estimated redshift of $z \sim 1.45$
   (see Sect.~1) for an absolute magnitude equal to $\sim M^{\star}+1.3$
   in H-band and $\sim M^{\star}$ in z-band, where $M^{\star}$ is
   the characteristic magnitude of the corresponding galaxy luminosity function
   at $z = 1.45$ (see Fassbender 2007).

\subsubsection{An over-density of very red galaxy at the location of XMMU\,J0338.7$+$0030}

   In Fig.~1 (Sect.~2.1.1), the projected distribution of galaxies
   with very red colours ($3.0 \le z - H \le 4.8~\mathrm{Vega~mag}$)
   exhibits a significant over-density: $4.6 \mathrm{\sigma}$
   above the mean background.
   The peak of this over-density is offest to the East
   by about $20^{\prime \prime}$ with respect to the X-ray position
   of XMMU\,J0338.7$+$0030.
   When considering the degradation of the angular resolution
   of XMM-{\em Newton} ($\sim 14^{\prime \prime}$ half-energy-width - HEW)
   at the off-axis angle of this X-ray source (Sect.~2.1.1),
   an association between the previous over-density of red galaxies
   and the cluster is very likely.

   Figure 3 shows the $z - H$ vs $H$ colour--magnitude diagram
   for the galaxies detected in the XMMU\,J0338.7$+$0030 region
   imaged with OMEGA2000.
   In particular, seven galaxies within $45^{\prime \prime}$
   from the X-ray position of the cluster
   exhibit $2.8 \le z - H \le 3.8~\mathrm{Vega~mag}$).
   A redshift $z \sim 1.45 \pm 0.15$ was initially inferred
   from comparison between the average $z - H$ colour
   of these galaxies and the synthetic $z - H$ colours
   expected for the fiducial SSP model described in Sect.~1.
   The assumption was that such a template described all those galaxies.

   \begin{figure*}
   \centering
   \vskip -0.75truecm
   \includegraphics[width=15cm]{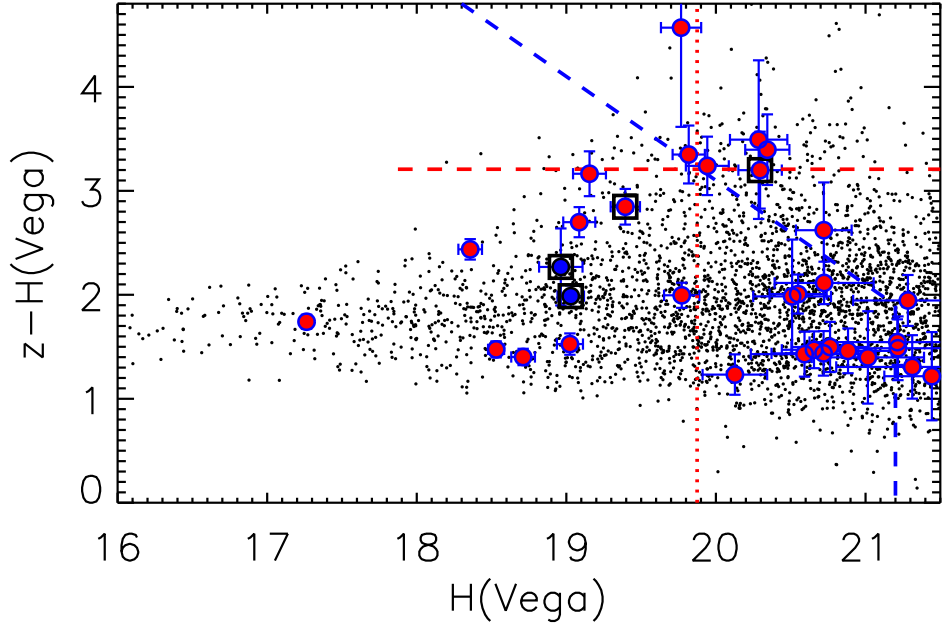}
   \vskip -0.25truecm
      \caption{$z - H$ vs $H$ colour--magnitude diagram
               for the region encompassing XMMU\,J0338.7$+$0030
               imaged with OMEGA2000.
               Red filled circles represent galaxies
               within $45^{\prime \prime}$
               from the X-ray position of the cluster.
               The blue dashed lines reproduce the 50\% completeness
               of the photometric catalogue extracted from the OMEGA2000 data.
               The horizontal red short-dashed line represents
               the $z - H$ colour expected for an SSP model corresponding
               to the evolutionary phase of an instantaneous burst
               of star formation observed at $z = 5$ with Salpeter (1955)
               stellar IMF and solar metallicity observed at $z = 1.45$.
               This SSP model describes the average colour
               of the seven galaxies associated with XMMU\,J0338.7$+$0030
               that exhibit $2.8 \le z - H \le 3.8~\mathrm{Vega~mag}$.
               The vertical red dotted line represents
               the characteristic magnitude $H^{\star}$ at $z = 1.45$
               (from Fassbender et al. 2007).
               Empty black squares mark the four spectroscopic cluster members
               that were successively established to be at $z = 1.1$
               through VLT-FORS2 observations (see Fig.~11 and Table 1
               in Sect.~2.4.2);
               blue filled circles highlight those with ID$=$12 and ID$=$15,
               which are at a cluster-centric distance
               greater than $45^{\prime \prime}$.
               The two spectroscopic cluster members
               without convincing evidence of ongoing star-formation activity
               (with ID$=$2 and ID$=$16) exhibit about as red $z - H$ colours
               as the previous SSP model (see Appendix B, however).
              }
         \label{FigZHhcolmagOmega2000}
   \end{figure*}

\subsection{Optical/near-infrared imaging with GROND}

\subsubsection{Simultaneous images of XMMU\,J0338.7$+$0030}

   \begin{figure*}
   \vskip 0.25truecm
   \centering
   \includegraphics[width=9cm]{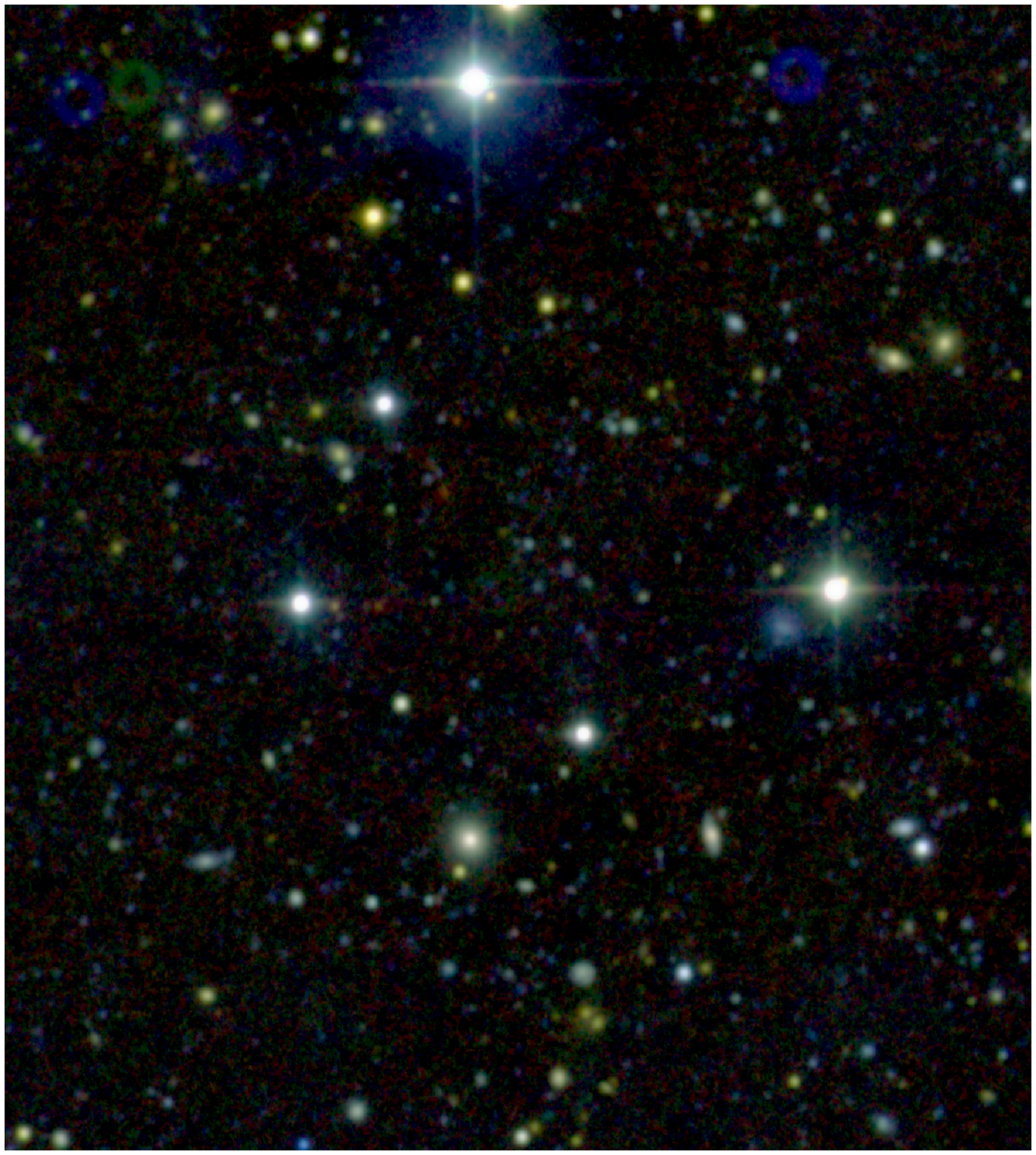}
   \includegraphics[width=9.05cm]{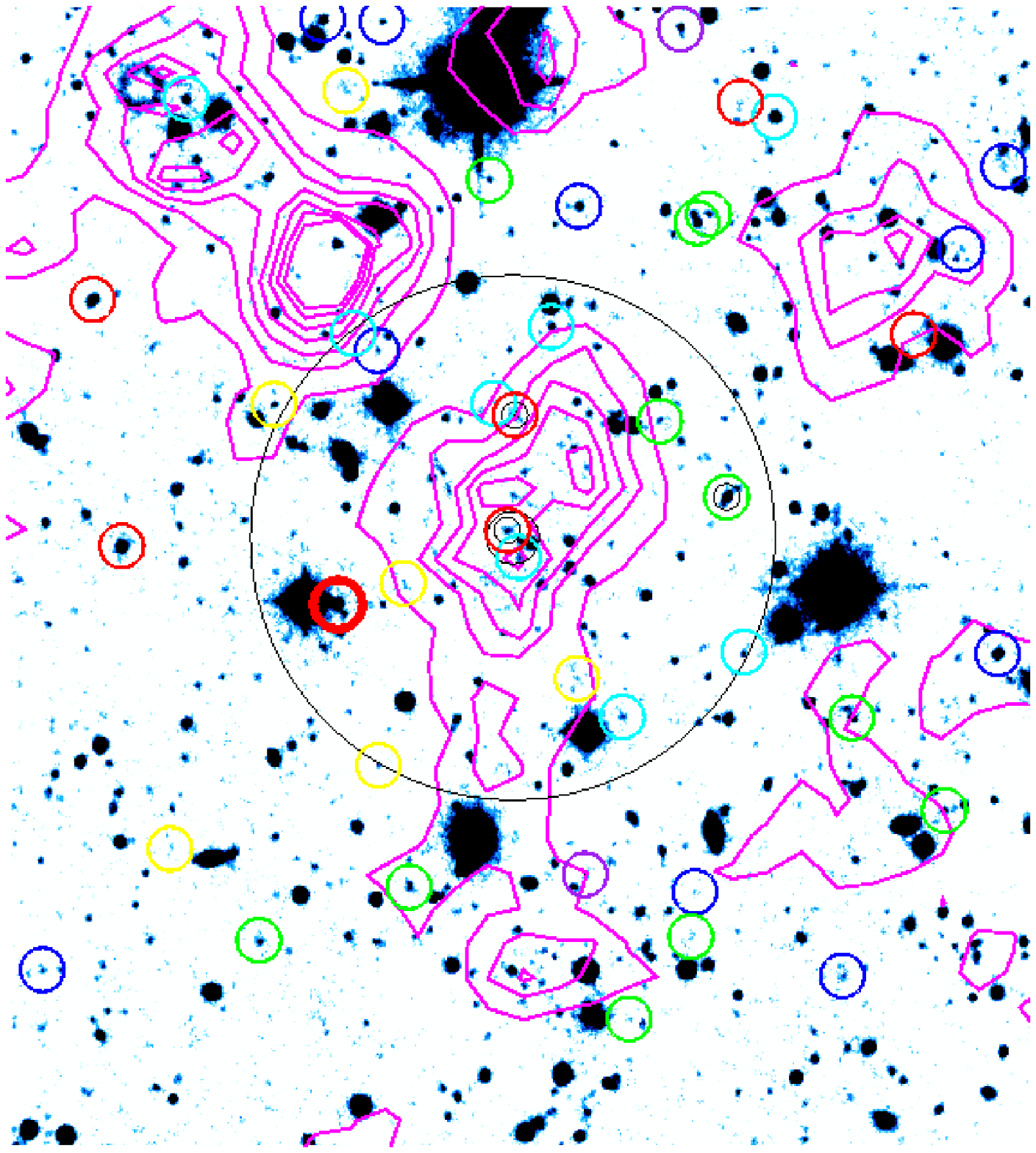}
   \vskip 0.35truecm
      \caption{$\mathrm{J, z^{\prime}, g^{\prime}}$ colour-composite
               image of the $3.9 \times 4.3~\mathrm{arcmin}^2$ region
               of XMMU\,J0338.7$+$0030 mapped with GROND at all seven bands
               (left panel).
               North is up and east to the left.
               The corresponding $\mathrm{z^{\prime}}$-image (right panel)
               illustrates the 2-D distribution of the 44 sources
               in the GROND photometric sample with a photo-$z$
               that is consistent with the spectroscopic redshift,
               i.e., at $1.01 \le z \le 1.23$ (empty circles).
               These 44 photometric members are colour-coded according
               to their spectro-photometric classification as E (red),
               Sbc (yellow), Scd (green), Im (cyan), SB (blue) or QSO (purple).
               The three spectroscopic members among them are marked
               with small, black empty circles.
               Two concentric black circles with radii of $6^{\prime \prime}$
               and $60^{\prime \prime}$ mark the X-ray position
               and bona-fide region of XMMU\,J0338.7$+$0030, respectively.
               Magenta contours represent the weak, marginally extended
               X-ray emission of the source determined in Sect.~2.1.2.
               This emission is consistent with the projected distribution
               of the photometric cluster members.
               However, the candidate BCG (thick red empty circle) is offset
               by $\sim 43^{\prime \prime}$ towards the east--south--east.
              }
         \label{FigRGBimage}
   \end{figure*}

   A region of the sky centred on the XMM-{\em Newton} X-ray position
   of XMMU\,J0338.7$+$0030 was imaged on four different nights
   (September 24, 2008, and January 27, 28, 29, 2009),
   in conditions of clear sky but variable seeing, with GROND.
   This is a seven-channel optical/near-IR imager
   primarily designed for rapid, simultaneous multi-wavelength observations
   of gamma-ray burst afterglows and built by the high-energy group of MPE
   in collaboration with the Landessternwarte Tautenburg and ESO.
   GROND was commissioned at the MPI/ESO 2.2m telescope at La Silla, Chile,
   in April 2007 and is operated as a PI-instrument
   (for its comprehensive description see Greiner et al. \cite{greiner08}).

   For the optical channels, the incoming telescope beam is split
   using four dichroics designed in such a way that
   their transmission functions are identical to those
   of the {\em Sloan} $\mathrm{g^{\prime}, r^{\prime}, i^{\prime}, z^{\prime}}$
   filter system, with the exception of the $\mathrm{i^{\prime}}$-band.
   Each of the four identical optical CCDs has a FoV
   of $5.4 \times 5.4~\mathrm{arcmin}^2$
   and a scale of $0.158^{\prime \prime}$/pixel.
   The near-IR part of GROND is designed as a focal reducer system
   yielding a FoV of $10 \times 10~\mathrm{arcmin}^2$
   with a scale of $0.6^{\prime \prime}$/pixel.
   The GROND $\mathrm{J, H, Ks}$ dichroic filters are close
   to other near-IR filter systems.
   Photometric system transformations between the GROND
   $\mathrm{g^{\prime}, r^{\prime}, i^{\prime}, z^{\prime}}$ bands
   and the {\em Sloan} filter system
   and between the GROND $\mathrm{J, H, Ks}$ bands
   and the 2MASS filter system are given in Greiner et al. (\cite{greiner08}).

   A total of 10 observations in the seven channels was executed
   through the GROND standard observing block (OB) named ``20min4TD''.
   By design, this OB corresponds
   to four single $369~\mathrm{s}$-long exposures
   in the $\mathrm{g^{\prime}, r^{\prime}}$ channels
   and four $369~\mathrm{s}$-long exposures out of four pairs of sub-exposures
   in the $\mathrm{i^{\prime}, z^{\prime}}$ channels,
   each long exposure being taken
   at one out of four different telescope dither positions.
   In parallel, the three near-IR channels
   are operated with $10~\mathrm{s}$ integrations.
   At each telescope dither position, a six-position dither pattern
   is executed through a flip mirror (internal dithering)
   only for the $\mathrm{Ks}$ channel;
   five $10~\mathrm{s}$-long exposures per mirror position
   are executed for the $\mathrm{J, H, Ks}$ channels.
   This produces a total of 120 single $10~\mathrm{s}$-long exposures per OB
   for each near-IR channel.
   Given the impact of overheads (about 50\% of the total exposure time
   of $4~\mathrm{hr}$ for each $\mathrm{J, H, Ks}$ channel),
   the total observing time amounted to about $6~\mathrm{hr}$.

   Pixel- and gain-corrected, astrometrized, stacked images for each channel
   and per OB were produced using the data reduction and photometry tools
   of the GROND pipeline (K\"{u}pc\"{u} Yolda\c{s} et al. \cite{kupkuyoldas08};
   Kr\"{u}hler et al. \cite{kruehler08}), based on standard {\em IRAF}\footnote{{\em IRAF} is distributed by the Na\-tio\-nal Opti\-cal Astro\-nomy Observa\-tories, which are operated by the Association of Universities for Research in Astronomy, Inc., under cooperative agreement with the National Science Foundation.}/{\em PyRAF} tools (Tody \cite{tody93}).
   Astrometry was achieved through the use of stars in common
   with the SDSS catalogue (Abazajian et al. \cite{abazajian09})
   for the optical bands and the 2MASS catalogue
   (Skrutskie et al. \cite{skrutskie06}) for the near-IR ones.
   For each image the FWHM of the average point-spread-function (PSF)
   was measured from the surface brightness profiles of randomly distributed
   non-saturated stars.
   Images corresponding to values of the average seeing FWHM 
   higher than $1.5^{\prime \prime}$ were not considered
   in the following analysis.
   For each band, all selected images were convolved
   to the worst PSF ($\sim 1.5^{\prime \prime}$ FWHM)
   and then stacked with a median algorithm.
   The total exposure times of the stacked
   $\mathrm{g^{\prime}, r^{\prime}, i^{\prime}, z^{\prime}, J, H, Ks}$ images
   are equal to $11808~\mathrm{s}$, $14760~\mathrm{s}$,
   $14760~\mathrm{s}$, $14760~\mathrm{s}$, $12000~\mathrm{s}$,
   $10800~\mathrm{s}$, and $9600~\mathrm{s}$, respectively.
   The stacked near-IR images were regridded onto the same scale
   as the optical ones ($0.158^{\prime \prime}$/pixel).
   All seven stacked images were re-mapped to the same aspect
   with the $\mathrm{z^{\prime}}$-image as reference.

   Figure 4 reproduces a $\mathrm{J, z^{\prime}, g^{\prime}}$ colour-composite
   (RGB) image of the region of XMMU\,J0338.7$+$0030 mapped with GROND
   (left panel) together with the deep $\mathrm{z^{\prime}}$-image
   (right panel).

\subsubsection{The photometric catalogue of XMMU\,J0338.7$+$0030: source extraction and photometry}

   A separate stack of seven individual images
   in the $\mathrm{z^{\prime}}$ channel, PSF-matched
   to the same angular resolution of $1.1^{\prime \prime}$ (FWHM),
   was used as a detection image with a total exposure time
   equal to $10332~\mathrm{s}$.
   This choice was the best compromise between image quality,
   photometric depth in the red/near-IR channels,
   robustness of detection and penalty in the source detection
   and extraction of photometry associated with deblending.
   It also enabled a better, independent separation between extended
   and point-like sources along the accessible flux domain.

   Automatic source detection and extraction of photometry
   within an area of $3.9 \times 4.3~\mathrm{arcmin}^2$
   was accomplished through the publicly available, standard software
   {\em SExtractor} (Bertin \& Arnouts \cite{bertin96}).
   A source was defined by the excess over a detection threshold
   equal to the standard deviation ($1 \mathrm{\sigma}$)
   of the global image background in at least two contiguous pixels;
   in addition, a Gaussian filter with a FWHM of $1.1^{\prime \prime}$
   was applied for detection.
   Photometry of individual sources was extracted in six circular apertures
   with diameters equal to 1, 2, 3, 4, 5 and 10 arcsec,
   in a Kron (\cite{kron80}) aperture of factor 2.5
   and minimum radius of 3.5 ({\em SExtractor} MAG\_AUTO)
   and in a Petrosian (\cite{petrosian76}) aperture of factor 2.0
   and minimum radius of 3.5.

   In the first step of extraction of source photometry,
   typical photometric zeropoints for the GROND channels\footnote{GROND photometric zeropoints in the AB magnitude system are available at http://www.mpe.mpg.de/$\sim$jcg/GROND/calib.html and in Greiner et al. (\cite{greiner08}) together with conversions between different filter systems or from the AB to the Vega magnitude system. These conversions are equal to $-0.008$, $-0.151$, $-0.386$, $-0.515$, $-0.910$, $-1.381$ and $-1.795~\mathrm{mag}$ for, respectively, the GROND $\mathrm{g^{\prime}, r^{\prime}, i^{\prime}, z^{\prime}, J, H, Ks}$ bands.} were assumed together with typical corrections
   for atmospheric absorption (or airmass),
   in order to aid identification of stars in common with
   the existing SDSS and 2MASS catalogues and enable the calibration
   of the zeropoints of the present
   $\mathrm{g^{\prime}, r^{\prime}, i^{\prime}, z^{\prime}, J, H, Ks}$ images.
   In particular, we made use of the seventh {\em Sloan} data release
   (DR7; Abazajian et al. 2009 and references therein)
   and the 2MASS All Sky Catalog of point sources (Cutri et al. 2003).
   After this photometric calibration,
   adjustments for the newly determined zeropoints were introduced
   and magnitudes were corrected for Galactic extinction
   according to the values tabulated in Schlegel et al. (\cite{schlegel98})
   in addition to the previous correction for atmospheric absorption.

   \begin{figure*}
   \centering
   \vskip -1.0truecm
   \includegraphics[width=6.5cm]{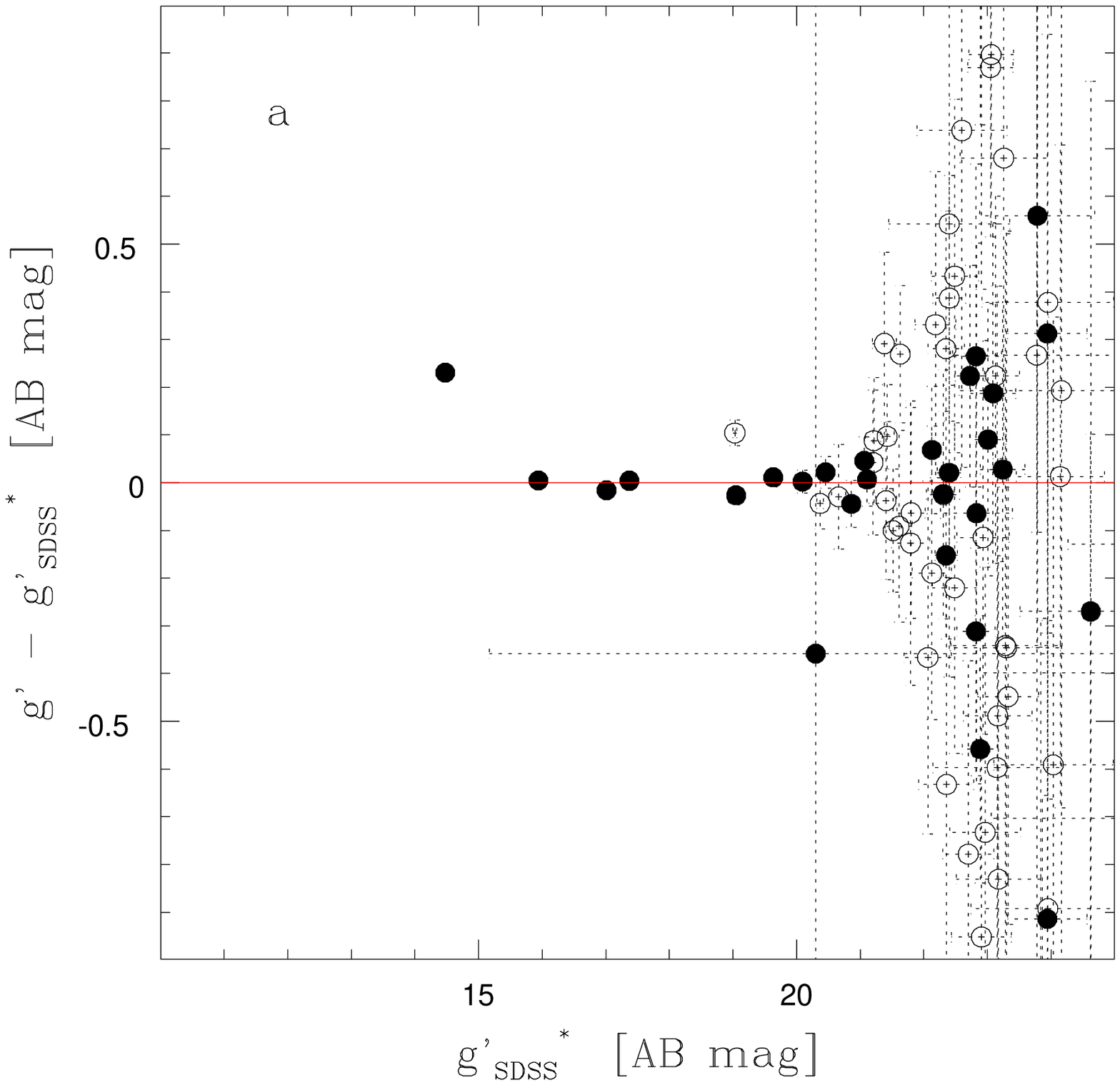}
   \vskip -0.75truecm
   \includegraphics[width=6.5cm]{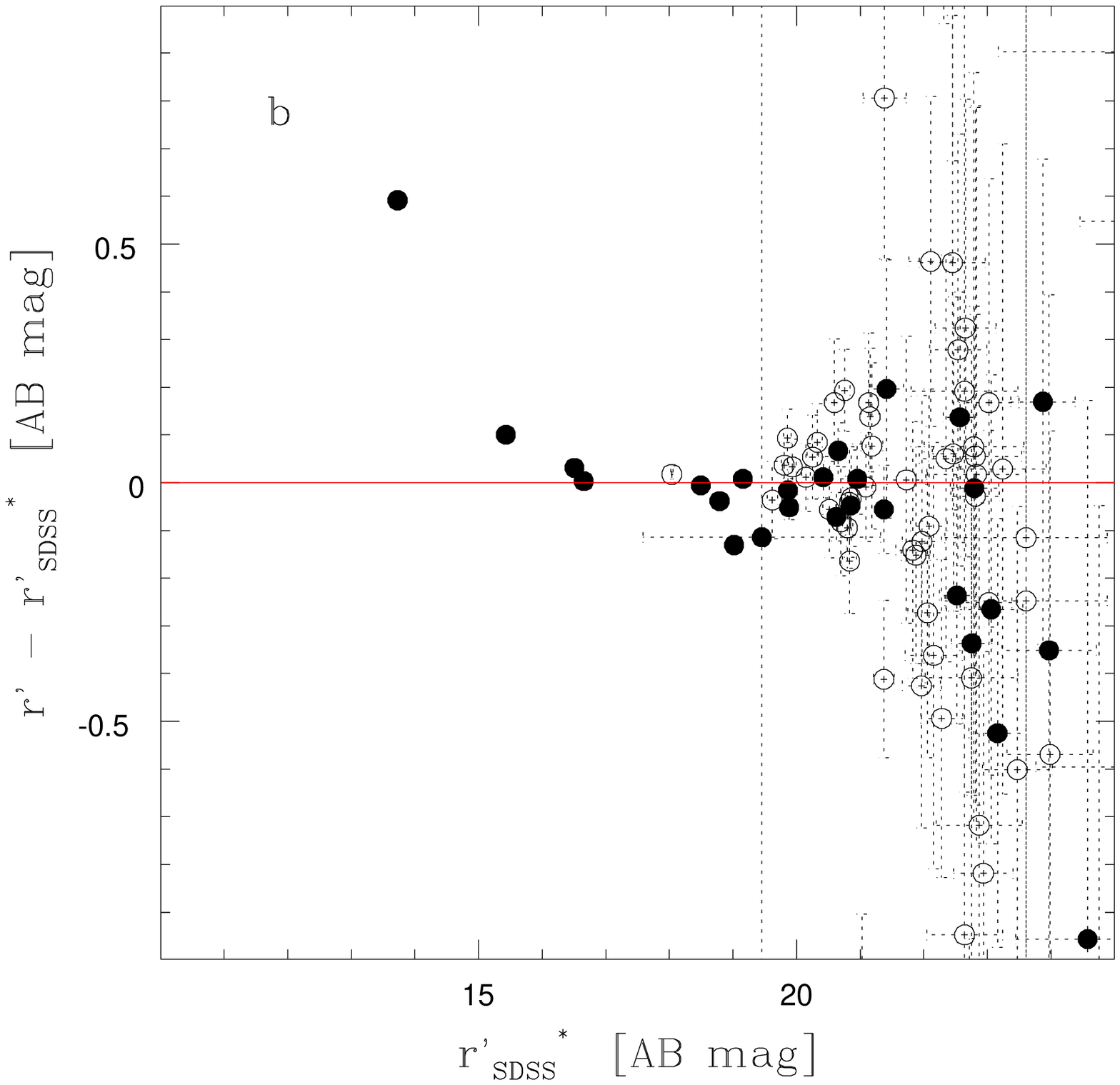}
   \vskip -0.75truecm
   \includegraphics[width=6.5cm]{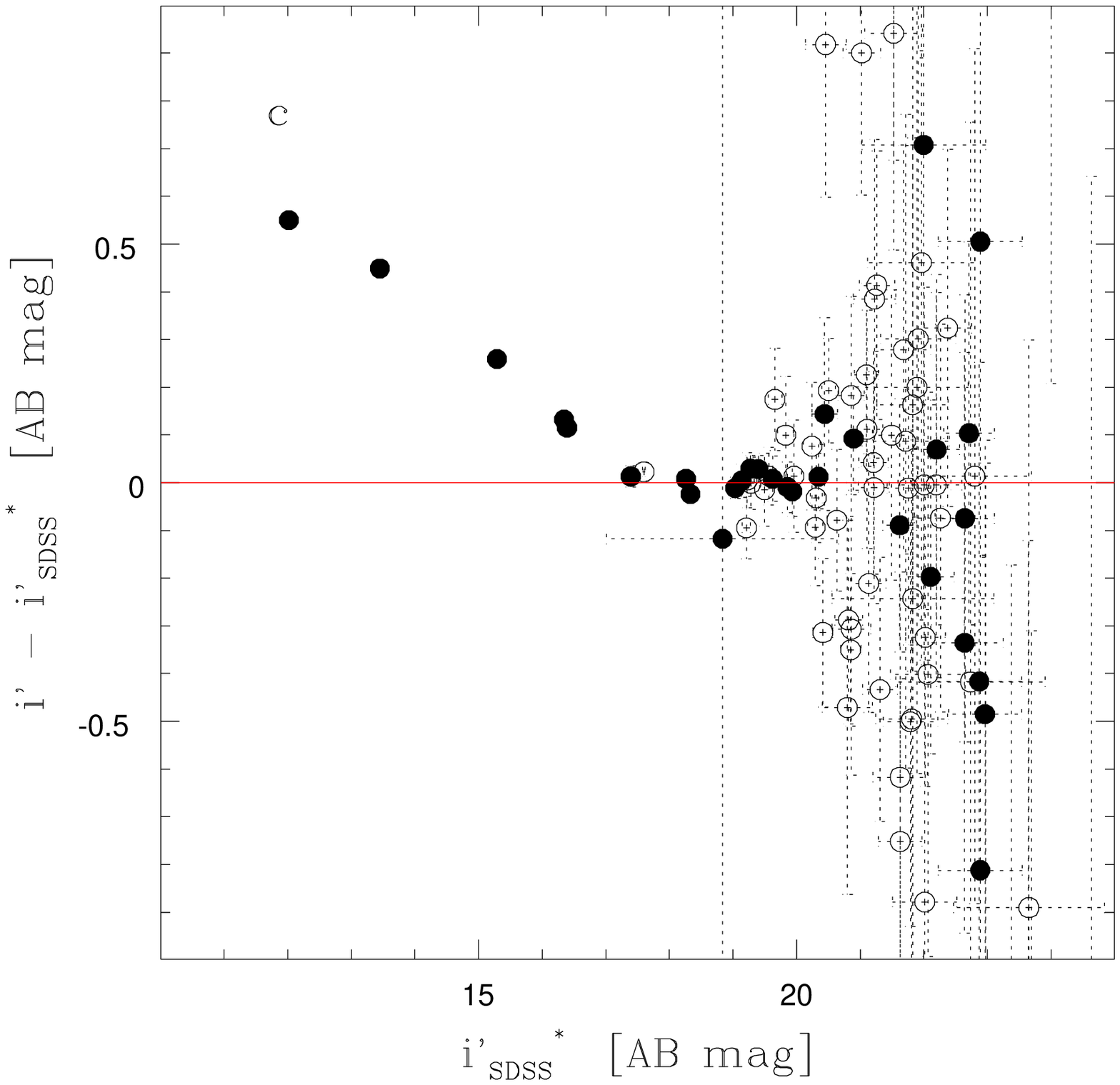}
   \vskip -0.75truecm
   \includegraphics[width=6.5cm]{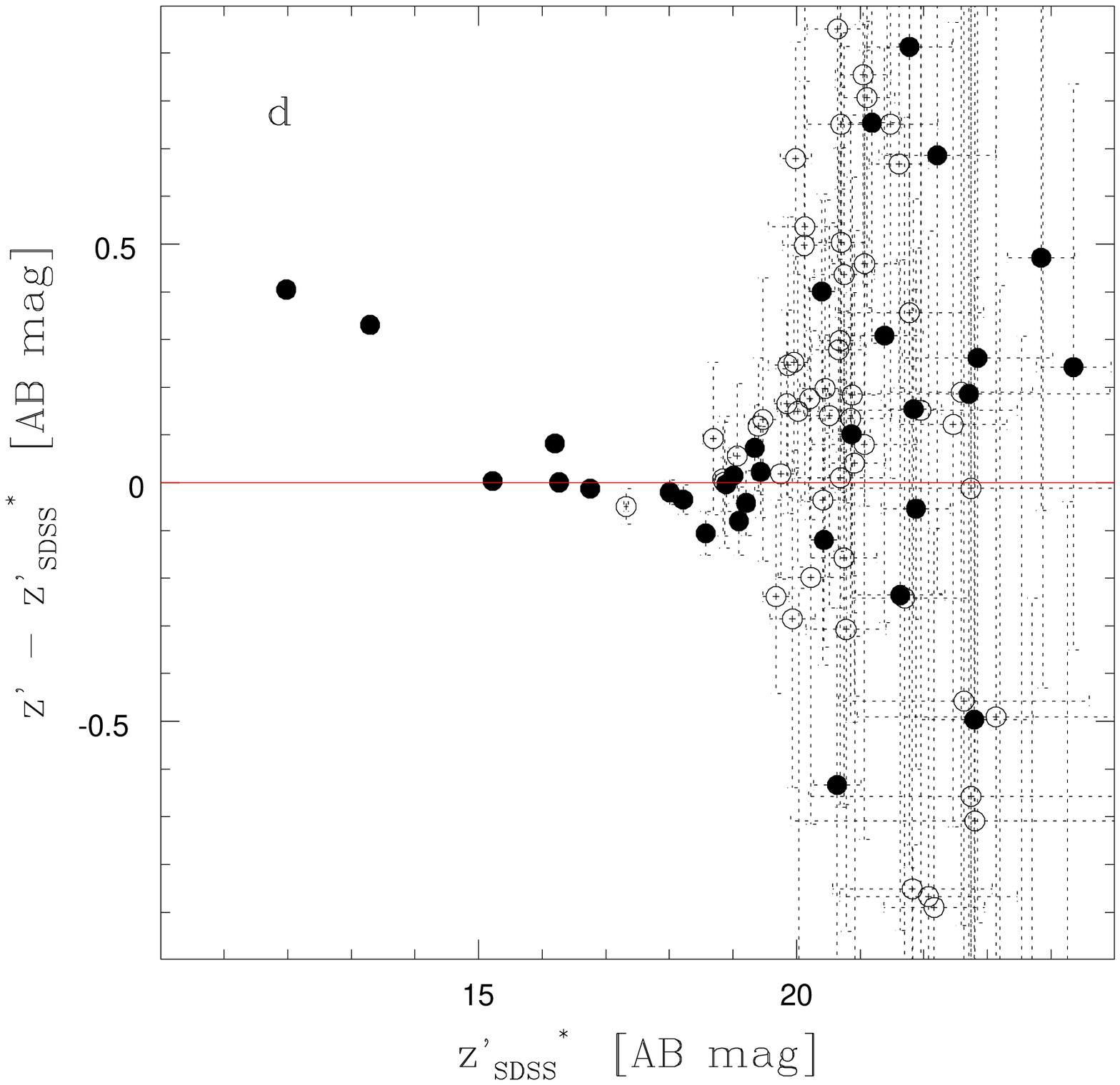}
   \vskip -0.75truecm
   \vskip 0.75truecm
   \caption{Comparison of the Petrosian magnitudes of stars (solid circles)
            and galaxies (empty circles) with GROND and SDSS photometry.
            The photometry in the GROND
            $\mathrm{g^{\prime}, r^{\prime}, i^{\prime}, z^{\prime}}$
            bands is calibrated with that of the stars in common with SDSS
            in the corresponding SDSS bands (panels a, b, c, d, respectively).
            Saturated stars and stars with poor photometry were flagged out.
            The asterisks indicate that the reference magnitudes
            are expressed in the GROND filter system.
            In each panel the red solid line represents equality.
            }
       \label{FigCalsdss}%
    \end{figure*}
%

   \begin{figure*}
   \centering
   \vskip -1.0truecm
   \includegraphics[width=6.5cm]{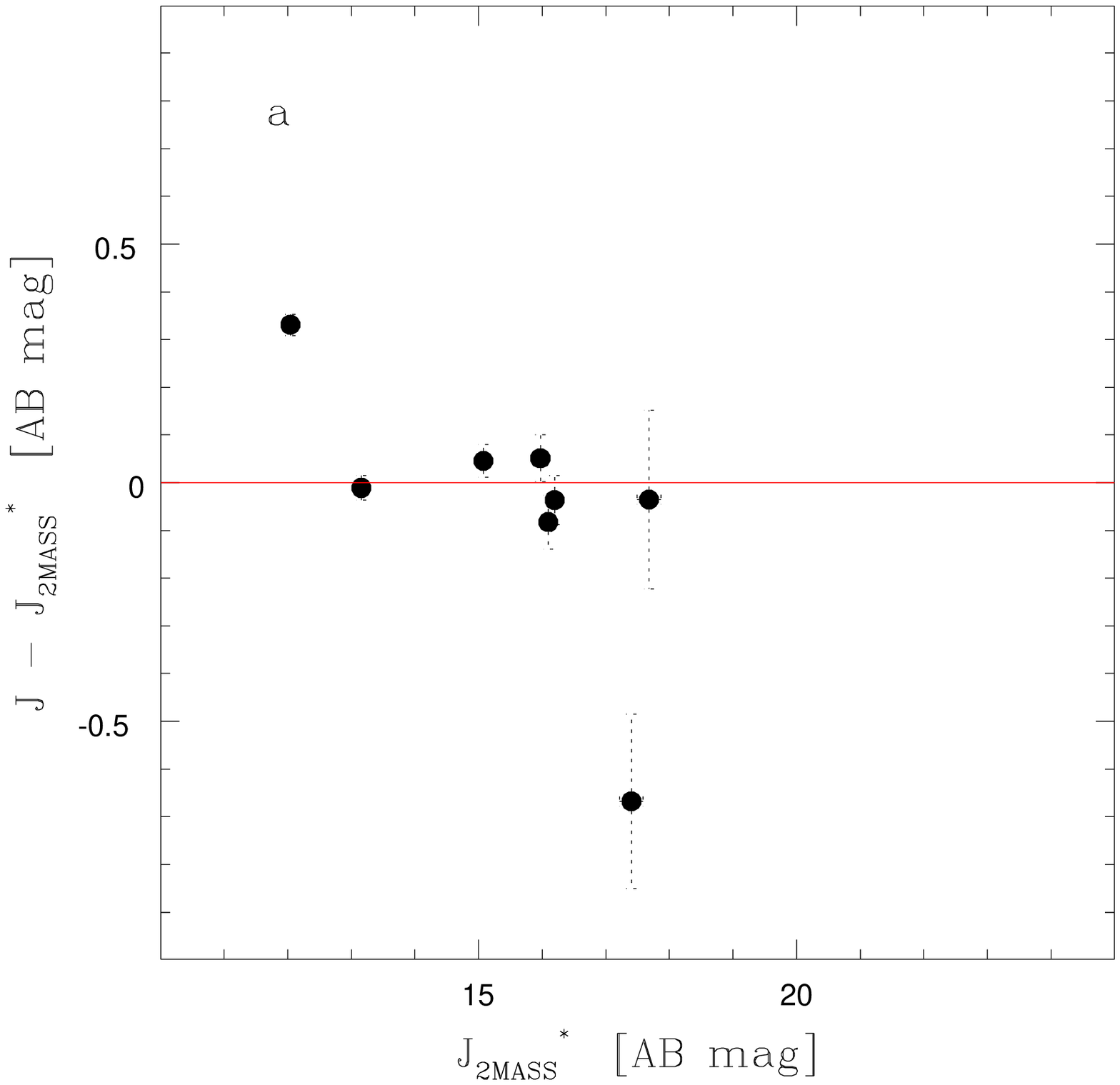}
   \vskip -0.75truecm
   \includegraphics[width=6.5cm]{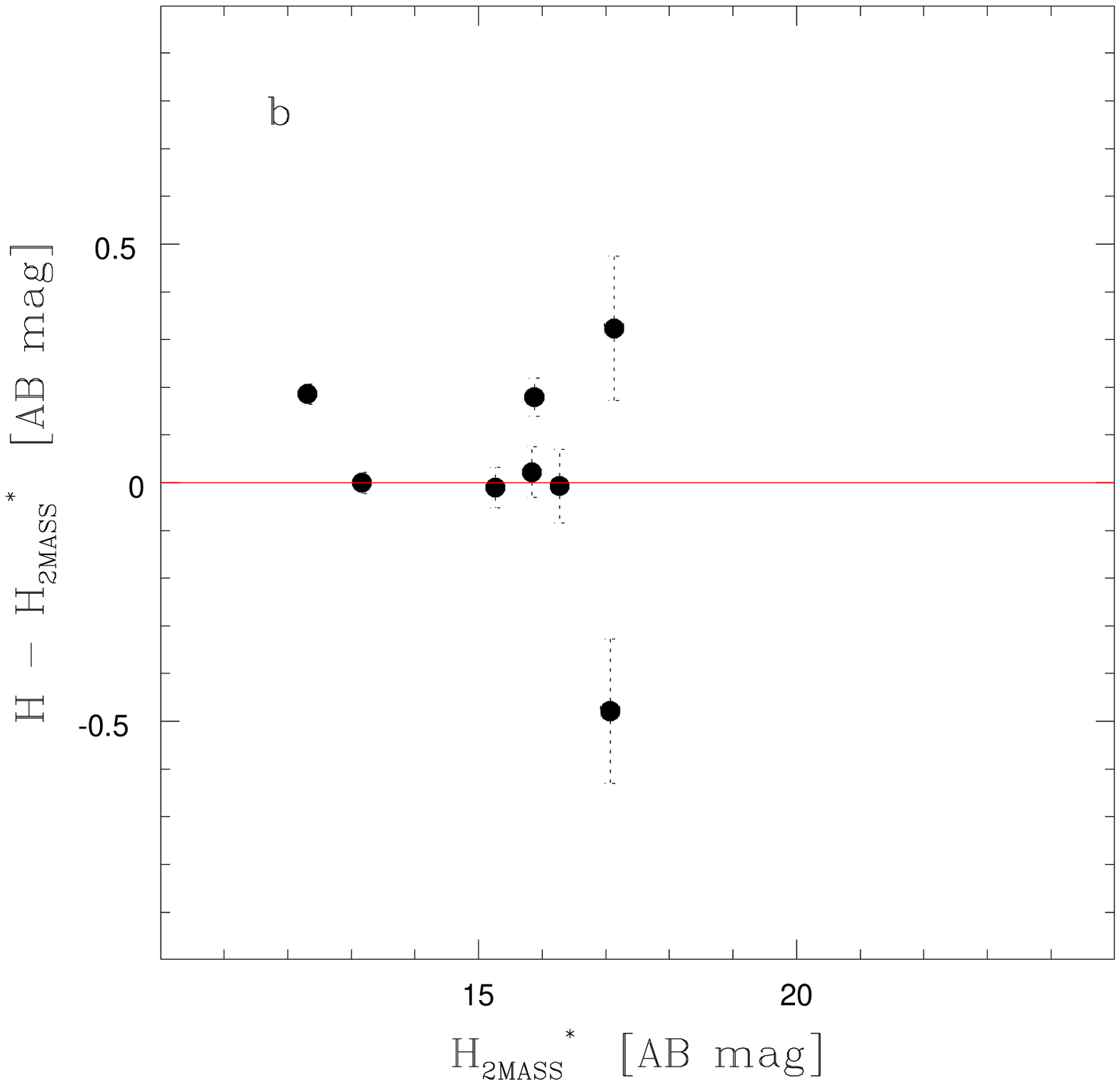}
   \vskip -0.75truecm
   \includegraphics[width=6.5cm]{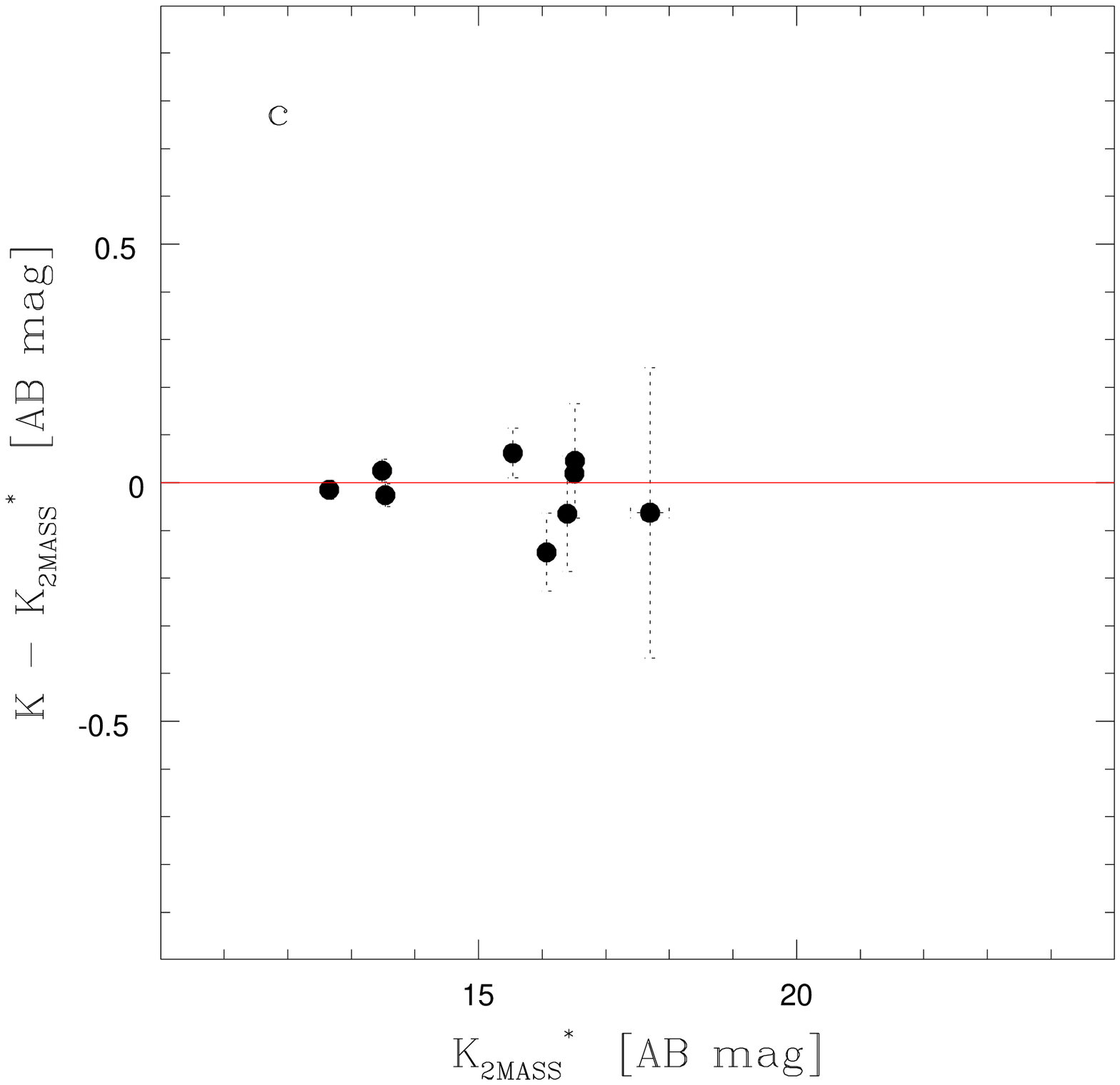}
   \caption{Comparison of the Petrosian magnitudes of stars (solid circles)
            after their photometry in the GROND $\mathrm{J, H, Ks}$ bands
            is calibrated with that in the corresponding 2MASS bands
            (panels a, b, c, respectively).
            Saturated stars and stars with poor photometry were flagged out.
            The asterisks indicate that the reference magnitudes
            are expressed in the GROND filter system.
            In each panel the red solid line represents equality.
            }
       \label{FigCal2mass}%
    \end{figure*}
%

   \begin{figure*}
   \centering
   \vskip -0.65truecm
   \includegraphics[width=6.5cm]{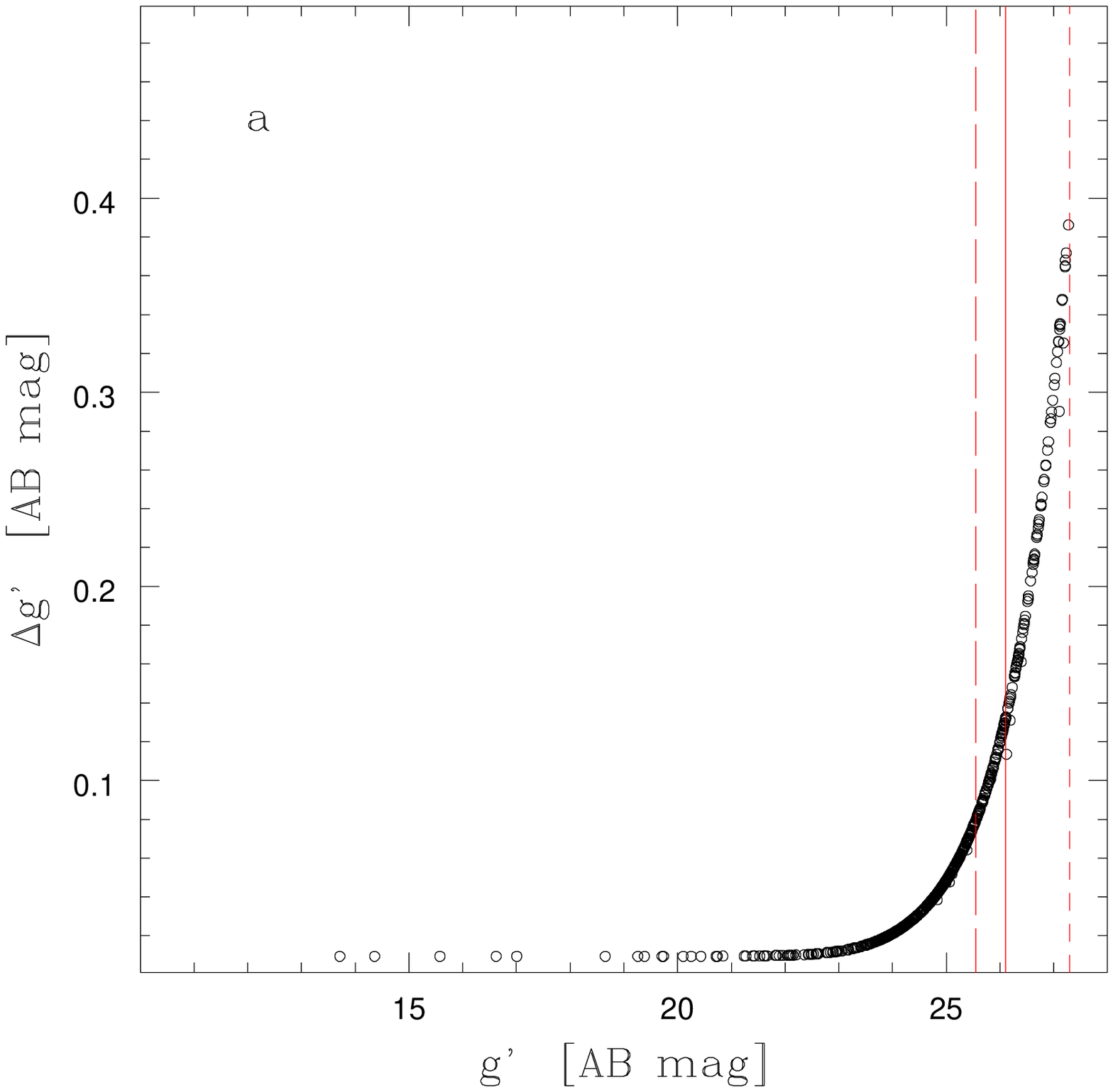}
   \includegraphics[width=6.5cm]{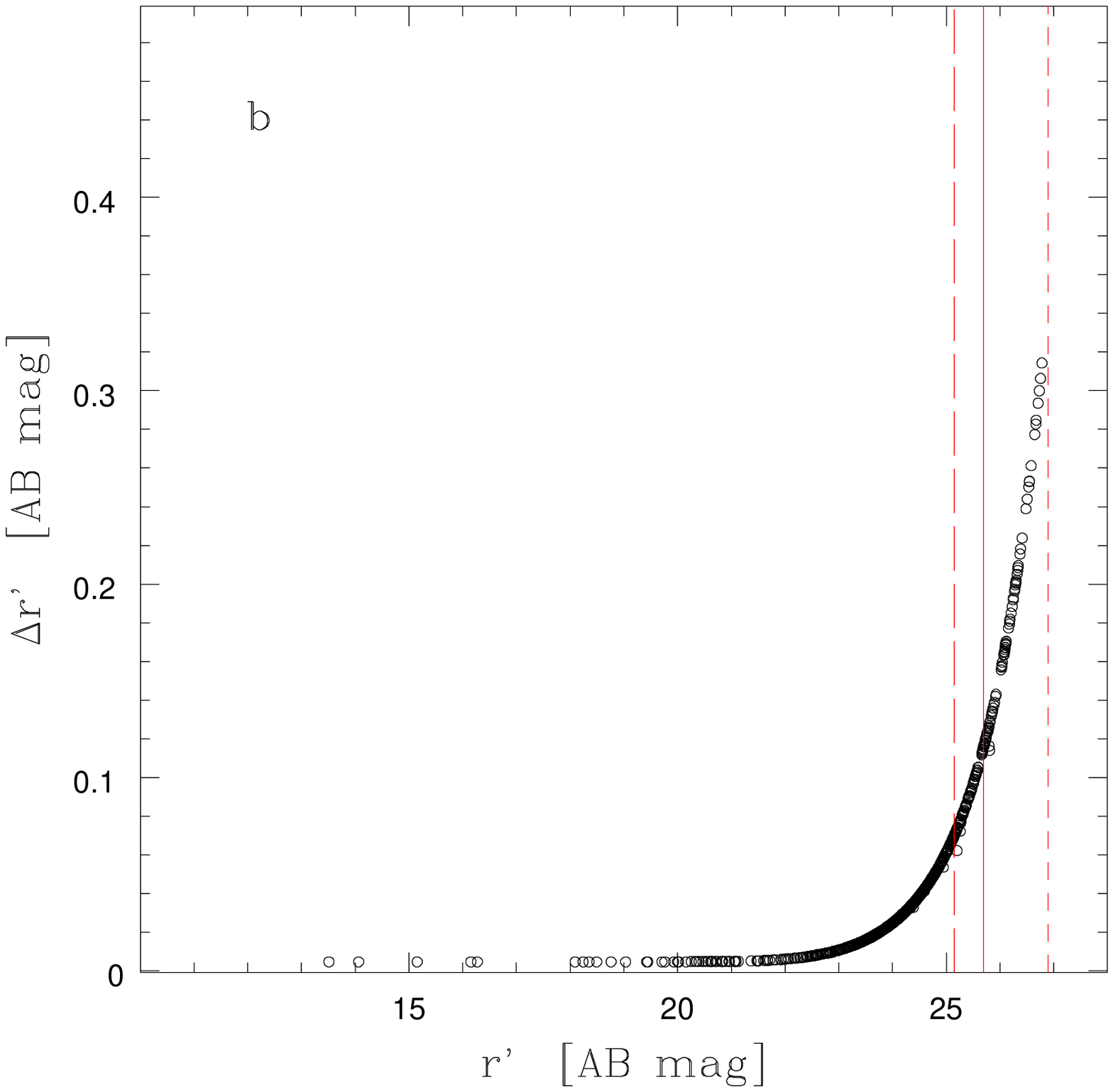}
   \vskip -0.75truecm
   \includegraphics[width=6.5cm]{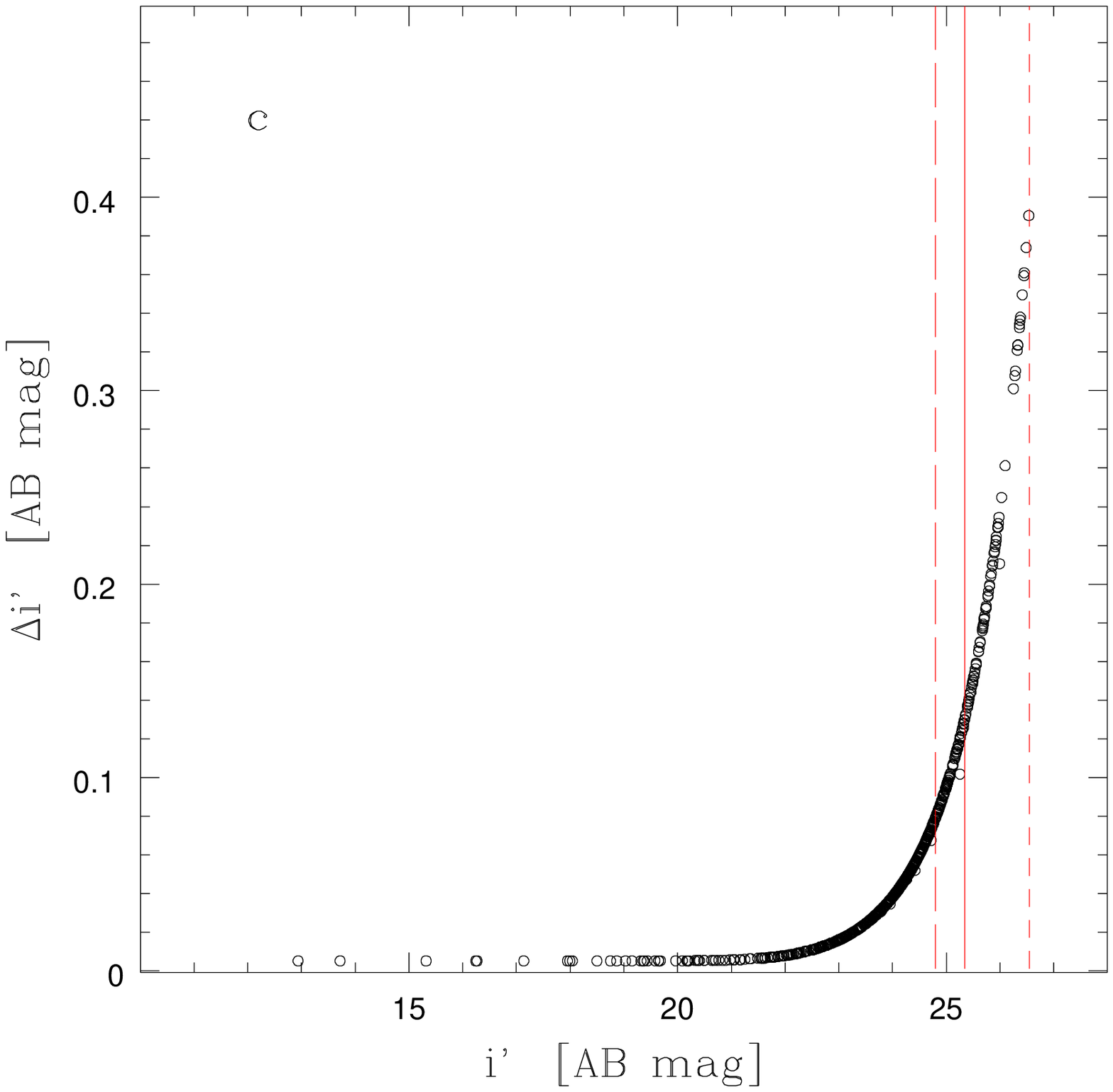}
   \includegraphics[width=6.5cm]{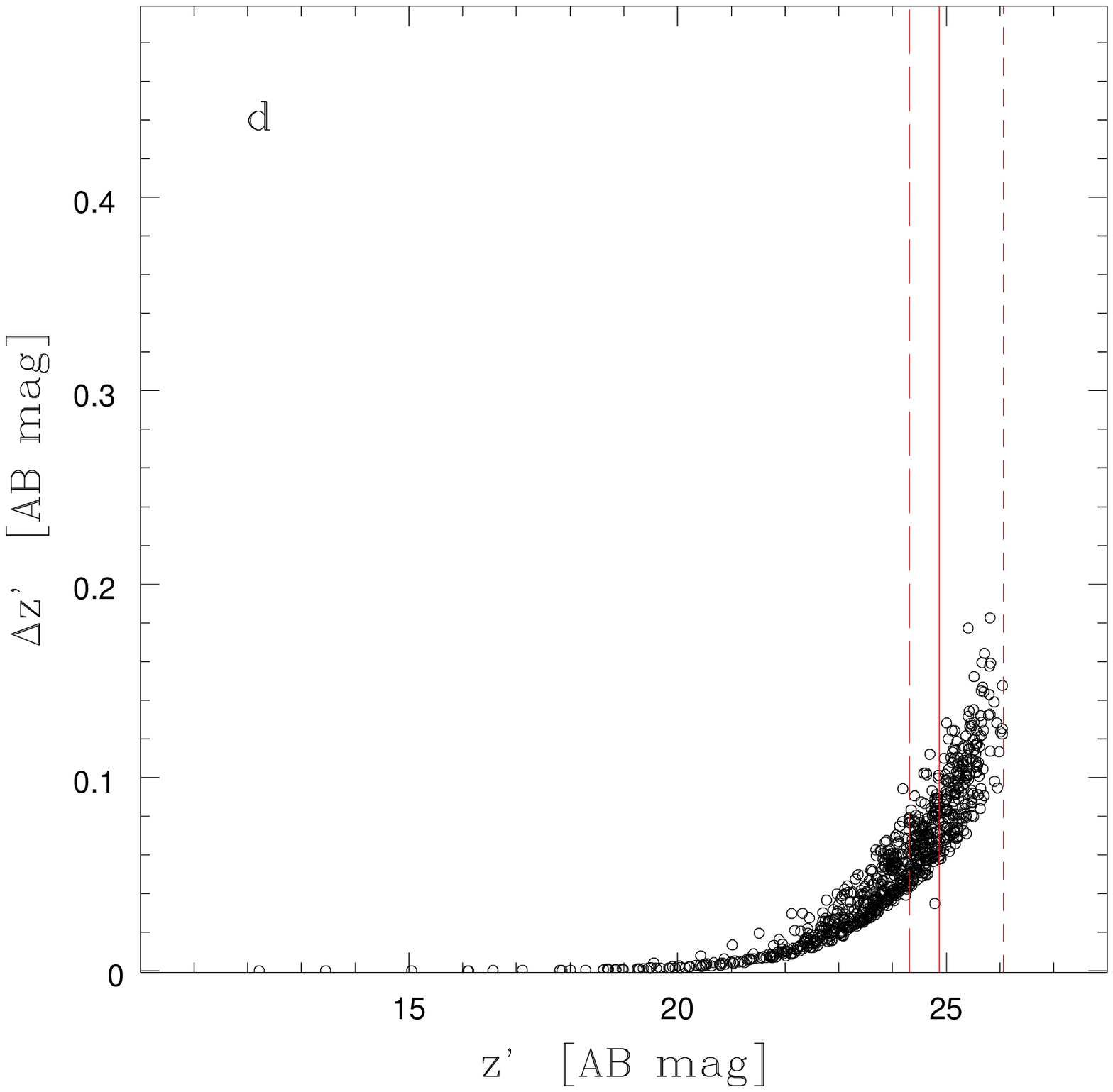}
   \vskip -0.75truecm
   \includegraphics[width=6.5cm]{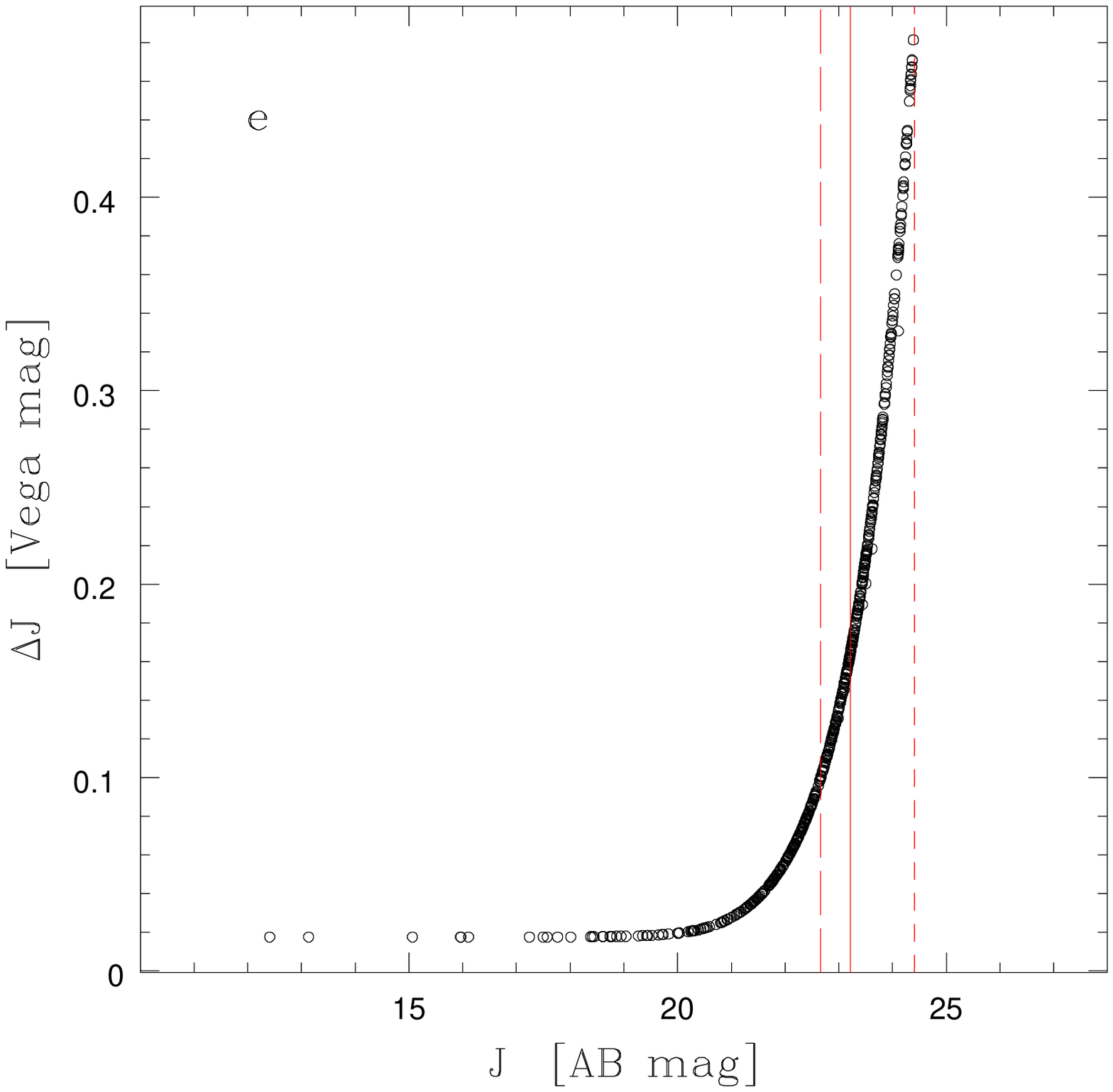}
   \includegraphics[width=6.5cm]{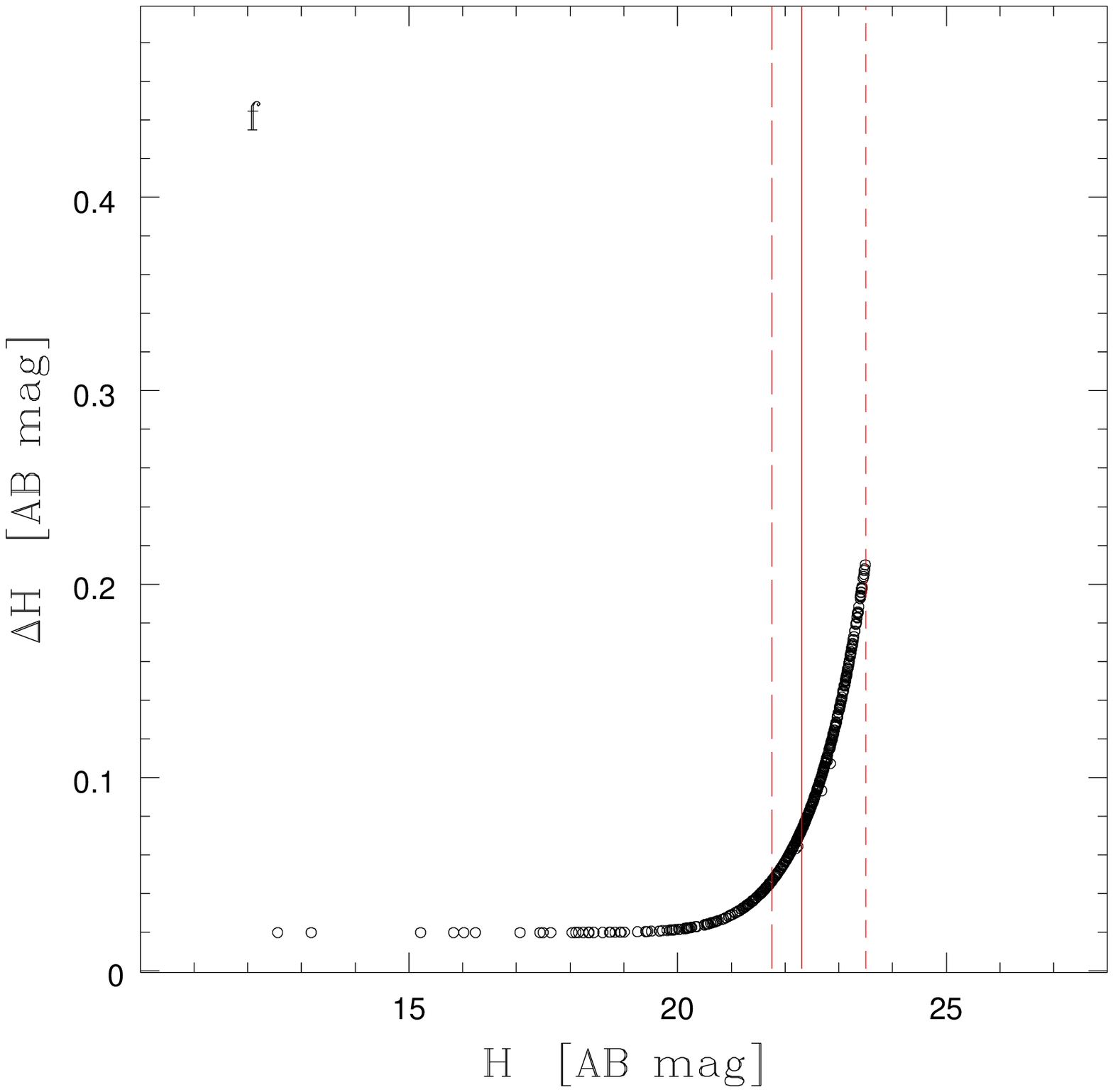}
   \vskip -0.75truecm
   \includegraphics[width=6.5cm]{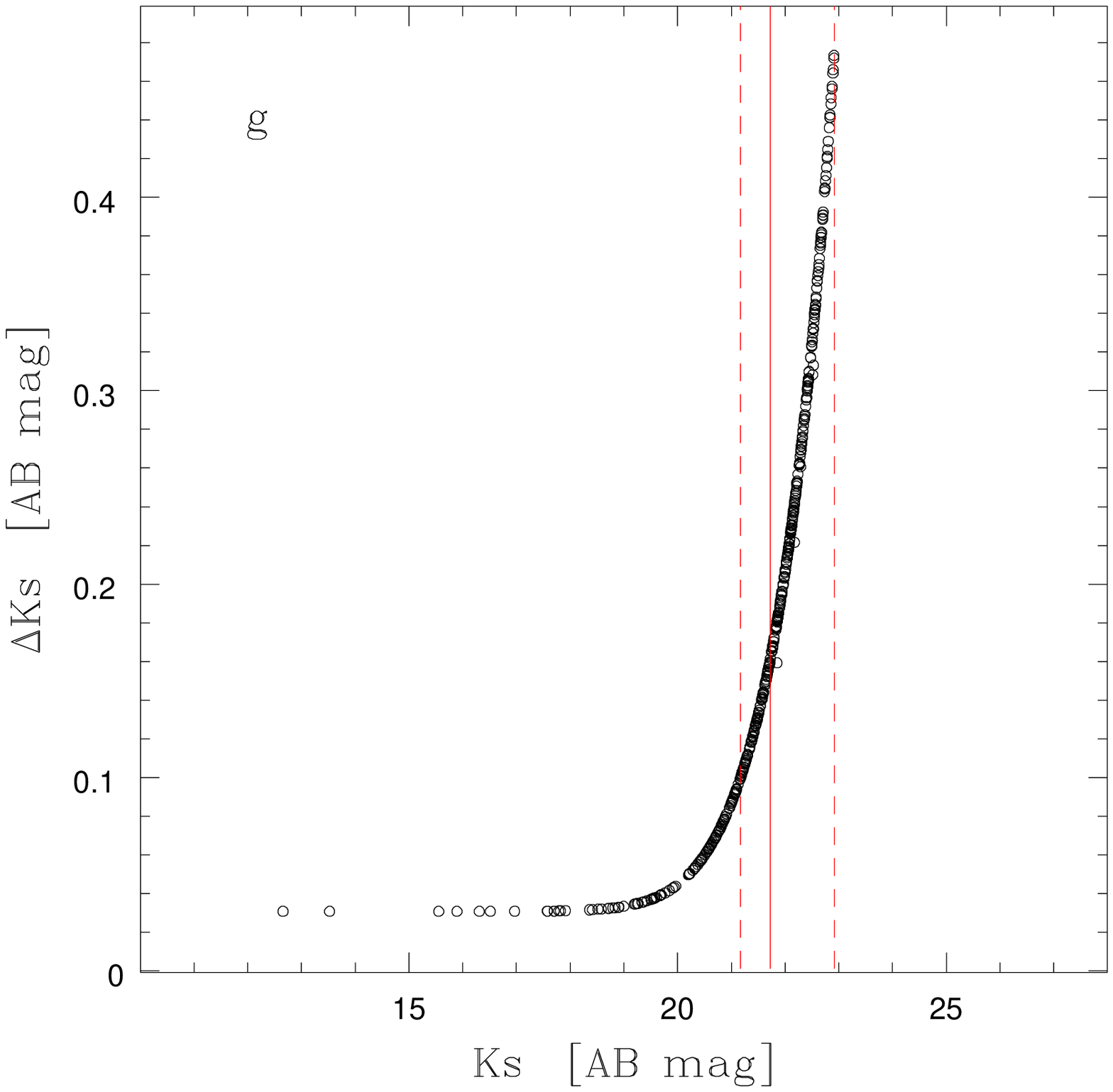}
   \caption{Uncertainty of the computed  total photometry
            in each of the
            $\mathrm{g^{\prime}, r^{\prime}, i^{\prime}, z^{\prime}, J, H, Ks}$
            broad-bands of GROND as a function of the total magnitude
            of a source (panels a, b, c, d, e, f, g, respectively).
            In each panel the short-dashed, solid and long-dashed red lines
            represent the $1 \mathrm{\sigma}$, $3 \mathrm{\sigma}$
            and $5 \mathrm{\sigma}$ limits, respectively.
            Only sources brighter than the limiting magnitude in a given band
            are reproduced.
            }
       \label{FigErrphot}%
    \end{figure*}

   For the photometric calibration
   of the $\mathrm{g^{\prime}, r^{\prime}, i^{\prime}, z^{\prime}}$
   ($\mathrm{J, H, Ks}$) images, the celestial coordinates
   of GROND sources and SDSS (2MASS) catalogue stars
   were first cross-correlated using a search radius of $3^{\prime \prime}$.
   ``Machine'' Petrosian magnitudes (corrected only for atmospheric absorption)
   and catalogue Petrosian magnitudes (not corrected for Galactic extinction)
   of non-saturated sources that are unambiguously identified as stars
   and have robust photometry were contrasted after taking into account
   eventual colour terms (see footnote 5).
   For a given band, the weighted mean of the magnitude differences
   was taken as the correction to the assumed zeropoint
   and the value of the weighted standard deviation
   as the uncertainty of the photometric zeropoint.
   These uncertainties are equal to $\pm 0.009$, $\pm 0.010$, $\pm 0.011$,
   $\pm 0.011$, $\pm 0.041$, $\pm 0.009$ and $\pm 0.036~\mathrm{mag}$
   in the $\mathrm{g^{\prime}, r^{\prime}, i^{\prime}, z^{\prime}, J, H, Ks}$
   channels of GROND, respectively.
   They reflect the quality of the reference photometry
   and the poor statistics of the matched stars used as calibrators.
   The good agreement between our set of calibrated magnitudes
   and that of SDSS (2MASS) for all the 38 (9) stars in common
   is shown in Fig.~5 (6).

   Our calibrated Petrosian magnitudes are compared to the SDSS ones
   (all not corrected for Galactic extinction)
   for the 65 sources in our photometric catalogue
   that are associated with galaxies identified in DR7
   using a search radius of $3^{\prime \prime}$.
   In addition, twenty-four of these 65 SDSS galaxies
   have a photometric redshift in DR7.
   The overall agreement between the two sets of Petrosian magnitudes
   is satisfying (Fig.~5).
   Unfortunately, no galaxy counterparts to our sources seem to exist
   in any 2MASS catalogue, according to NED.
   This is no surprise given the shallow limiting magnitudes of 2MASS
   (see Jarrett et al. 2000; Skrutskie et al. 2006).

   A calibrated photometric catalogue with 832 sources was built
   as previously described, where magnitudes are corrected
   for atmospheric extinction and Galactic extinction.
   Limiting magnitudes ($1 \mathrm{\sigma}$ values) were computed
   as the median values of magnitudes obtained in $3^{\prime \prime}$ apertures
   on 10 separated regions of blank sky distributed across the individual
   $\mathrm{g^{\prime}, r^{\prime}, i^{\prime}, z^{\prime}, J, H, Ks}$ images.
   They correspond to $27.30$, $26.89$, $26.54$, $26.06$, $24.40$,
   $23.50$ and $22.91~\mathrm{mag}$, respectively.
   We note that the corresponding $5 \sigma$ values
   in the $\mathrm{z^{\prime}, J, H, Ks}$ channels are about the depths
   in analogous bands that the ESO public survey VIKING
   ({\em VISTA Kilo-Degree Infrared Galaxy Survey}; P.I.: W. Sutherland)
   is expected to reach (see e.g., Arnaboldi et al. 2007).

   Corrections to the {\em SExtractor} output photometric errors
   for a given band (in part caused by correlated noise in the stacked images)
   were computed from the ratio of the median value
   of the uncertainties of the previous 10 aperture magnitudes
   and the median {\em SExtractor} uncertainty
   of the $3^{\prime \prime}$ aperture photometry of sources
   with magnitudes within $\pm 1 \mathrm{\sigma}$ from the limiting magnitude.
   These correction factors amount to 1.74, 1.67, 1.82, 1.28, 2.15, 1.36,
   and 1.78 for
   the $\mathrm{g^{\prime}, r^{\prime}, i^{\prime}, z^{\prime}, J, H, Ks}$
   bands, respectively.

   Extended and point-like sources were initially identified on the basis of
   surface brightness distribution
   (i.e., the {\em SExtractor} parameters stellarity index and flux radius)
   and apparent brightness.
   The photometry of point-like sources enabled us to estimate corrections
   from the $3^{\prime \prime}$ aperture magnitudes
   to magnitudes in the $\mathrm{z^{\prime}}$-band Kron aperture for all bands.
   These aperture corrections (in the standard {\em SExtractor} format)
   amount to
   \begin{displaymath}
      g^{\prime}    = g^{\prime}(3^{\prime \prime}) - 0.1143 (\pm 0.0094)\, , \;
   \end{displaymath}
   \begin{displaymath}
      r^{\prime}    = r^{\prime}(3^{\prime \prime}) - 0.1366 (\pm 0.0046)\, , \;
   \end{displaymath}
   \begin{displaymath}
      i^{\prime}    = i^{\prime}(3^{\prime \prime}) - 0.1219 (\pm 0.0051)\, , \;
   \end{displaymath}
   \begin{displaymath}
      z^{\prime}    = z^{\prime}(3^{\prime \prime}) - 0.1254 (\pm 0.0040)\, , \;
   \end{displaymath}
   \begin{displaymath}
      J             = J(3^{\prime \prime}) - 0.1613 (\pm 0.0175)\, , \;
   \end{displaymath}
   \begin{displaymath}
      H             = H(3^{\prime \prime}) - 0.1510 (\pm 0.0197)\, , \;
   \end{displaymath}
   \begin{displaymath}
      Ks            = Ks(3^{\prime \prime}) - 0.1463 (\pm 0.0307).
   \end{displaymath}

   We have then adopted two operational definitions of total magnitudes
   and investigated their performances given the characteristics
   of the available imaging.
   As a first prescription (method {\em A}),
   we adopted the {\em Kron} magnitude
   in the $\mathrm{z^{\prime}}$ band as the reference total magnitude
   and computed total magnitudes
   in the $\mathrm{g^{\prime}, r^{\prime}, i^{\prime}, J, H, Ks}$ bands
   from the reference total magnitude plus the corresponding colour term
   measured within a $3^{\prime \prime}$ aperture
   and corrected to the {\em Kron} aperture.
   As a second prescription (method {\em B}),
   we adopted the {\em Kron} magnitude as the total magnitude
   in the $\mathrm{z^{\prime}}$ band but computed total magnitudes
   in the $\mathrm{g^{\prime}, r^{\prime}, i^{\prime}, J, H, Ks}$ bands
   from the corresponding $3^{\prime \prime}$ aperture magnitudes,
   corrected to the {\em Kron} aperture.
   This second set of total magnitudes yielded photometric redshifts
   that are better consistent with the spectroscopic ones
   for the three confirmed members of XMMU\,J0338.7$+$0030
   with GROND photometry (Sect.~3.2.1).
   At the same time the distribution of these three spectroscopic members
   in the $i^{\prime} - z^{\prime}$ vs $z^{\prime}$
   colour--magnitude diagram is much better consistent with the red-sequence
   of the cluster RDCS\,J0910$+$5422 at $z=1.106$ (Mei et al. 2006a),
   as discussed in Sect.~3.2.2.
   For these reasons, we will make use
   of the {\em aperture-corrected} magnitudes
   in the $\mathrm{g^{\prime}, r^{\prime}, i^{\prime}, J, H, Ks}$ bands
   and the {\em Kron} magnitude in the $\mathrm{z^{\prime}}$ band
   for all sources (method {\em B}), and refer to them
   as {\em total} magnitudes hereafter.

   \begin{figure}
   \centering
      \vskip -1.0truecm
      \includegraphics[width=9cm]{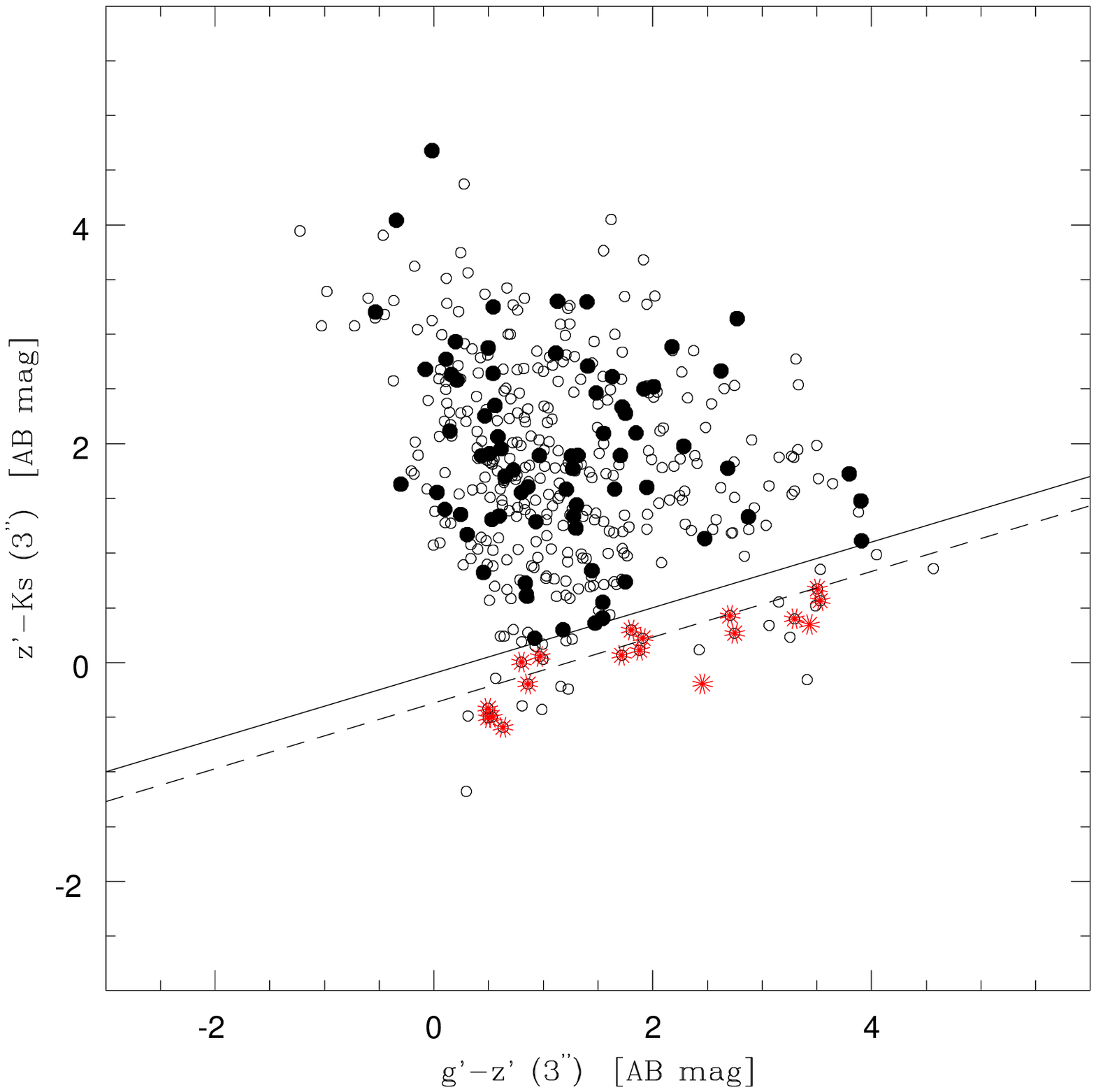}
      \vskip -0.15truecm
      \caption{Distribution of all sources with fluxes
               in the GROND $\mathrm{g^{\prime}, z^{\prime}, Ks}$ channels
               above the respective $1 \mathrm{\sigma}$ thresholds
               in the $z^{\prime} - Ks$ vs $g^{\prime} - z^{\prime}$
               colour--colour diagram.
               Magnitudes within a $3^{\prime \prime}$-wide circular aperture
               are used.
               Sources identified as point-like on the basis
               of a {\em SExtractor} stellarity index larger than 0.98
               are marked with red asterisks.
               Conversely, extended sources within (beyond) $1^{\prime}$
               from the position of XMMU\,J0338.7$+$0030
               are marked with filled (empty) circles.
               The short-dashed and solid lines represent, respectively,
               the computed and adopted thresholds to select bona-fide stars
               (see text).
              }
         \label{Figstargalsep}
   \end{figure}
%

   \begin{figure}
   \centering
      \vskip -1.0truecm
      \includegraphics[width=9cm]{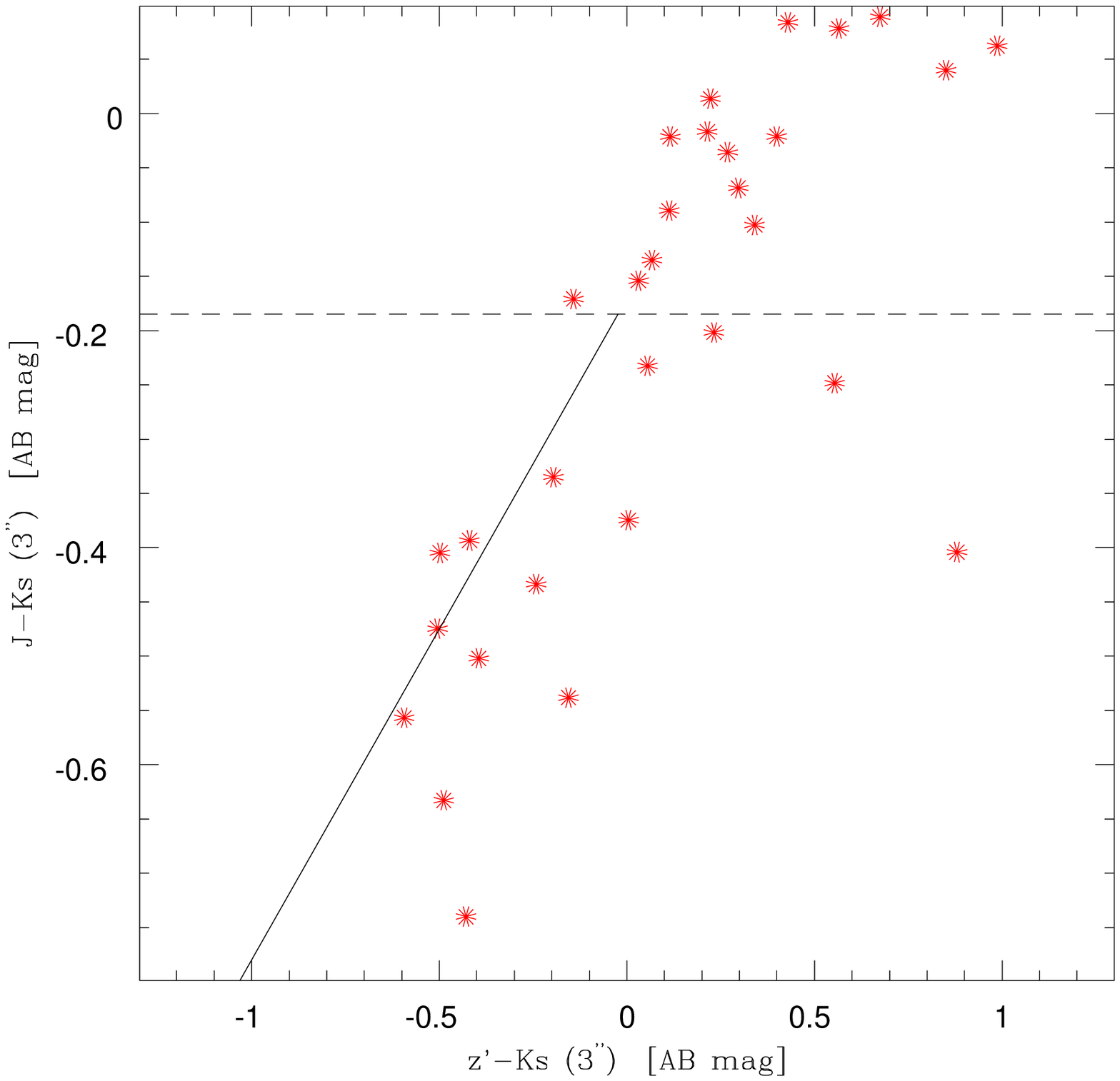}
      \vskip -0.15truecm
      \caption{Distribution of all bona-fide stars,
               selected according to Eq.~1,
               in the $J - Ks$ vs $z^{\prime} - Ks$ colour--colour diagram,
               where $3^{\prime \prime}$-aperture photometry is used.
               This distribution is consistent with the loci occupied
               by spectrally classified stars with SDSS and 2MASS photometry
               (cf. fig. 1 in Galametz et al. \cite{galametz09}).
               As shown by Galametz et al., stars with a spectral type
               later than K5 populate the region above the short-dashed line
               and tend to exhibit $z^{\prime} - Ks > 0$.
               Conversely, stars with a spectral type earlier than K5
               populate the region blueward of the horizontal short-dashed line
               and define a linear colour--colour relation (solid line)
               with extensions towards very red $z^{\prime} - Ks$ colours.
              }
         \label{Figstartest}
   \end{figure}

   Figure 7 illustrates the relation between the computed total magnitude
   and its uncertainty in each of the GROND
   $\mathrm{g^{\prime}, r^{\prime}, i^{\prime}, z^{\prime}, J, H, Ks}$
   channels.
   The regular behaviour observed for the GROND
   $\mathrm{g^{\prime}, r^{\prime}, i^{\prime}, J, H, Ks}$ broad bands
   is due to the fact that two sources hardly exhibit exactly
   the same measured value for their aperture fluxes.
   For the GROND
   $\mathrm{g^{\prime}, r^{\prime}, i^{\prime}, J, H, Ks}$ channels
   the uncertainty on the total magnitude is dominated by the uncertanty
   on the aperture correction at bright magnitudes
   and by the measurement error at faint magnitudes.
   The transition between these two regimes corresponds to a total magnitude
   that depends on the filter.

\subsubsection{The photometric catalogue: star/galaxy separation}

   As an alternative to the previous selection criteria,
   galaxies (extended sources) can be separated from bona-fide stars
   (point-like sources) on the basis of their $g^{\prime} - z^{\prime}$
   and $z^{\prime} - Ks$ colours, in analogy with the $B z K$ criterion
   (Daddi et al. \cite{daddi04}; cf. e.g. Galametz et al. \cite{galametz09}).
   We adopted the formulation of the colour criterion
   given by Galametz et al. (their fig. 5)
   and applied the empirical colour transformation
   between the SDSS photometry and the Johnson--Cousins photometric system
   of Jordi, Grebel \& Ammon (\cite{jordi06})
   \begin{displaymath}
     g - B = 0.370 \times (B - V) - 0.124
   \end{displaymath}
   plus suitable corrections for the different filter transmission functions
   and magnitude systems.
   The ensuing threshold corresponding
   to the GROND $g^{\prime} z^{\prime} Ks$ criterion
   (short-dashed line in Fig.~8) was contrasted against the behaviour
   of our catalogue sources classified as stars
   on the basis of the {\em SExtractor} stellarity index in Sect.~2.3.2.
   This comparison prompted the introduction of a shift
   equal to $+ 0.27~\mathrm{mag}$ in $z^{\prime} - Ks$ (solid line in Fig.~8).
   Therefore, the $g^{\prime} z^{\prime} Ks$ criterion
   hereafter adopted to single out bona-fide stars is
   \begin{equation}
     z^{\prime} - Ks~(3^{\prime \prime}) \le 0.3 \times [g^{\prime} - z^{\prime}~(3^{\prime \prime})] -0.1,
   \end{equation}
   where we adopted $3^{\prime \prime}$-aperture measurements.
   Comparison with the source classification provided by the use
   of the photometric redshift technique (Fig.~13a in Sect.~3.2.1)
   reveals the presence of a few extragalactic point-like sources
   with $g^{\prime} - z^{\prime}~(3^{\prime \prime}) \sim 1$
   and $z^{\prime} - Ks~(3^{\prime \prime}) \sim 0$
   that are misclassified as stars in Fig.~8.

   As an independent test of the overall goodness of the previous criterion,
   in Fig.~9, we compared the distribution of the bona-fide stars
   identified through Eq.~1
   in the $J - Ks$ vs $z^{\prime} - Ks$ colour--colour diagram
   with the loci occupied by spectrally classified stars with SDSS and 2MASS
   photometry (Galametz et al. \cite{galametz09}, their fig. 1)
   after introducing suitable corrections
   for the different filter transmission functions.
   As discussed by Galametz et al., most of the stars
   in a matched 2MASS/SDSS catalogue populate either the region
   of the $J - Ks$ vs $z^{\prime} - Ks$ colour--colour diagram
   above the short-dashed line and tend to be redder
   than $z^{\prime} - Ks = 0$ if their individual spectral types
   are later than K5, or the region below the short-dashed line
   and across/to the right of the solid line if they are earlier than K5.
   The consistent behaviour we find confirms the validity of Eq.~1.

\subsection{Optical spectroscopy with FORS2}

\subsubsection{Observations and data reduction}

   The SDSS DR7 archive contains no spectroscopy
   for the region of the sky encompassing XMMU\,J0338.7$+$0030.
   However, sparse spectroscopic information is available from XDCP,
   since the neighboring distant cluster XMMU\,J0338.5$+$0029
   (Fassbender et al. 2011c) was observed on November 7, 2007
   (programme ID 079.A$-$0634, service mode) for $3~\mathrm{hr}$
   with FORS2 (Appenzeller et al. 1998).
   This is a multi-mode (imaging, polarimetry, long slit, MOS) instrument
   working at wavelengths of 3300--$11000~\mathrm{\AA}$,
   mounted on the Cassegrain focus of VLT UT1 in Paranal, Chile.

   A single mask with 32 slits of $1^{\prime \prime}$-width each
   was designed and used in the FORS2 MXU-mode (Mask eXchange Unit)
   with the standard resolution collimator,
   which yields a $6.8^{\prime} \times 6.8^{\prime}$ FoV
   with a scale of $0.25^{\prime \prime}$/pixel.
   Coupled to the 300I grism and the OG590 order blocking filter,
   this provided a resolution $\mathrm{R} = 660$
   and a wavelength coverage of $6000$--$10000~\mathrm{\AA}$.
   Six single exposures were taken during the night for the same set-up,
   yielding a total net integration time of $8400~\mathrm{s}$;
   the seeing ranged between $1.2^{\prime \prime}$ and $1.3^{\prime \prime}$
   (FWHM).
   Because XMMU\,J0338.7$+$0030 was the secondary target
   of these FORS2 observations, only 30\% of the slits were used to probe
   galaxies in its region.
   As primary and secondary spectroscopic targets were chosen, respectively,
   very red and bluer objects in the $z - H$ vs $H$ colour--magnitude diagram
   built out of the OMEGA2000 photometry (Fig.~3).
   As a result, only five slits out of a total of 32 turned out to correspond
   to objects in the region of XMMU\,J0338.7$+$0030 later imaged with GROND,
   for which photometry was extracted.

   A new version of the {\em VIMOS Interactive Pipeline
   and Graphical Interface} ({\em VIPGI}; Scodeggio et al. 2005),
   modified to make it usable on FORS2 multi-slit spectra, was used.
   A detailed discussion of this new multi-spectra reduction pipeline
   is deferred to a forthcoming paper (Nastasi et al. in preparation).
   In brief, the stacked 2-D raw data were flat-fielded, bias-subtracted
   and calibrated in wavelength by means of a helium--argon lamp
   used as a reference line spectrum.
   The final dispersion solutions had a root-mean-squares
   calibration uncertainty lower than $1~\mathrm{\AA}$/pixel.
   The continuum of each spectrum was corrected
   for the sensitivity function of the FORS2 instrument,
   and the 1-D reduced spectrum was extracted from each slit
   for the redshift determination.
   Each 1-D reduced spectrum was both visually inspected
   by means of the graphical tools available in {\em VIPGI}
   and cross-correlated with template galaxy spectra
   using the {\em IRAF} package {\em RVSAO} (Kurtz \& Mink 1998).

\subsubsection{The spectroscopic redshift of XMMU\,J0338.7$+$0030}

   \begin{figure}
   \centering
      \includegraphics[width=9cm]{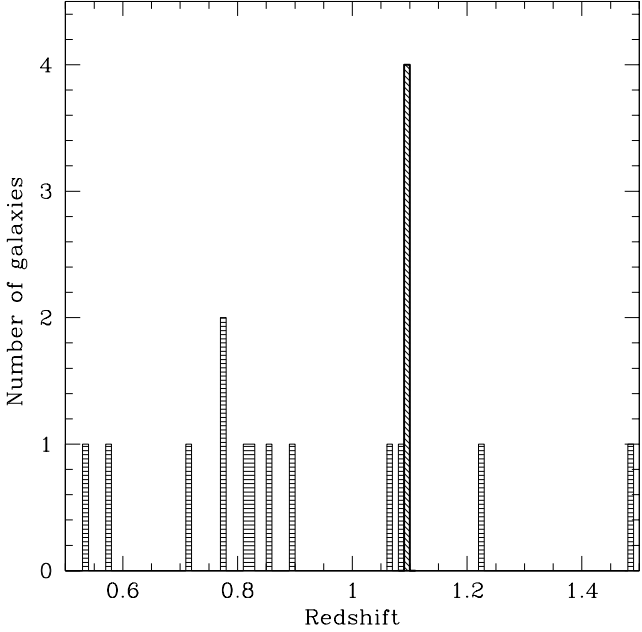}
      \caption{Histogram of the measured galaxy spectroscopic redshifts
               in bins of $\mathrm{\Delta} z = 0.01$.
               An overdensity at $z = 1.1$ is evident.
               Only five of these redshifts belong to objects
               in the GROND photometric catalogue.
              }
         \label{FigNspecz}
   \end{figure}
%

   \begin{figure}
   \centering
      \includegraphics[width=9cm]{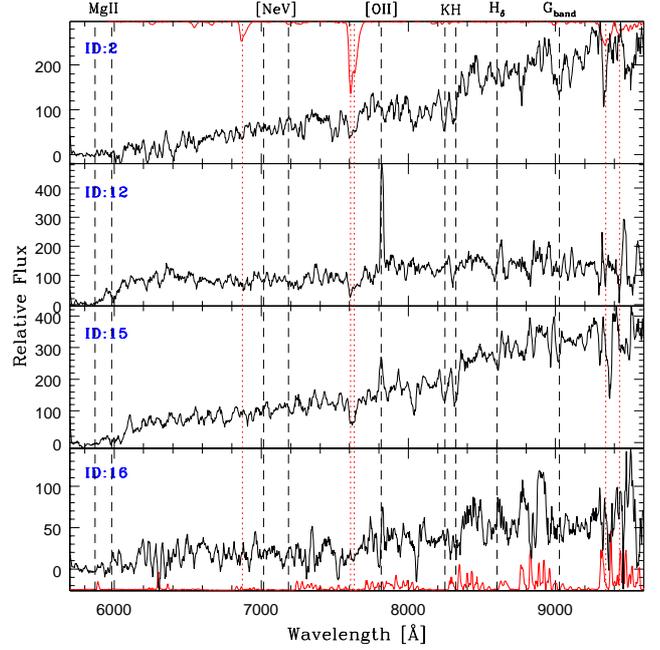}
      \caption{Spectra of the four cluster members
               smoothed with a seven-pixel boxcar filter.
               Black, dashed lines indicate the expected positions
               of prominent spectral features at $z = 1.1$.
               Features caused by telluric absorption (top panel)
               or sky emission (bottom panel) are plotted in red.
              }
         \label{FigMemspecz}
   \end{figure}
\begin{table*} 
\caption{Summary of the main properties of the four galaxies observed
   with OMEGA2000 and FORS2 identified as spectroscopic cluster members.}
\label{table:1}
\centering
\begin{tabular}{r c c c c c c c}
\hline\hline
ID&RA (J2000)&Dec (J2000)&H&z$-$H&z&D4000&[OII] EW\\
 &hh mm ss&dd mm ss&Vega mag&Vega mag& & &$\mathrm{\AA}$\\
\hline
2	&03 38 44.136 &+00 30 29.95 &19.39 &2.84 &$1.09437 \pm 0.00090$&$1.901 \pm 0.052$&\\
12	&03 38 43.026 &+00 32 18.55 &19.03 &1.98 &$1.09787 \pm 0.00040$&$1.194 \pm 0.036$&$-32.908 \pm 1.380$\\
15	&03 38 40.912 &+00 30 10.66 &18.96 &2.27 &$1.09641 \pm 0.00027$&$1.633 \pm 0.028$&$-7.952 \pm 0.962$\\
16	&03 38 44.299 &+00 30 03.71 &20.29 &3.20 &$1.09930 \pm 0.00044$&$1.690 \pm 0.131$&$-9.291 \pm 3.873$\\
\hline
\end{tabular}
\end{table*}

   Figure 10 shows the redshift distribution of the galaxies at $z \ge 0.5$
   that were observed with FORS2 in a $6.8^{\prime} \times 6.8^{\prime}$
   region of the sky that contains both XMMU\,J0338.5$+$0029 (primary target)
   and XMMU\,J0338.7$+$0030 and that are not members of XMMU\,J0338.5$+$0029
   (see Fassbender et al. 2011c).
   Four galaxies exhibit consistent redshifts and are therefore identified
   as spectroscopic cluster members.
   As a consequence, XMMU\,J0338.7$+$0030 is at a mean spectroscopic redshift
   $< z > = 1.097 \pm 0.002$ ($1 \mathrm{\sigma}$, where the uncertainty
   is determined from a boostrap analysis combined with the biweight location
   estimator.
   These four spectroscopic members lie within $2.3^{\prime}$
   from the X-ray position of their host cluster, which corresponds
   to a projected distance of $\sim 1.1~\mathrm{Mpc}$
   for the assumed cosmology (Sect.~1).
   Their spectra are shown in Fig.~11,
   where the most prominent spectral features are labelled;
   their main properties are listed in Table 1.

   Out of these four spectroscopic members,
   three (with ID$=$2, ID$=$15 and ID$=$16) have counterparts
   in the GROND photometric catalogue and lie within $1^{\prime}$
   from the original X-ray position of XMMU\,J0338.7$+$0030 (see Fig.~4).
   The object with ID$=$2 can be identified as a passively evolving galaxy
   (likely an early-type galaxy) on the basis of its spectral properties.
   Conversely, the spectroscopic member with ID$=$15 exhibits robust evidence
   of on-going star-formation activity,
   as traced by the (unresolved) [O\,II] doublet
   at $\mathrm{\lambda}=3727~\mathrm{\AA}$.
   This activity is not as vigorous as in the spectroscopic member
   with ID$=$12 (modulo dust attenuation) though.
   The object with ID$=$16 has the least robust determination of photo-$z$
   and a controversial evidence of [O\,II] line emission.
   It is located in correspondence of the X-ray centroid
   of XMMU\,J0338.7$+$0030 determined in Sect.~2.1.1.

   The absence/presence of the [O\,II] line and its equivalent width ($EW$)
   are consistent with the measured values of the strength
   of the $4000~\AA$ break ($D4000$) as shown in Table 1.
   In particular, a value of $D4000 \sim 1.9$ is consistent with a minimum
   light-weighted age of the stellar populations of $\sim 3~\mathrm{Gyr}$,
   according to the set of models with different star-formation histories
   and metallicities in Gallazzi et al. (2005)
   (A. Gallazzi private communication).
   Together with the lack of detected [O\,II] line emission,
   this confirms the spectroscopic member with ID$=$2
   as a galaxy dominated by old, passively evolving stellar populations
   (i.e., an early-type galaxy).
   Minimum light-weighted ages of the stellar populations
   equal to $\sim 1~\mathrm{Gyr}$ and $\sim 1.4~\mathrm{Gyr}$
   seem to characterize the spectroscopic members with ID$=$15 and ID$=$16,
   respectively.
   These galaxy are likely evolved systems
   with ongoing star-formation activity.
   Finally, the spectroscopic member with ID$=$12 exhibits not only
   the strongest ongoing star-formation activity but also, consistently,
   the lowest value of $D4000$, which sets the minimum light-weighted age
   of its stellar populations to be equal to $\sim 0.3~\mathrm{Gyr}$.


\section{Results}   

\subsection{Galaxy number counts}

   In order to illustrate depth and completeness of the photometric catalogue
   extracted from the GROND optical/near-IR imaging (Sect.~2.3.3),
   the number of galaxies per 0.5 mag-bin and unit area
   for each of the $\mathrm{i^{\prime}, z^{\prime}, Ks}$ bands is reproduced
   in Fig.~12\footnote{Corrections for masked areas are introduced here; they correspond to 0.8\% of the total area, which is equal to $4.64 \times 10^{-3}~\mathrm{deg}^2$. Since these corrections are negligible, they will not be introduced later on.}.
   Galaxies are selected as in Sect.~2.3.3.
   Statistical uncertainties are determined from the approximate formulae
   for confidence limits based on the Poisson and binomial statistics
   presented in Gehrels (\cite{gehrels86}).

   The GROND $\mathrm{i^{\prime}}$-, $\mathrm{z^{\prime}}$-
   and $\mathrm{Ks}$-band galaxy number counts are compared with those
   determined from observations of deep fields or survey areas like
   the Hawaii Hubble Deep Field North (Hawaii HDF-N,
   Suprime-Cam $\mathrm{I}$- and $\mathrm{z^{\prime}}$-band --
   Capak et al. \cite{capak04}), 
   the Cosmic Evolution Survey (COSMOS,
   ACS WFC F814W filter -- Leauthaud et al. \cite{leauthaud07};
   Suprime-Cam $\mathrm{i^+}$ band -- Capak et al. \cite{capak07};
   WIRCam $\mathrm{Ks}$ band -- McCracken et al. \cite{mccracken10}),
   the Calar Alto Deep Imaging Survey (CADIS, OMEGA Prime
   $\mathrm{K^{\prime}}$ band -- Huang et al. \cite{huang01}),
   the Subaru Deep Field (SDF, CISCO $\mathrm{K^{\prime}}$ band --
   Maihara et al. \cite{maihara01}\footnote{We plot the completeness-corrected, $\mathrm{K^{\prime}}$-band galaxy number counts provided by Maihara et al. (\cite{maihara01}).}),
   the infrared complement of the VIMOS-VLT deep survey
   (VIRMOS, SOFI $\mathrm{Ks}$ band -- Iovino et al. \cite{iovino05}),
   the FLAMINGOS Extragalactic Survey (FLAMEX, FLAMINGOS $\mathrm{Ks}$ band --
   Elston et al. \cite{elston06}), and the UKIRT Infrared Deep Sky Survey
   Ultra-deep survey (UKIDSS UDS, UKIRT $\mathrm{Ks}$ band -- Hartley et al.
   \cite{hartley08}).

   The present GROND $\mathrm{z^{\prime}}$-selected sample of galaxies
   has a 50\% completeness limit
   equal to $\mathrm{z^{\prime}}^{\mathrm{compl}} \sim 25.3$,
   as determined from the extrapolation of the galaxy number counts
   in Capak et al. (\cite{capak04}).
   This completeness magnitude corresponds
   to a $\sim 2 \mathrm{\sigma}$ flux detection (cf. Fig.~7d).
   The 50\% completeness limit is equal to $\sim 24.8~\mathrm{AB~mag}$
   in $\mathrm{i^{\prime}}$-band, which corresponds
   to a $\sim 5 \mathrm{\sigma}$ flux detection.
   Finally, the GROND photometric sample is 50\% complete
   down to $Ks^{\mathrm{compl}} \sim 22.9$,
   which corresponds to a $\sim 1 \mathrm{\sigma}$ flux detection.
   This completeness magnitude also corresponds to $\sim Ks^{\star}+2.4$,
   where $Ks^{\star} = 20.5^{+0.4}_{-1}$ is the characteristic magnitude
   (Schechter \cite{schechter76}) of the $\mathrm{Ks}$-band luminosity function
   of cluster galaxies at $z \sim 1.2$ (Strazzullo et al. \cite{strazzullo06}).

   Comparison of the different galaxy number counts in Fig.~12
   shows an overall agreement between our results
   and those in the literature down to the completeness limit
   of the GROND photometric sample of galaxies in each band.
   This is important when considering the differences in the sky area
   that is probed by different observations (leading to cosmic variance),
   the operational definition of total magnitude
   and the adopted galaxy/star separation criteria.
   These differences mostly affect the bright end of the galaxy number counts,
   which corresponds to magnitudes approximately brighter
   than $20~\mathrm{AB~mag}$ in Fig.~12.
   Furthermore, our galaxy number counts refer to the same sample
   extracted in the $\mathrm{z^{\prime}}$-band,
   whereas the galaxy number counts in the literature refer
   to individual samples extracted in the $\mathrm{i^{\prime}}$-,
   $\mathrm{z^{\prime}}$- or $\mathrm{K}$-band.

   In spite of the overall agreement, our number counts exhibit an excess
   (by a factor of almost two) in the magnitude range between
   $Ks^{\mathrm{compl}} - 2$ and $Ks^{\mathrm{compl}} - 0.5$
   (i.e., $20.8 \le Ks \le 22.3$) with respect to all others.
   This magnitude range corresponds
   to $\sim Ks^{\star}+0.3 \le Ks \le Ks^{\star}+1.8$ for cluster galaxies
   at $z \sim 1.2$ (Strazzullo et al. \cite{strazzullo06}).
   Consistently, a slight excess across a broader magnitude range
   (i.e., approximately at magnitudes between 21.5 and $23.5~\mathrm{AB~mag}$)
   is visible from comparison of the galaxy number counts
   for the $\mathrm{z^{\prime}}$ band.
   No clear excess is apparent in the galaxy number counts
   for the $\mathrm{i^{\prime}}$-band.
   This behaviour is interpreted as an indication for the presence
   of an over-density of galaxies associated with the weak, marginally extended
   X-ray emission corresponding to XMMU\,J0338.7$+$0030.

   \begin{figure}
   \centering
      \vskip -0.85truecm
      \includegraphics[width=8.5cm]{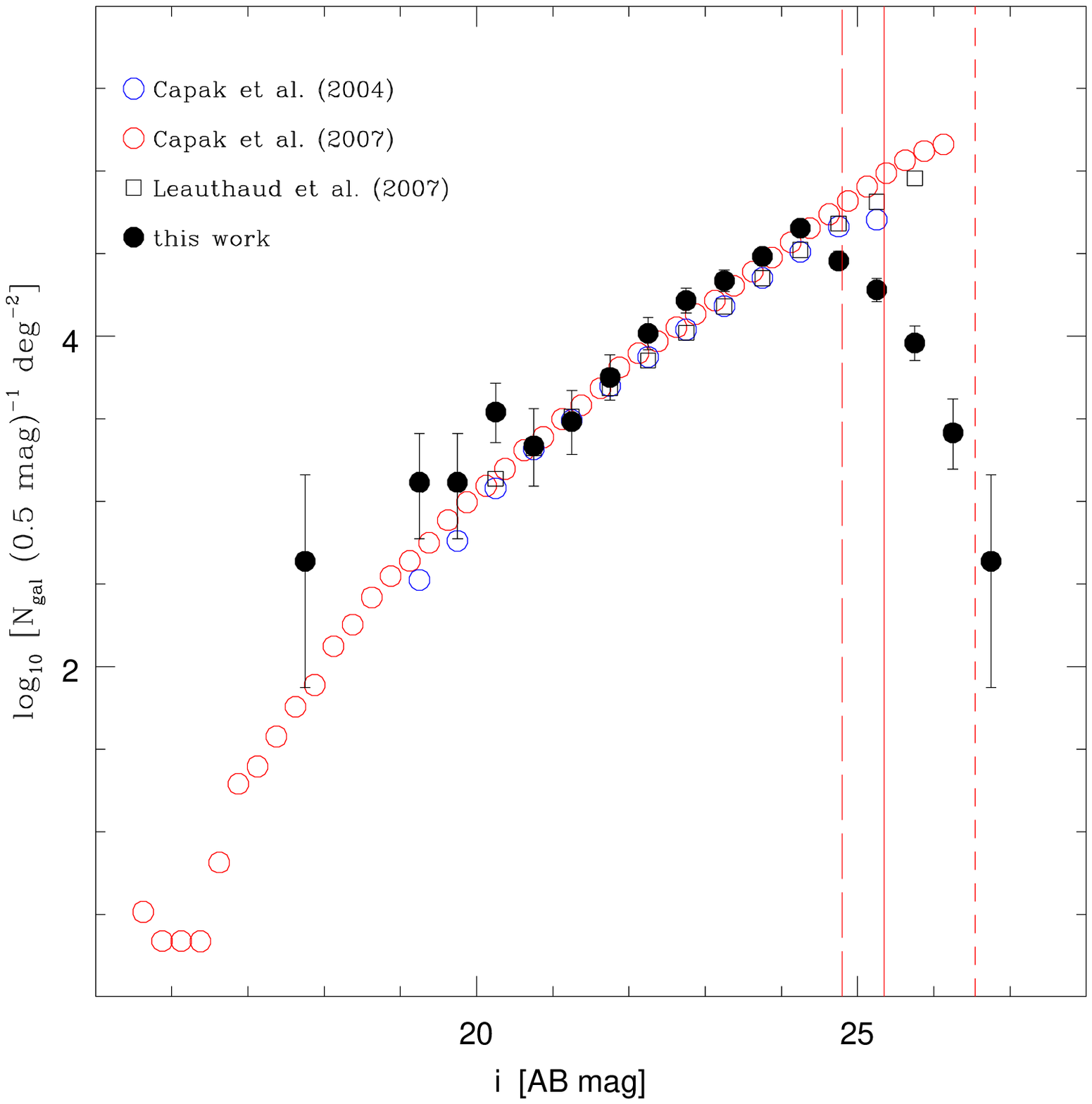}
      \vskip -1.15truecm
      \includegraphics[width=8.5cm]{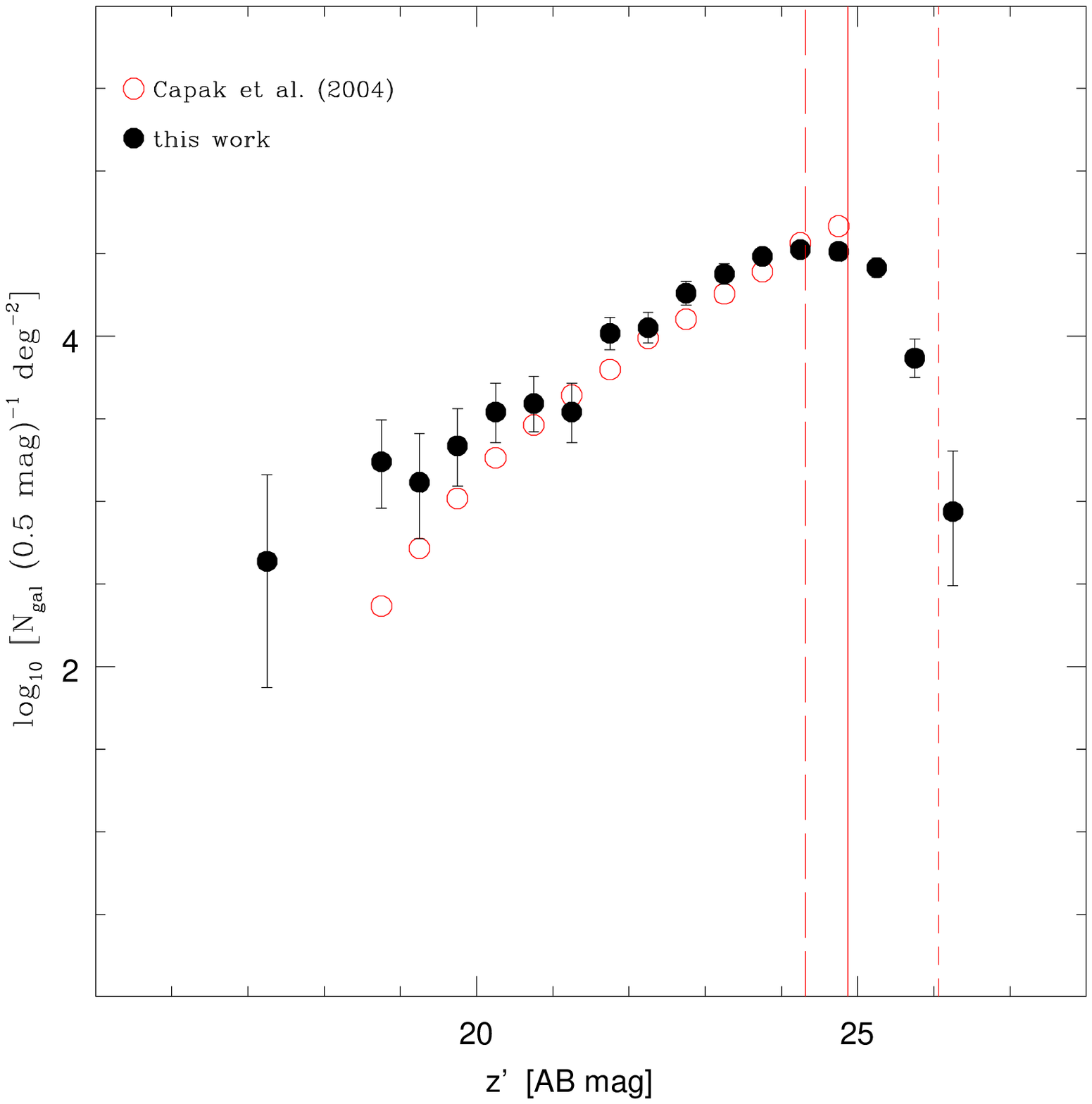}
      \vskip -1.15truecm
      \includegraphics[width=8.5cm]{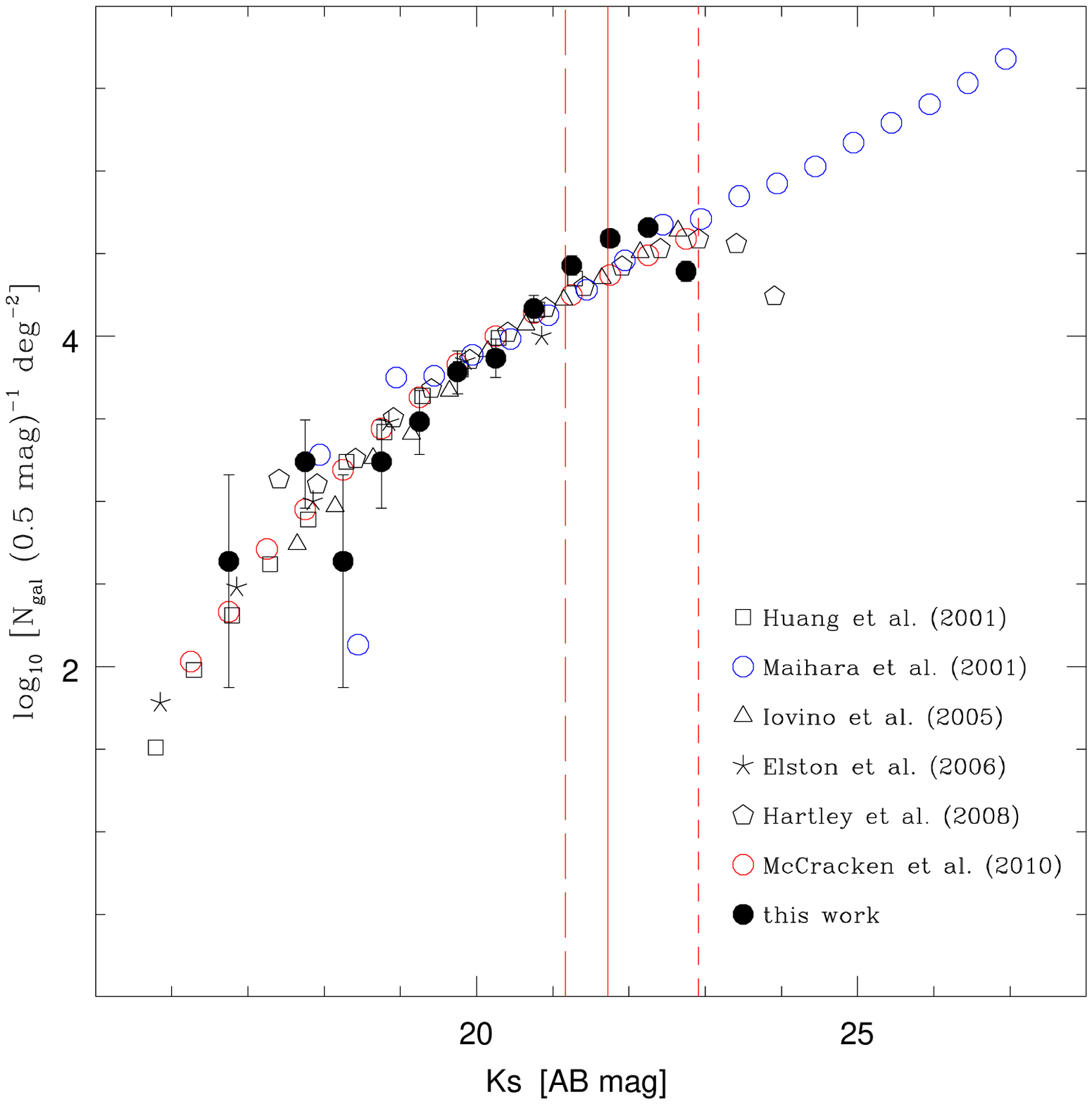}
      \vskip -0.15truecm
      \caption{Comparison of the galaxy number counts determined
               in the region of XMMU\,J0338.7$+$0030 imaged with GROND 
               for the $\mathrm{i^{\prime}}$ (top),
               $\mathrm{z^{\prime}}$ (centre) and $\mathrm{Ks}$ band (bottom)
               and the corresponding ones from deep fields or survey areas.
               Error bars are shown only for our points.
               In each panel, the short-dashed, solid and long-dashed lines
               represent the $1~\mathrm{\sigma}$, $3~\mathrm{\sigma}$
               and $5~\mathrm{\sigma}$ thresholds of our photometry,
               respectively.
               The GROND galaxy number counts exhibit a clear excess
               for $20.8 \le Ks \le 22.3$.
              }
         \label{Figizkcounts}
   \end{figure}

\subsection{Photometric-redshifts}

   The photometric redshift technique enables us
   to assess the membership to XMMU\,J0338.7$+$0030
   for the 832 sources in the GROND photometric sample.
   Distances to individual sources were estimated
   through the publicly available package
   for {\em PHotometric Analysis for Redshift Estimations},
   {\em le Phare}\footnote{www.oamp.fr/people/arnouts/LE\_PHARE.html}
   (S. Arnouts \& O. Ilbert), which is a set of Fortran programmes
   to compute photo-$z$'s using the standard SED fitting technique.

   We selected the large pool of observed SEDs of stars (254)
   and the mixture of observed and synthetic SEDs for QSOs (28)
   that are provided by {\em le Phare}, as it is standard.
   However, we initially adopted two sets of templates
   to describe observed SEDs of galaxies and investigated their performances.

   The first set was used for the COSMOS photo-$z$ paper
   (Ilbert et al. \cite{ilbert09}).
   It contains 31 templates: the SEDs numbered from 1 to 7 correspond
   to galaxies that can be morphologically classified as elliptical (E),
   the SED number 8 corresponds to a lenticular (or S0) galaxy,
   SEDs numbered from 9 to 12 correspond to bulge-dominated disc galaxies
   (i.e, Sa and Sb), SEDs numbered from 13 to 19 correspond to bulge-less
   disc galaxies (i.e, Sc, Sd, and Sdm), SEDs numbered from 20 to 31
   describe starburst (SB) galaxies.
   We followed the indicated prescriptions
   for the model-dependent attenuation\footnote{We note that the extinction curve describes the combined absorption and out-of-the beam scattering properties of a mixture of dust grains of given size distribution and chemical composition in a screen geometry as a function of wavelength; the attenuation function is the combination of the extinction curve with the geometry of a dusty stellar system, in which a substantial fraction of the scattered light is returned to the line of sight (e.g. Pierini et al. 2004b).} of galaxy templates, so that
   a ``Prevot extinction law'' (Prevot et al. 1984) was applied
   to SEDs numbered from 13 to 23,
   a so-called ``Calzetti attenuation law'' (Calzetti et al. 2000)
   and two ``modified'' Calzetti attenuation laws
   (see Ilbert et al. \cite{ilbert09}) were applied
   to SEDs numbered from 23 to 31.
   This choice accounts for the fact that high-$z$ star-forming galaxies
   are characterized by extinction laws that differ at rest-frame ultraviolet
   wavelengths, as it is in the Local Group
   (Noll et al. 2009 and references therein).
   In all cases, the reddening $E(B-V)$ was enabled to range
   between 0 and $0.5~\mathrm{mag}$ in steps of $0.05~\mathrm{mag}$.

   The second set of galaxy templates corresponds to the 66 SEDs
   that are used in Ilbert et al. (2006) for the analysis
   of the CFHTLS Deep fields.
   In the CFHTLS set, the ratio between the numbers
   of E and SB templates is reversed with respect to the COSMOS set:
   E, Sbc, Scd, Im and SB galaxies are described by 22, 17, 12, 11,
   and 4 SEDs, respectively.
   For Scd, Im and SB galaxy templates, a Prevot et al. (1984) extinction law
   is adopted, with values of $E(B-V)$ ranging from 0 to 0.25
   with a step of 0.05 consistently with the indications
   in the {\em le Phare} documentation.

   For each library of galaxy SEDs, we considered two separate families
   of models, by including the contribution to the total flux
   from line emission or not.
   We opted to do this because there is independent evidence that
   the bright end of the galaxy luminosity function of a high-$z$ cluster
   can be populated by [O\,II] line emitters (as for XMMU\,J0302.2$-$0001
   at $z = 1.186$, \u{S}uhada et al. 2011), sometimes with vigorous
   starburst activity (as for XMMU\,J1007.4$+$1237 at $z = 1.555$,
   Fassbender et al. 2011b).

   Finally, for galaxy templates as well as for QSO templates,
   we considered the redshift range 0--6 with a step of 0.02 in $z$
   and the range $-24 \le M_{\mathrm{z^{\prime}}} \le -8$
   in the $\mathrm{z^{\prime}}$-band absolute magnitude.

   For all stellar, QSO and galaxy templates,
   synthetic photometry was computed for all GROND bands
   in the AB magnitude system and confronted with the observed photometry.
   Both systematic (e.g., caused by calibration uncertainties)
   and statistical (i.e., caused by measurement uncertainties) errors
   were taken into account, the latter being magnified by 50\% (in mag).

   The photometric redshift code {\em le Phare},
   based on a simple $\chi^2$ fitting method, 
   yields the best-fitting template in each class, the best-fit value of $z$,
   the 68\% range of the photo-$z$ solutions and the probability
   per redshift bin, in particular.
   All ouputs for individual detected sources were inspected by eye.
   A source was classified as a star if the best-fitting stellar template
   gave the lowest $\chi^2$ value.
   Conversely, it was classified as a QSO if the best-fitting QSO template
   gave a $\chi^2$ value that was better by a factor of two
   than those given by the best-fitting stellar and galaxy templates.

   \begin{figure*}
   \centering
      \vskip -0.75truecm
      \includegraphics[width=5.5cm]{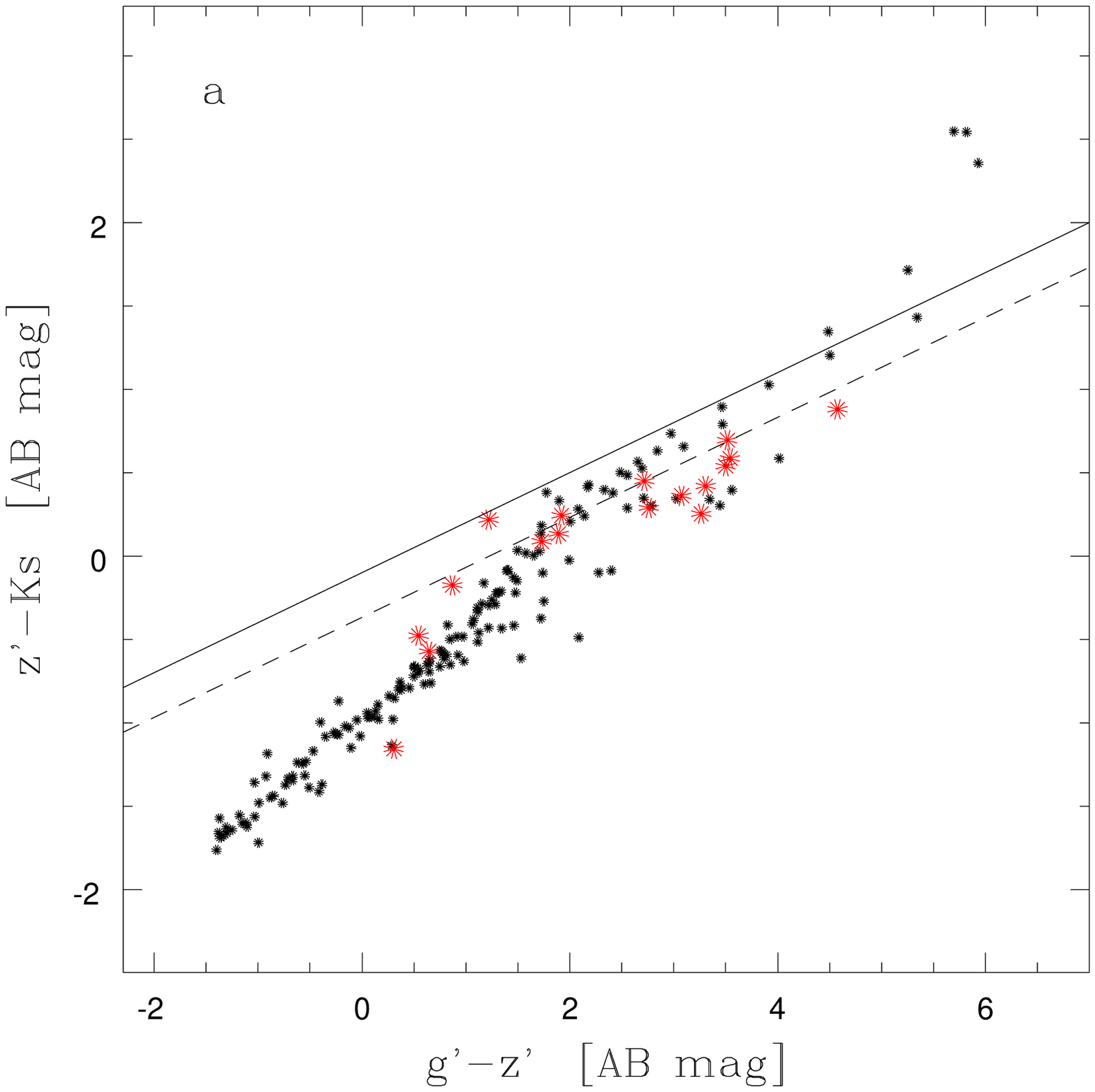}
      \vskip -0.75truecm
      \includegraphics[width=5.5cm]{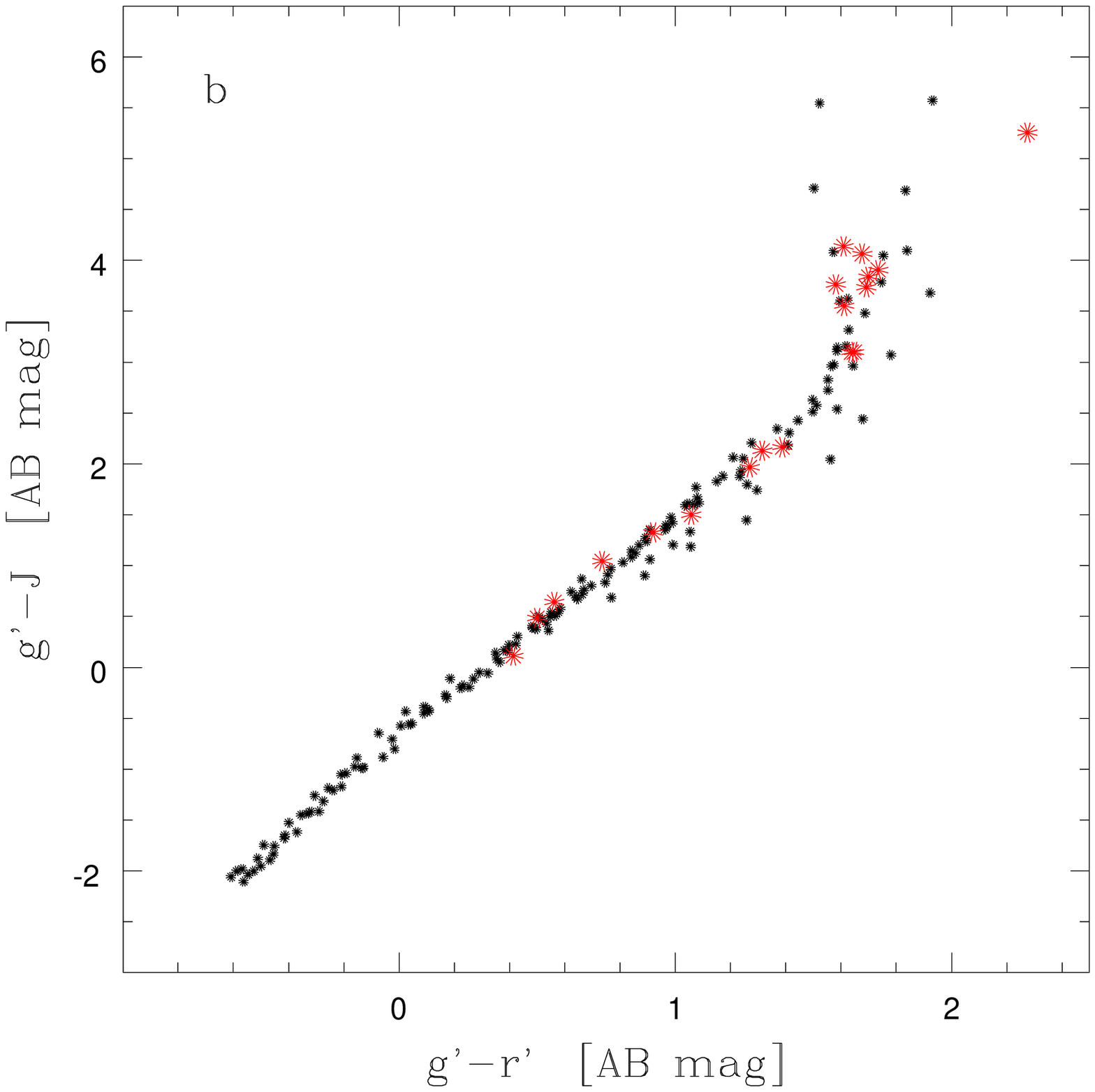}
      \vskip -0.75truecm
      \includegraphics[width=5.5cm]{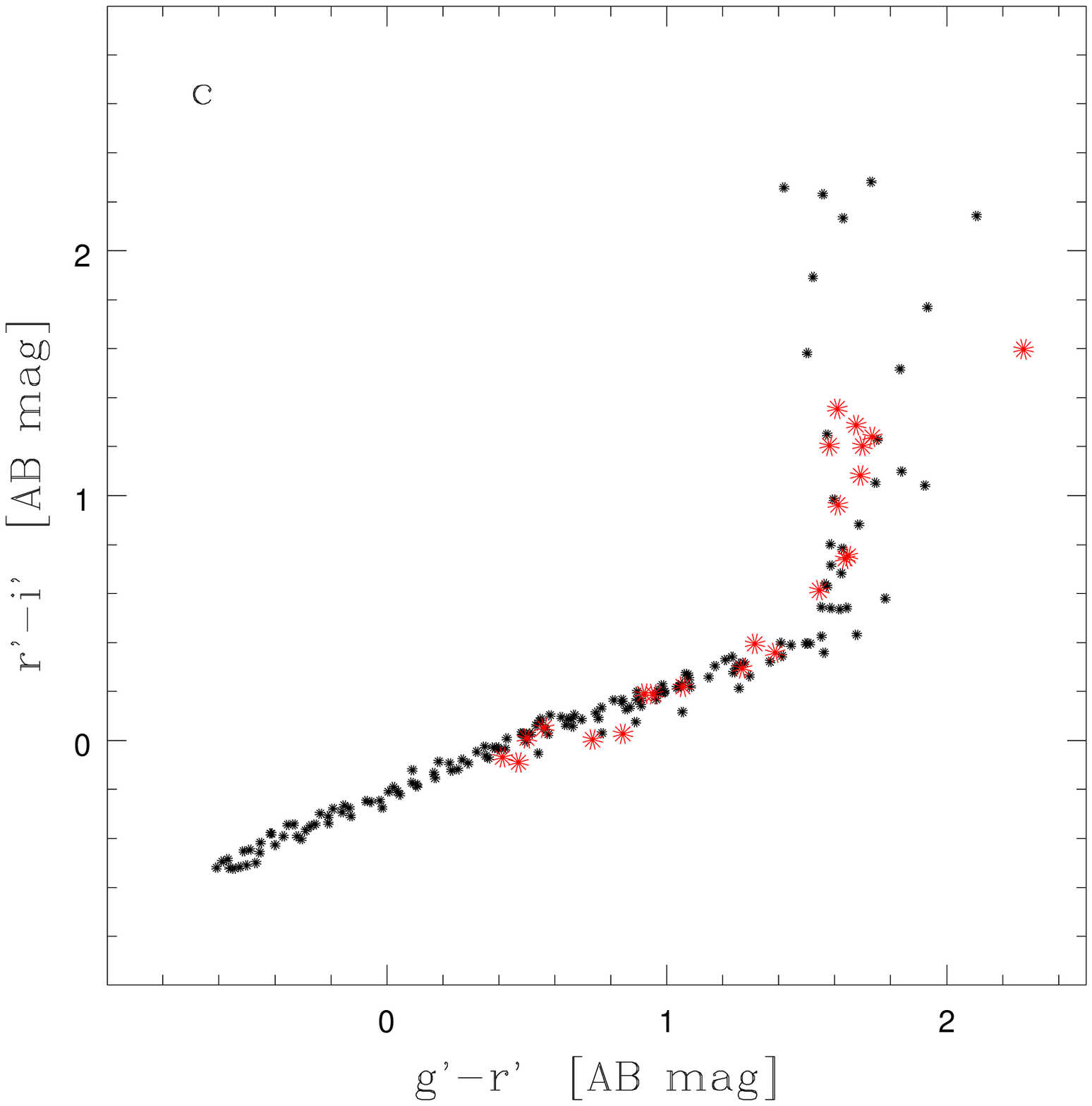}
      \vskip -0.75truecm
      \includegraphics[width=5.5cm]{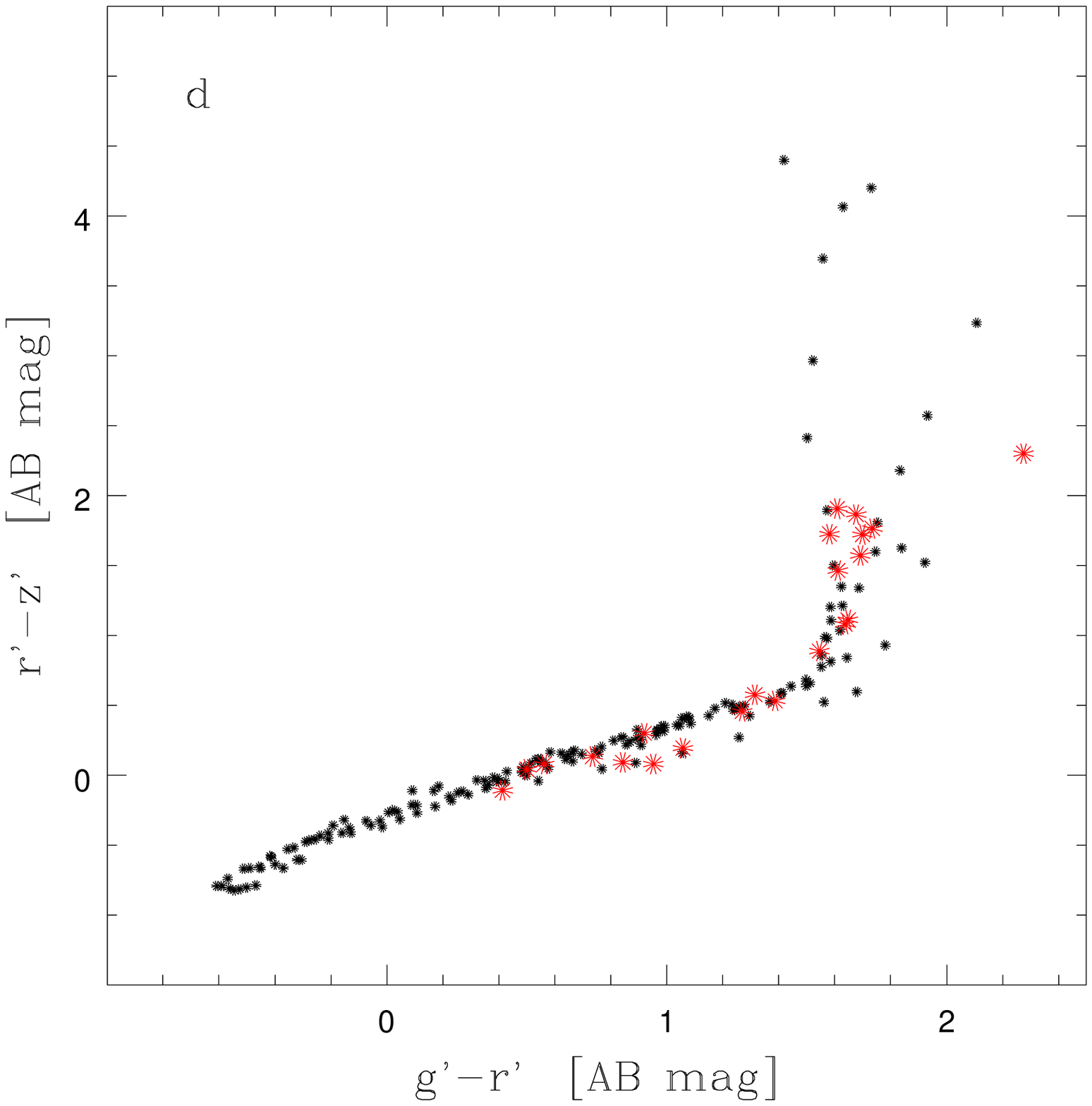}
      \vskip -0.75truecm
      \includegraphics[width=5.5cm]{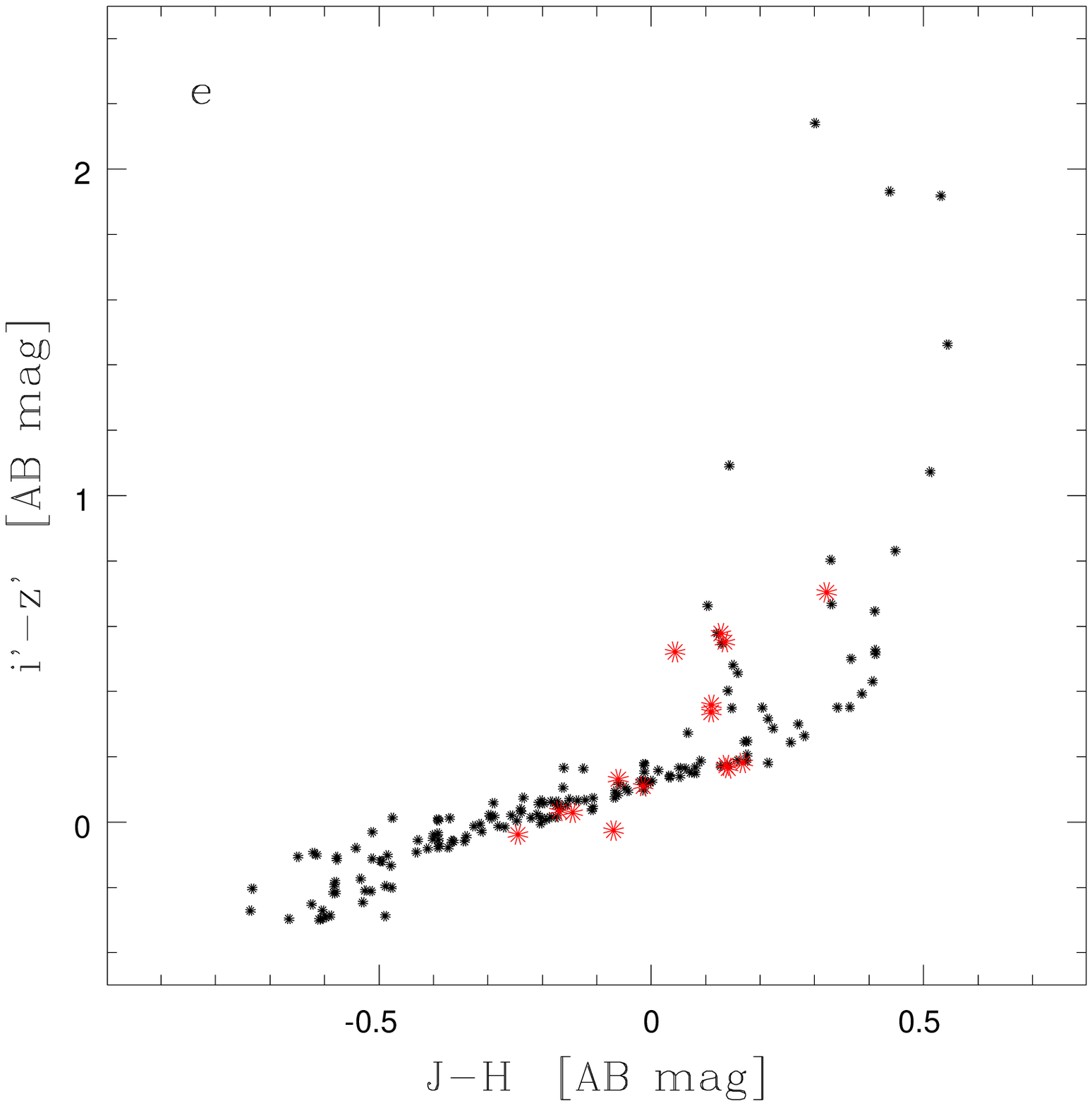}
      \vskip -0.75truecm
      \vskip 0.5truecm
      \caption{Five colour--colour diagrams for stars
               in the {\em le Phare} library (black filled circles)
               and the 15 to 23 objects detected at all GROND bands,
               with the most accurate photometry in the relevant bands
               and robustly classified as stars by {\em le Phare}
               using the full seven-channel photometry (red asterisks).
               These observed stars nicely fall
               onto the synthetic stellar sequences,
               after the synthetic photometry is slighlty corrected
               by -0.05, +0.025 and $+0.025~\mathrm{mag}$
               in the $r^{\prime}, i^{\prime}$ and $H$ bands, respectively.
               We recall that photometric redshifts
               are particularly sensitive to the $4000~\mathrm{\AA}$ break
               in the spectrum of a galaxy, which spans
               the entire wavelength range of the $z^{\prime}$
               broad-band filter for $z = 1.1$--1.6.
              }
         \label{FigFilterstest}
   \end{figure*}

   As a first step, we checked the reliability
   of the publicly available transmission functions
   of the GROND dichroic filters uploaded into {\em le Phare}
   by comparing the distributions of the library stars and the stars
   observed by us with the most accurate photometry
   in a large suite of colour--colour diagrams (see Fig.~13 for examples).
   Eventual problems affecting the GROND photometry of each star
   in individual bands (e.g., owing to poor S/N and deblending)
   were inspected: if there were compromised photometric data,
   they were excluded.
   As a result, we found an overall good agreement between observed
   and synthetic stellar tracks, which required the addition
   of very small shifts ($-0.05~\mathrm{mag}$, $+0.025~\mathrm{mag}$,
   $+0.025~\mathrm{mag}$) to the synthetic photometry
   in the $\mathrm{r^{\prime}, i^{\prime}, H}$ bands, respectively.
   Furthermore, the colour criterion adopted to statistically select stars
   from $\mathrm{g^{\prime}, z^{\prime}, Ks}$ photometry only (Eq.~1)
   was nicely confirmed (cf. Figs.~8 and 13a).

   As a second step, we computed photometric redshifts
   for two photometric catalogues with the same sources
   as part of the test to determine the best operational definition
   of total magnitude for our data.
   This test was executed with the COSMOS set of galaxy templates
   excluding the line emission contribution; in particular, it addressed
   the five galaxies with GROND photometry and a spectroscopic redshift
   (Sect. 2.4.1).
   Overall the uncertainty on the ensuing two sets of photometric redhifts
   obtained for these five galaxies is similar, as well as the best-fit
   solutions yielded for the two foreground galaxies.
   However, a mean photometric redshift equal to $\sim 1.22 \pm 0.25$
   ($1 \mathrm{\sigma}$) or $\sim 1.02 \pm 0.15$($1 \mathrm{\sigma}$)
   is obtained for XMMU\,J0338.7$+$0030 when total magnitudes were computed
   using the operational definitions {\em A} and {\em B}
   in Sect.~2.3.2, respectively.
   Because this cluster is spectroscopically confirmed to be at $z = 1.1$
   (Sect.~2.4.2), the operational definition of total magnitude {\em B}
   has to be favoured.
   This choice is also supported by the distribution
   of the three spectroscopic members in the $i^{\prime} - z^{\prime}$
   vs $z^{\prime}$ colour--magnitude diagram (Fig.~15).
   Therefore, we only discuss results from the fiducial photometric catalogue
   where total magnitudes are computed with method {\em B} hereafter.

   Finally, we compared the performances
   of the COSMOS and CFHTLS sets, whether the emission line contribution
   to the templates of star-forming galaxies is included or not.
   We based our judgement on the comparison of the best-fit solutions
   and their uncertainties but also on the capability
   of the set of choice to recover the redshifts
   of the five galaxies with available spectroscopy (see Sect.~2.4.1).
   This figure corresponds to only $\sim 0.6$\% of the GROND photometric sample
   of 832 sources; the ensuing span in redshift is very narrow:
   $z = 0.85$--1.1.
   Nevertheless, the results obtained for the three spectroscopic members
   of XMMU\,J0338.7$+$0030 (see Sect.~2.4.2) gave us useful indications.
   The use of the COSMOS galaxy templates yielded a mean photometric redshift
   equal to $\sim 1.02 \pm 0.15$ for this cluster, 
   whether emission lines were included or not.
   Conversely, the use of the CFHTLS galaxy templates yielded
   values of $\sim 1.14 \pm 0.11$ and $\sim 1.12 \pm 0.09$, respectively.
   Given the better agreement with the spectroscopic redshift
   of XMMU\,J0338.7$+$0030 ($< z > = 1.097 \pm 0.002$)
   and the lower uncertainty, hereafter we discuss results obtained
   from the use of the CFHTLS set of templates without
   emission line contribution\footnote{This suggests that the COSMOS galaxy templates are less adequate to represent the characteristics of the galaxy populations of XMMU\,J0338.7$+$0030. An analogous point was made by Guennou et al. (2010), who discussed photometric redshifts along lines of sight to 10 clusters at spectroscopic redshifts between 0.48 and 0.79 obtained with the same COSMOS galaxy templates and photo-$z$ code used here.}.

   As a consequence of this choice, out of the 832 entries
   in the GROND photometric catalogue, 174 sources do not have
   a (reliable) photo-$z$ estimate (owing to very poor photometry
   or multiple peaks in the photo-$z$ probability distribution function),
   64 sources are identified as stars, 158 sources exhibit SEDs
   that are best-fit by a QSO template, 436 sources are identified as galaxies.
   On one hand, the past large percentage of objects identified as QSOs
   reflects our arbitrary classification criterion of the stellar, galaxy
   and QSO templates that best-fit the SED of a given object.
   On the other hand, it reflects the reduced capability
   of discriminating between best-fit templates
   for the numerous objects with incomplete photometric information.
   For the majority of the objects identified as QSOs,
   the photometric redshift is reassuringly consistent, if not similar,
   to the value associated with the best-fit galaxy template.

   The quality of the best-fit photometric redshifts
   obtained with CFHTLS galaxy templates not including emission lines
   is discussed in Appendix A.
   The distribution of these photo-$z$ solutions is often single peaked
   and narrow when accurate photometry is available
   for all seven GROND channels.

\subsection{Optical identification of the high-$z$ cluster}

\subsubsection{The photometric-redshift distribution function}

   We computed the surface density of extragalactic sources
   within or beyond a cluster-centric distance of $1^{\prime}$,
   as well as across the entire $3.9 \times 4.3~\mathrm{arcmin}^2$ region
   of the sky imaged with GROND, as a function of the photometric redshift
   (Fig.~14a).
   A simple visual inspection reveals that for the redshift bins
   1.0--1.1, 1.2--1.3 and 1.5--1.6, the density of sources
   within the bona fide cluster region
   exceeds that computed in the adjacent field by a factor of 2.
   An equivalent over-density of $i^{\prime} - Ks$-selected extremely red objects (EROs)\footnote{Colours redder than $I - K = 4~\mathrm{Vega~mag}$ statistically select EROs at $z \sim 1$--2 (Pozzetti \& Mannucci \cite{pozzetti00} and references therein; see also Pierini et al. \cite{pierini04a,pierini05}). For the GROND channels, this corresponds to $i^{\prime} - Ks = 2.8$ in the AB magnitude system.} exists in each of the last two bins (Fig.~14b).
   These EROs are found in the redshift range where they are expected to be,
   which gives additional confidence in the quality
   of our photo-$z$'s, at least in a  statistical sense.

   \begin{figure}
   \centering
      \vskip -1.0truecm
      \includegraphics[width=9cm]{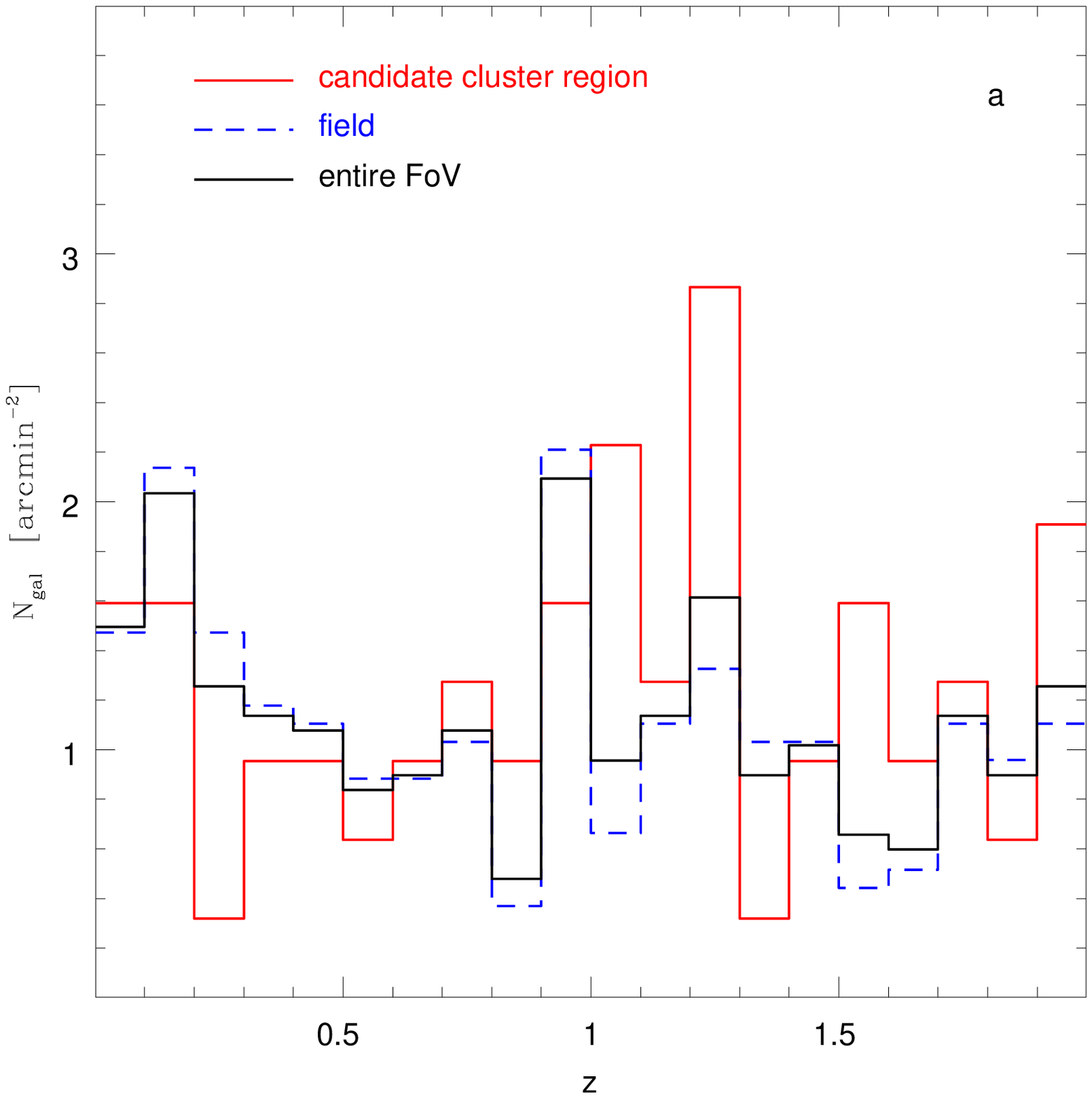}
      \vskip -1.25truecm
      \includegraphics[width=9cm]{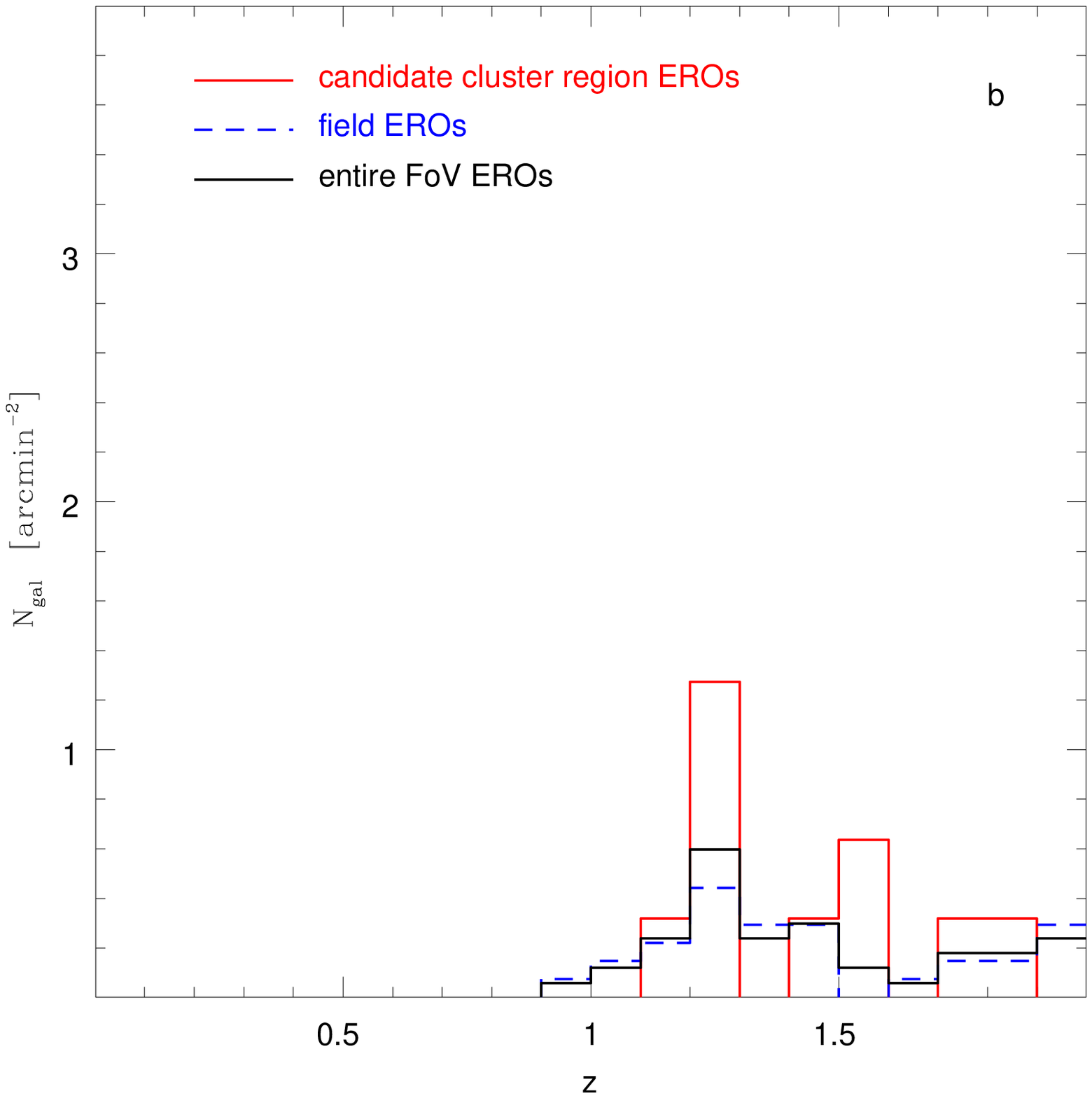}
      \vskip -0.25truecm
      \caption{Number density as a function of the photometric redshift
               for the extragalactic sources with a photo-$z$ less than 2
               in the entire $3.9 \times 4.3~\mathrm{arcmin}^2$
               region of the sky imaged with GROND (black solid line),
               for those among them inside the bona fide cluster area
               (red solid line) or in the surrounding field
               (blue short-dashed line) - panel a.
               The analogous distribution functions for the sub-sample of EROs,
               expected to be at $z \sim 1$--2 (see text), are also reproduced
               (panel b).
               For the redshift bins 1.0--1.1, 1.2--1.3 and 1.5--1.6,
               the density of sources within a cluster-centric distance
               of $1^{\prime}$ exceeds
               that computed in the adjacent field by a factor of 2.
               In correspondence of the last two redshift bins,
               there is a clear excess of EROs.
              }
         \label{FigNPhotonowel}
   \end{figure}
%

   \begin{figure*}
      \vskip -4.85truecm
   \centering
      \includegraphics[width=15cm]{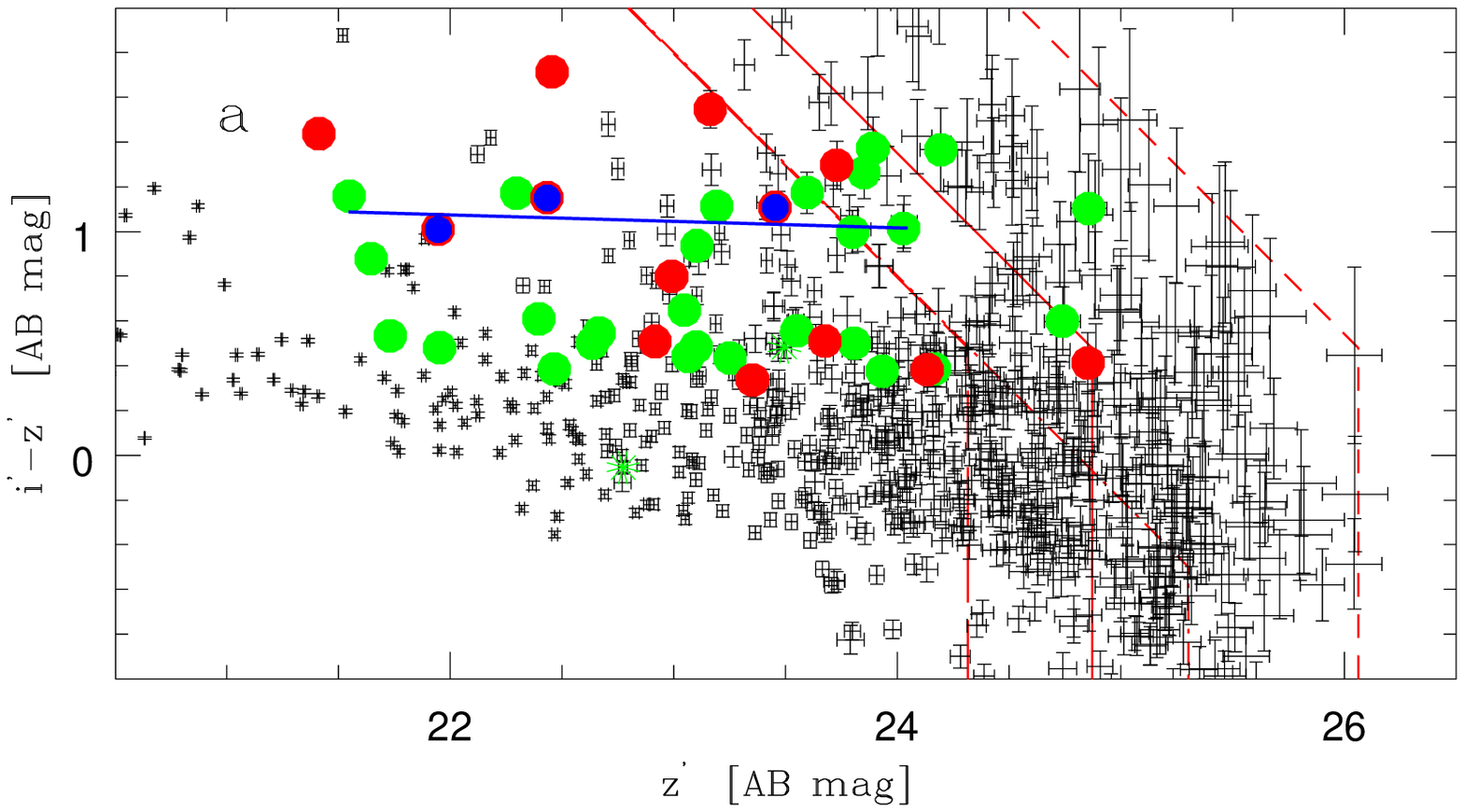}
      \vskip -7.5truecm
      \includegraphics[width=15cm]{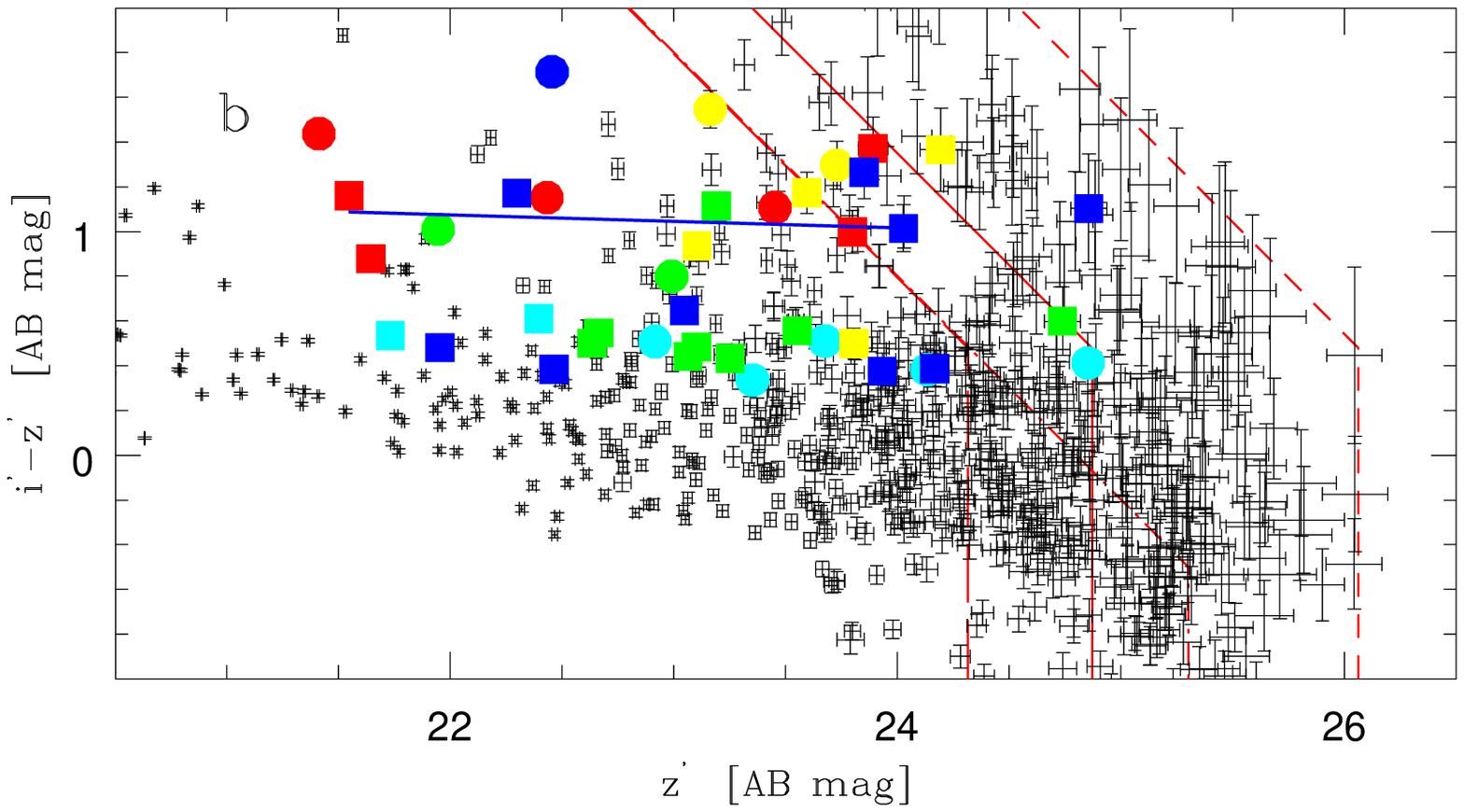}
      \vskip -2.8truecm
      \caption{$i^{\prime} - z^{\prime}$ vs $z^{\prime}$
               colour--magnitude diagram for all GROND sources
               in the XMMU\,J0338.7$+$0030 region
               with flux detections in the $\mathrm{i^{\prime}, z^{\prime}}$
               bands at a significance level $\ge 1 \mathrm{\sigma}$,
               whatever the classification of the source by {\em Le Phare}
               (i.e., stars are included).
               Photometric errors ($1 \mathrm{\sigma}$) are also shown.
               Candidate members of XMMU\,J0338.7$+$0030 are reproduced
               with red (green) filled circles if they are within (outside)
               the bona fide cluster region (panel a).
               The three spectroscopic cluster members
               in the GROND photometric catalogue are marked
               with blue filled circles there.
               Green asterisks represent the two QSO at a cluster-centric
               distance greater than $1^{\prime}$.
               Alternatively (panel b), photometric cluster members classified
               as galaxies are colour-coded according to
               their spectro-photometric type, which is E (red), Sbc(yellow),
               Scd (green), Im (cyan) or SB (blue).
               In each panel, the red short-dashed, solid and long-dashed lines
               represent the $1 \mathrm{\sigma}$, $3 \mathrm{\sigma}$,
               and $5 \mathrm{\sigma}$ flux thresholds
               of our photometry (cf. Sect.~2.3.2).
               The red dotted--short-dashed line represents
               the completeness limits
               in the $\mathrm{i^{\prime}, z^{\prime}}$ bands,
               as obtained from Fig.~12.
               As a reference, the blue solid line
               reproduces the red-sequence defined by the inner regions
               of the 31 E$+$S0$+$S0/a galaxies that are located
               within $1^{\prime}$ from the X-ray centroid
               of the cluster RDCS\,J0910$+$5422 at $z=1.106$
               (Mei et al. 2006a).
              }
         \label{FigColmagizz}
   \end{figure*}

   In support of this conclusion, the spectroscopic members
   with ID$=$2, ID$=$15 and ID$=$16 have photometric redhifts\footnote{Here we give the best-fit photo-$z$ together with the values that bracket the 68\% range of the photo-$z$ solutions.}
   equal to $1.0110_{0.9918}^{1.0311}$, $1.1711_{1.1490}^{1.1939}$
   and $1.1820_{1.1378}^{1.2461}$, respectively.
   Furthermore, the morphological type of the best-fitting template
   corresponds to an E galaxy for the spectroscopic members with ID$=$2
   and ID$=$16 and to an Scd galaxy for the spectroscopic member with ID$=$15.
   This spectro-photometrically determined classification
   is nicely consistent with the spectroscopic classification in Sect.~2.4.2.
   Furthermore, it is reasonable to conclude that the two peaks
   at $z \sim 1.0$--1.1 and $z \sim 1.2$--1.3 in Fig.~14a
   are caused by true cluster members, at least in part.

   Therefore we selected as photometric (also candidate or fiducial hereafter)
   cluster members those sources with a photo-$z$ in the range 1.0110--1.2317,
   which is centred on the mean photometric redshift of XMMU\,J0338.7$+$0030
   ($< z > = 1.1214 \pm 0.0957$, $1 \mathrm{\sigma}$),
   contains the previous three spectroscopic members
   and, thus, has a width equal to $\pm 1.15 \mathrm{\sigma}$
   (or $\pm 0.1 < z >$).
   Forty-four sources satisfy our criterion:
   fourteen (36) are within $1^{\prime}$ ($2^{\prime}$) from the X-ray position
   of XMMU\,J0338.7$+$0030.
   Only two of these 44 candidate cluster members are identified as QSOs
   according to their best-fit templates; they are at cluster-centric distances
   greater than $1^{\prime}$.
   The remaining 42 fiducial cluster members
   are spectro-photometrically classified as E (7), Sbc (6), Scd (11), Im (8)
   or SB (10) according to the templates that best fit their individual SEDs
   and provide their fiducial photometric redshifts.
   As discussed in Appendix A, the photometric redshifts associated
   with these 42 galaxies are, in general, well-behaved and exhibit
   low uncertainties.

\subsubsection{The $i^{\prime} - z^{\prime}$ vs $z^{\prime}$ colour--magnitude diagram: evidence for a red sequence}

   We now investigate if the previously selected candidate cluster members
   consistently define a red sequence in a suitable colour-magnitude diagram.
   We considered the $\mathrm{i^{\prime}, z^{\prime}}$ bands
   because the rest-frame $4000~\mathrm{\AA}$ break
   spans the entire wavelength range of the $\mathrm{z^{\prime}}$ band
   as the redshift of a source increases from 1.1 to 1.6.
   Figure 15 shows the distribution of all sources
   of the GROND photometric sample in the $i^{\prime} - z^{\prime}$
   vs. $z^{\prime}$ colour--magnitude diagram.
   In the upper panel, the photometric members of XMMU\,J0338.7$+$0030
   are marked in red or green if they fall inside or outside
   the bona-fide cluster region, respectively.
   In the lower panel, the same sources are colour-coded according to
   the best-fit template (if they are identified as galaxies).
   About a dozen candidate cluster members identified as galaxies
   (including the three spectroscopic cluster members) suggest the existence
   of a red locus
   in the $i^{\prime} - z^{\prime}$ vs. $z^{\prime}$ colour--magnitude diagram:
   $i^{\prime} - z^{\prime}$ slightly decreases from $\sim 1.1~\mathrm{AB~mag}$
   to $\sim 1.0~\mathrm{AB~mag}$ as $z^{\prime}$ increases
   from $\sim 21.5~\mathrm{AB~mag}$ to $\sim 24.0~\mathrm{AB~mag}$.
   Furthermore, half of these galaxies are identified as ellipticals.
   This is consistent with the existence of a red sequence
   (Arimoto \& Yoshii 1987).

   As a reference, the red sequence of RDCS\,J0910$+$5422 at $z = 1.106$
   (Mei et al. 2006a) is reproduced in both panels of Fig.~15.
   Mei et al. obtained high-resolution imaging of this massive cluster
   with the {\em Hubble Space Telescope} Advanced Camera for Surveys
   ({\em HST} ACS) in the F775W and F850LP bandpasses.
   They computed total $z_{850}$ magnitudes ({\em SExtractor} MAG\_AUTO)
   and $i_{775} - z_{850}$ colours within the effective radii
   of the individual galaxies in order to select candidate cluster members.
   The use of these aperture colours was meant to avoid a selection bias
   introduced by the potential presence of radial colour gradients.
   Furthermore, Mei et al. fitted different $i_{775} - z_{850}$
   vs $z_{850}$ colour--magnitude relations to different sub-samples
   of morphologically classified early-type galaxies.
   We have transformed their best-fit relation
   for the 31 E$+$S0$+$S0/a galaxies within $1^{\prime}$
   from the X-ray centroid of RDCS\,J0910$+$5422
   (see table 1 in Mei et al. 2006a) into a red sequence
   in the $i^{\prime} - z^{\prime}$ vs $z^{\prime}$ colour--magnitude diagram
   by introducing corrections for the different transmission functions
   of the corresponding {\em HST} ACS and GROND filters.
   In spite of the uncertainties on these colour transformations
   (likely of the order of $0.1~\mathrm{mag}$), the red-sequence
   of RDCS\,J0910$+$5422 describes the red locus of XMMU\,J0338.7$+$0030
   rather well.

   However, galaxies with very different spectro-photometric types
   populate the red locus of XMMU\,J0338.7$+$0030
   within a cluster-centric distance of $1^{\prime}$
   and not just early-type galaxies as in RDCS\,J0910$+$5422.
   This was also observed in e.g. XMMU\,J1230.3$+$1339 at $z = 0.975$
   (see fig.~7 in Lerchster et al. 2011).
   Nevertheless the two coeval clusters XMMU\,J0338.7$+$0030
   and RDCS\,J0910$+$5422 seem to host
   also analogus populations of star-forming galaxies that exhibits bluer
   $i^{\prime} - z^{\prime}$ colours with respect to
   the red, passively evolving members by about $0.5~\mathrm{AB~mag}$
   (cf. fig.~7 in Mei et al. 2006a).
   For XMMU\,J0338.7$+$0030, these galaxies constitute about half
   of the candidate members and are identified as Scd, Im or SB.
   Furthermore, those within the bona-fide cluster region
   span the magnitude range $22.9 \le z^{\prime} \le 24.9$, whereas
   those farther out can be as bright as $z^{\prime} \sim 21.7$.
   These bright, star-forming, candidate cluster-member galaxies
   might be analogous to the spectroscopic member with ID$=$12
   at a cluster-distance of $2.3^{\prime}$ (cf. Table 1).
   Evidence of a population of bright star-forming disc galaxies
   was found in RDCS\,J0910$+$5422 as well and within a cluster-centric
   distance of $2^{\prime}$ (Mei et al. 2006a).

   In conclusion, the photometric redshifts determined in Sect.~3.2
   confirm that XMMU\,J0338.7$+$0030 is a cluster at $z = 1.1$
   and suggest that its galaxy population is qualitatively similar
   to that of the coeval cluster RDCS\,J0910$+$5422.

\subsubsection{Projected galaxy distribution}

   \begin{figure}
   \centering
   \vskip -1.0truecm
      \includegraphics[width=9cm]{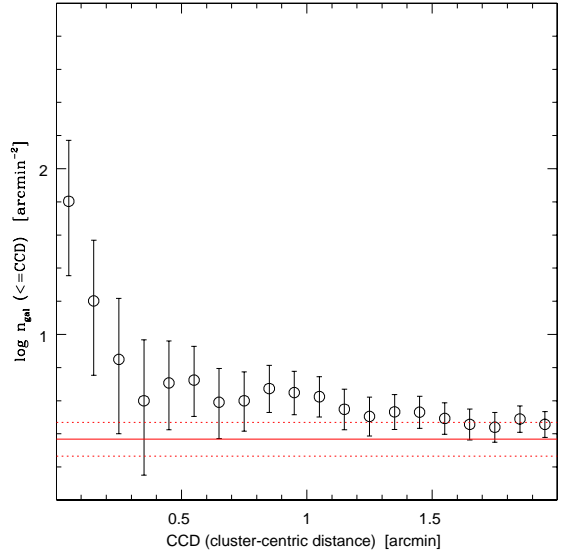}
      \vskip -0.25truecm
      \caption{Cumulative number density of the 44 photometric members
               of XMMU\,J0338.7$+$0030 as a function of their cluster-centric
               distances.
               The mean number density of those members within a concentric,
               circular annulus delimited by radii of $1^{\prime}$
               and $2^{\prime}$ is reproduced together with
               upper and lower limits (red solid and short-dashed lines,
               respectively) computed as in Gehrels (1986).
              }
         \label{FigGalDen}
   \end{figure}

   In the entire region of XMMU\,J0338.7$+$0030 imaged with GROND
   at all seven bands, there are 44 candidate cluster members,
   including three spectroscopic members (Sect.~3.2.1).
   Their projected distribution as a function of cluster-centric distance
   (up to $2^{\prime}$) is reproduced in Fig.~16.
   There are 14 photometric members in the bona-fide cluster region
   and 22 in the concentric annulus of inner and outer radii
   equal to $1^{\prime}$ and $2^{\prime}$
   that has a three times larger area.
   If one assumes this annulus to represent the surrounding low-density region
   (i.e., the ``field''), the over-density of galaxies within $1^{\prime}$
   from the X-ray position of XMMU\,J0338.7$+$0030 has a significance
   of $4.3 \mathrm{\sigma}$.
   The significance rises to $5.3 \mathrm{\sigma}$ if one assumes that
   the region external to the circular area of $2^{\prime}$ radius
   centred on the X-tay position of the cluster represents
   the coeval field environment.
   Considering that a value of $R_{200}$ equal to $88.8^{\prime \prime}$
   (equivalent to $\sim 725~\mathrm{kpc}$) can be estimated
   from the scaling relations in Fassbender et al. (2011c),
   we conclude that the existence of an over-density
   of galaxies at the same photometric redshift of XMMU\,J0338.7$+$0030
   is highly significant.
   This is consistent with the excess by a factor of two
   in the $\mathrm{Ks}$-band galaxy number counts (for the entire region
   imaged with GROND) at $\sim 20.5$--$22.5~\mathrm{AB~mag}$
   with respect to several computations for the field
   from survey areas/deep fields (Fig.~12 in Sect.~3.1).

   In addition, the projected distribution of the 44 photometric members
   is not concentrated towards the X-ray position of the cluster.
   This is consistent only in part with the original inference
   from the OMEGA2000 $\mathrm{z, H}$ imaging that the 2-D distribution
   of the galaxies with very red $z - H$ colours is offset
   by about $20^{\prime \prime}$ (see Fig.~1 and Sect.~2.2.1).
   Indeed, the 2-D distribution of the 44 candidate cluster members
   in the $\mathrm{z^{\prime}}$-image of XMMU\,J0338.7$+$0030
   (right panel of Fig.~4 in Sect.~2.3.1) is broadly consistent
   with the weak, marginally extended X-ray emission of the cluster.
   A large part of the candidate members is distributed
   in the north--south direction.
   This is consistent with the location of the spectroscopic member
   with ID$=$12, that lays $2.3^{\prime}$ away from the X-ray position
   of the cluster and to the north (Sect.~2.4.2).
   The elongation of the X-ray emission from the ICM suggests
   that XMMU\,J0338.7$+$0030 is still in an assembly phase.

   We identify as the brightest central galaxy (BCG)
   the brightest photometric member, which sits on the red locus
   of the $i^{\prime} - z^{\prime}$ vs. $z^{\prime}$ colour--magnitude diagram
   of XMMU\,J0338.7$+$0030 (Fig.~15).
   The BCG is offset by $\sim 43^{\prime \prime}$,
   which corresponds to a projected distance of $\sim 350~\mathrm{kpc}$
   from the X-ray position of the cluster for $z = 1.1$
   and the adopted cosmology (Sect.~1).
   This galaxy is consistently classified as elliptical by {\em le Phare};
   however, it is very close to a bright star, which can affect its photometry.
   For this reason, the identified BCG was not selected as a target
   for spectroscopy with FORS2 (see Sect.~2.4.1).

   Conversely, the spectroscopic member with ID$=$16
   is only $2.1^{\prime \prime}$ away from the X-ray position
   of XMMU\,J0338.7$+$0030: this separation is smaller than
   the on-axis HEW ($\sim 14^{\prime \prime}$) of XMM-{\em Newton}.
   This galaxy exhibits a weak evidence for [O\,II] line emission
   (cf. Fig.~11 and Table 1 in Sect.~2.4.2) but is identified
   as an E galaxy by {\em le Phare}.
   With $Ks = 21.19 \pm 0.10$ and $i^{\prime} - Ks = 3.36 \pm 0.12$
   (i.e., $Ks = 19.40 \pm 0.10$ and $i^{\prime} - Ks = 4.77 \pm 0.12$
   in the Vega magnitude system) it also meets the selection criterion of EROs
   (footnote 11).
   Its classification as a galaxy dominated by old, passively evolving
   stellar populations (i.e., an elliptical galaxy) or a dusty starburst
   is ambiguous because it exhibits $J - Ks = 1.28 \pm 0.13$
   (i.e., $J - Ks = 2.17 \pm 0.13$ in the Vega magnitude system;
   see Pozzetti \& Mannucci 2000; see also fig.~13 in Pierini et al. 2004a).
   This spectroscopic cluster member might be associated
   with part of the emission detected through XMM-{\em Newton}
   because about 15\% of the $I - K$-selected EROs
   with $K \le 20.1~\mathrm{Vega~mag}$ exhibit X-ray properties
   consistent with those expected from luminous, obscured
   active-galactic-nuclei (e.g., Alexander et al. \cite{alexander02}).
   If so, the true X-ray emission centroid of XMMU\,J0338.7$+$0030
   might be different.


\section{Discussion}

   \begin{figure}
   \centering
   \vskip -1.0truecm
      \includegraphics[width=9cm]{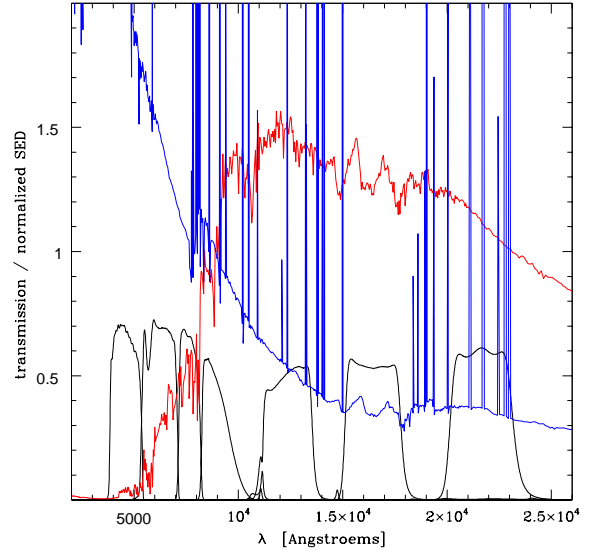}
      \vskip -0.25truecm
      \caption{Transmission curves of the GROND
               $\mathrm{g^{\prime}, r^{\prime}, i^{\prime}, z^{\prime}, J, H, Ks}$ dichroic filters (black)
               plus normalized SEDs that represent a galaxy dominated by
               an old, passively evolving stellar population (red)
               and a very young, dusty starburst galaxy (blue).
               See the text for details.
              }
         \label{FigTranscrves}
   \end{figure}
%

   \begin{figure*}
      \vskip -4.85truecm
   \centering
      \includegraphics[width=15cm]{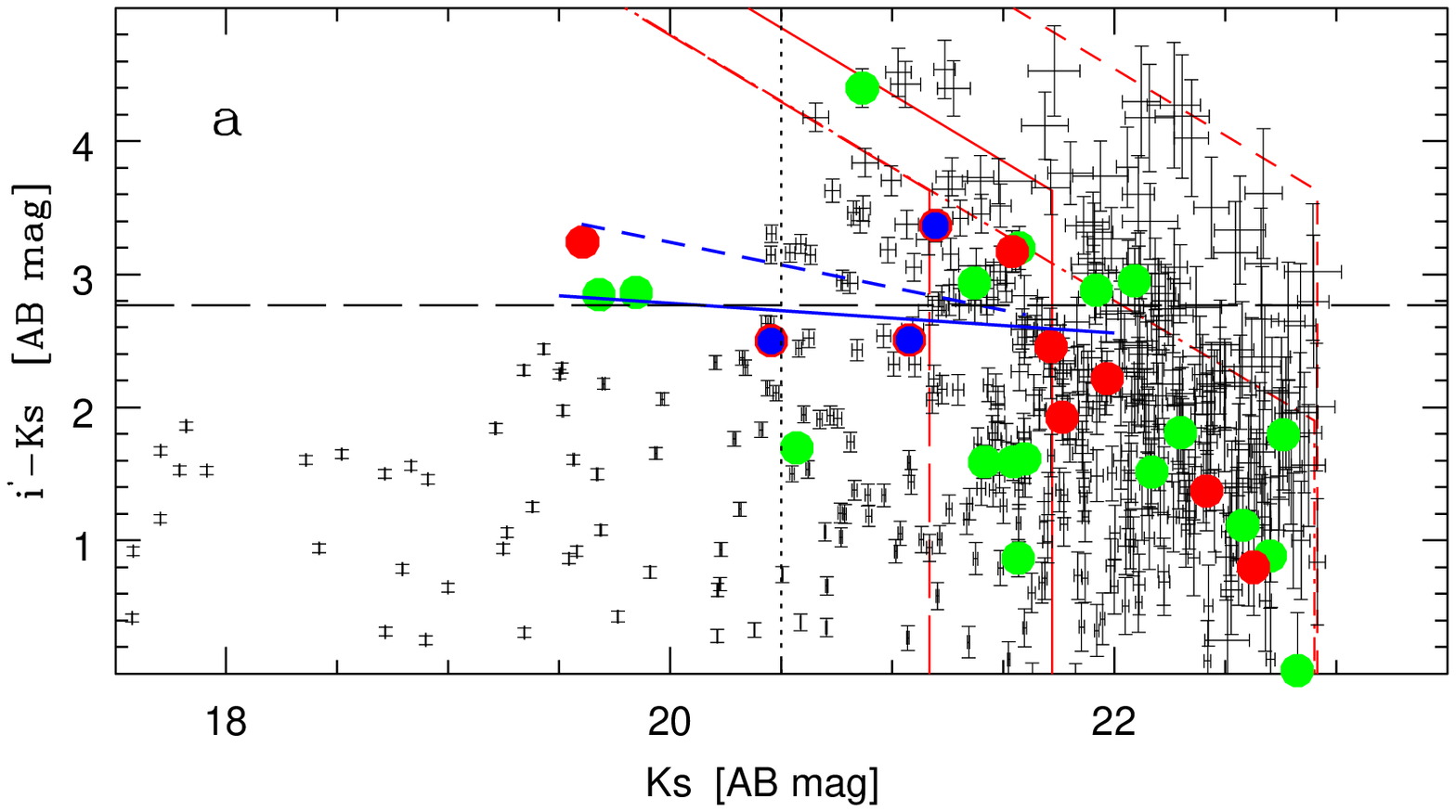}
      \vskip -7.5truecm
      \includegraphics[width=15cm]{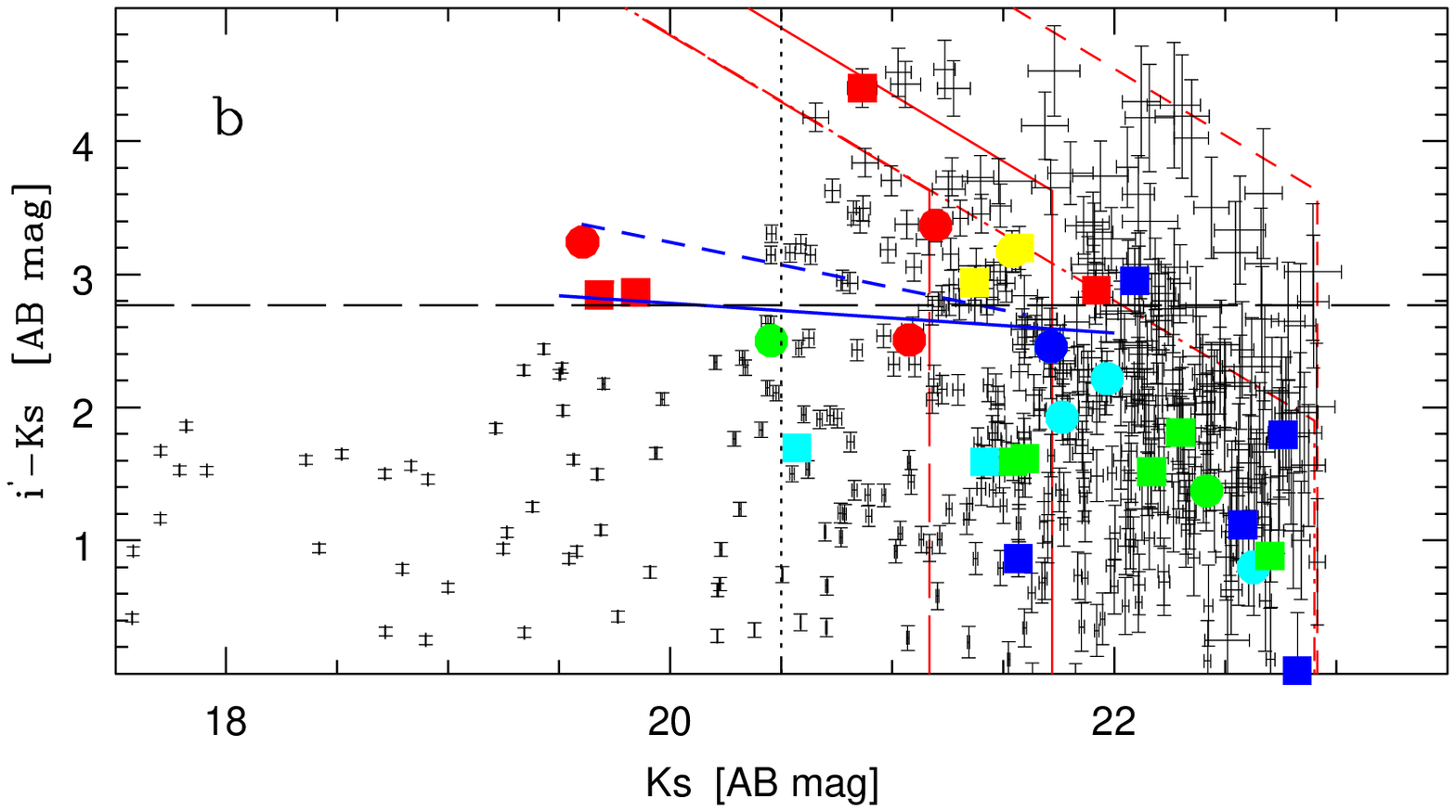}
      \vskip -2.8truecm
      \caption{$i^{\prime} - Ks$ vs $Ks$ colour--magnitude diagram
               for all GROND sources in the XMMU\,J0338.7$+$0030 region
               with flux detections
               in the $\mathrm{i^{\prime}, z^{\prime}, Ks}$ bands
               at a significance level $\ge 1 \mathrm{\sigma}$,
               whatever the classification of the source by {\em Le Phare}
               (i.e., stars are included).
               Photometric errors ($1 \mathrm{\sigma}$) are also shown.
               In each panel, candidate members of XMMU\,J0338.7$+$0030
               are reproduced according to the corresponding panel of Fig.~15.
               Here the red short-dashed, solid and long-dashed lines
               represent the $1 \mathrm{\sigma}$,
               $3 \mathrm{\sigma}$ and $5 \mathrm{\sigma}$ flux thresholds
               in the $\mathrm{i^{\prime}, Ks}$ bands, respectively.
               The red dotted--short-dashed line represents
               the completeness limits in the $\mathrm{i^{\prime}, Ks}$ bands,
               as obtained from Fig.~12.
               The black vertical dotted and horizontal long-dashed lines
               reproduce the characteristic magnitude
               of the $\mathrm{Ks}$-band luminosity function
               of cluster galaxies at $z \sim 1.2$
               (Strazzullo et al. \cite{strazzullo06})
               and the $i^{\prime} - Ks$ colour--selection threshold for EROs
               (Pozzetti \& Mannucci \cite{pozzetti00}).
               As a reference, the blue solid and short-dashed lines
               reproduce the red sequences of the clusters
               RDCS\,J0910$+$5422 at $z = 1.106$
               (Tanaka et al. \cite{tanaka08})
               and RDCS\,J1252.9$-$2927 at $z = 1.24$
               (Tanaka et al. \cite{tanaka09}), respectively.
               The candidate members of XMMU\,J0338.7$+$0030
               that are classified as elliptical galaxies (red symbols
               in panel b) populate a locus that is broadly consistent
               with the red sequences of the other two clusters.
              }
         \label{FigColmagikk}
   \end{figure*}
%

   \begin{figure*}
      \vskip -4.85truecm
   \centering
      \includegraphics[width=15cm]{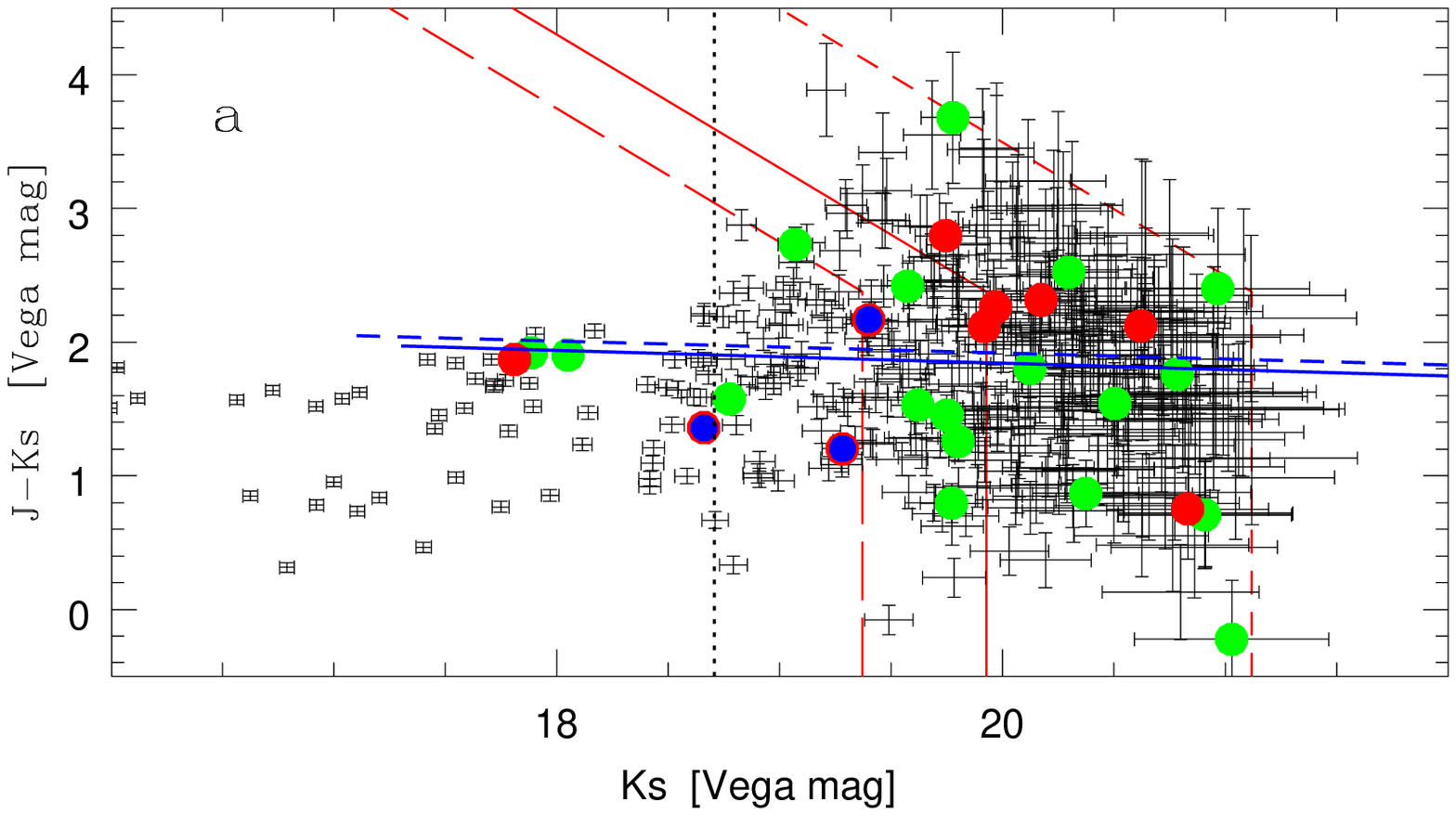}
      \vskip -7.5truecm
      \includegraphics[width=15cm]{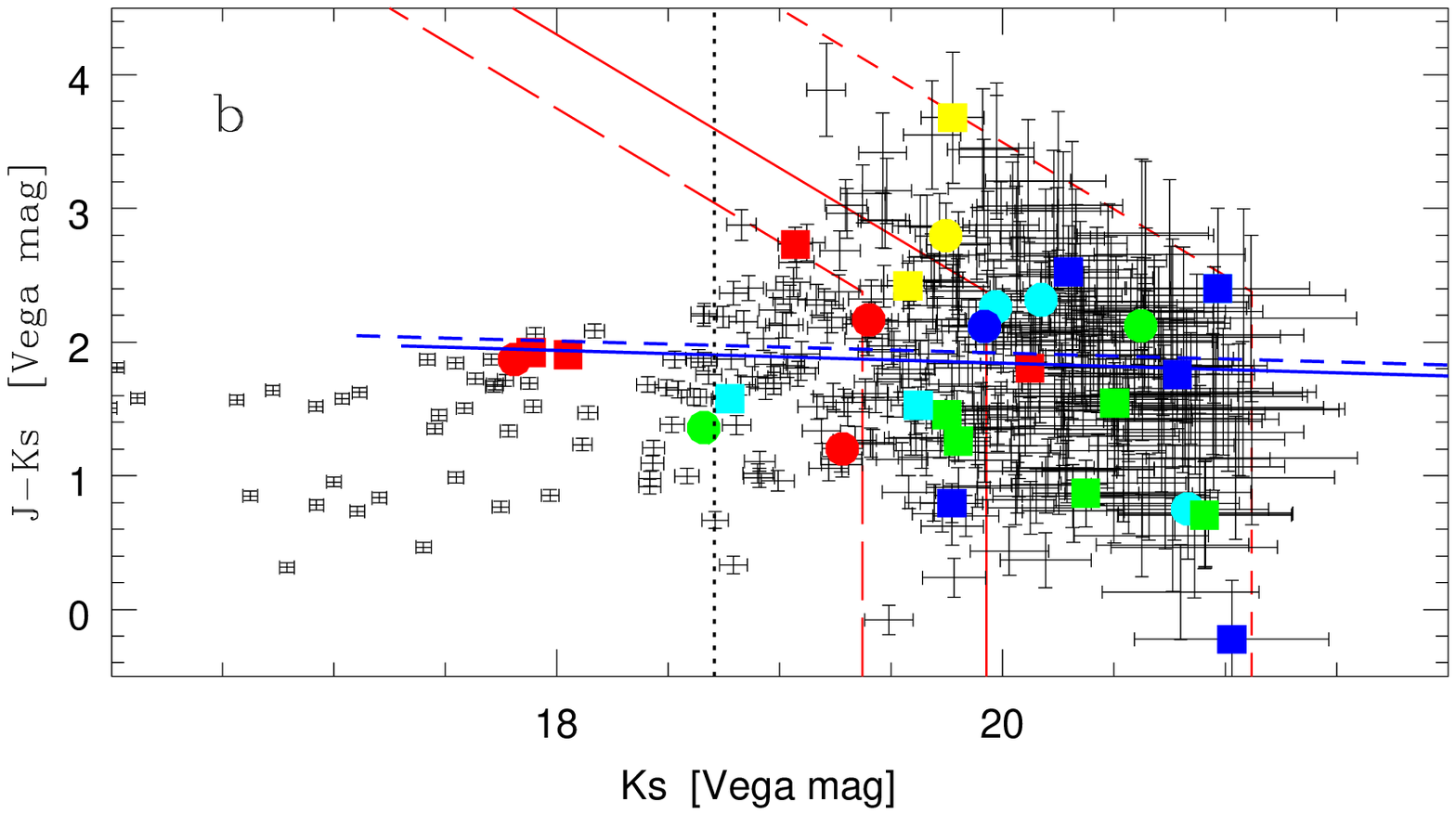}
      \vskip -2.8truecm
      \caption{$J - Ks$ vs $Ks$ colour--magnitude diagram
               for all GROND sources in the XMMU\,J0338.7$+$0030 region
               with flux detections
               in the $\mathrm{z^{\prime}, J, Ks}$ bands
               at a significance level $\ge 1 \mathrm{\sigma}$,
               whatever the classification of the source by {\em Le Phare}
               (i.e., stars are included).
               Photometric errors ($1 \mathrm{\sigma}$) are also shown.
               In each panel, candidate members of XMMU\,J0338.7$+$0030
               are reproduced according to the corresponding panel of Fig.~15.
               Here the red short-dashed, solid and long-dashed lines
               represent the $1 \mathrm{\sigma}$,
               $3 \mathrm{\sigma}$ and $5 \mathrm{\sigma}$ flux thresholds
               in the $\mathrm{J, Ks}$ bands.
               The black vertical dotted and horizontal dotted--long-dashed
               lines reproduce the characteristic magnitude
               of the $\mathrm{Ks}$-band luminosity function
               of cluster galaxies at $z \sim 1.2$
               (Strazzullo et al. \cite{strazzullo06})
               and the $J - Ks$ colour--selection threshold for DRGs
               (Franx et al. \cite{franx03}, see text).               
               As a reference, the blue solid and short-dashed lines
               reproduce the red sequences of the clusters
               RDCS\,J1252.9$-$2927 at $z = 1.24$
               (Lidman et al. \cite{lidman04})
               and XMMU\,J2235.3$-$2557 at $z = 1.39$
               (Lidman et al. \cite{lidman08}), respectively.
               The candidate members of XMMU\,J0338.7$+$0030
               that are classified as elliptical galaxies (red symbols
               in panel b) populate a locus that is broadly consistent
               with the red sequences of the other two clusters.
               However, candidate members that are classified as star forming
               span the same broad range of colours, especially at magnitudes
               fainter than $\sim Ks^{\star}+1$.
              }
         \label{FigColmagjkk}
   \end{figure*}

   Sections 2.1 and 2.2 have described data reduction and analysis
   for the archival XMM-{\em Newton} X-ray observations
   and the near-IR follow-up imaging with OMEGA2000
   that led to the selection of XMMU\,J0338.7$+$0030
   as a candidate cluster at $z > 0.8$.
   Available FORS2 observations indicate four galaxies within $2.3^{\prime}$
   from the X-ray position of this cluster as spectroscopic members.
   This establishes the existence of a bound system of galaxies and hot plasma
   at $< z > = 1.097 \pm 0.002$ ($1 \mathrm{\sigma}$)
   as illustrated in Figs.~10 and 11 and Table 1 (Sect.~2.4.2).
   With an estimated total mass $M_{200} \sim 10^{14}~\mathrm{M}_{\sun}$,
   XMMU\,J0338.7$+$0030 is in between X-ray selected groups and clusters
   (Sect.~2.1.2).

   There is some tension between the spectroscopic redshift
   and the original estimate of $z \sim 1.45 \pm 0.15$ ($1 \mathrm{\sigma}$)
   that was based on the tentative identification of red, passively evolving,
   member galaxies in the $z - H$ vs $z$ colour--magnitude diagram
   of XMMU\,J0338.7$+$0030 (Fig.~3 in Sect.~2.2.2).
   A honest but wrong assumption (see Sect.~2.2.2) as well as systematics
   (see Appendix B) are at the origin of the discrepancy.

   In addition to the spectroscopic observations with FORS2,
   we performed simultaneous optical/near-IR imaging
   of a region of the sky centred on XMMU\,J0338.7$+$0030 with GROND.
   Several pieces of evidence consistently point to the existence
   of an optical counterpart to the weak, marginally extended
   X-ray emission detected with XMM-{\em Newton} in the area imaged at all
   GROND $\mathrm{g^{\prime}, r^{\prime}, i^{\prime}, z^{\prime}, J, H, Ks}$
   bands:
   \begin{itemize}
   \item
   the presence of an excess by a factor of two in the $\mathrm{Ks}$-band
   galaxy number counts at $\sim 20.5$--$22.5~\mathrm{AB~mag}$
   with respect to analogous computations for the field
   from survey areas/deep fields (Fig.~12 in Sect.~3.1);
   \item
   the consistency between the spectroscopic redshift of the cluster
   and the mean value of the photometric redshifts
   of the three spectroscopic members imaged with GROND,
   which is equal to $1.12 \pm 0.09$ ($1 \mathrm{\sigma}$);
   \item
   the selection of 44 sources in the photometric redshift range
   1.01--1.23 that contains the three aforementioned cluster members,
   out of which fourteen are within a cluster-centric distance
   of $1^{\prime}$ and signpost a galaxy over-density at a significance
   of $4.3 \mathrm{\sigma}$ (Sect.~3.2.1);
   \item
   the existence of a red locus in the distribution
   of these 44 photo-$z$ selected galaxies in the $i^{\prime} - z^{\prime}$
   vs $z^{\prime}$ colour--magnitude diagram, which is consistent with
   the red sequence of the cluster RDCS\,J0910$+$5422 at $z = 1.106$
   (Fig.~15 in Sect.~3.2.2);
   \item
   the consistency between the projected spatial distribution
   of the 44 photometric cluster members and the weak, marginally extended
   emission of the X-ray source XMMU\,J0338.7$+$0030 (Fig.~4 in Sect.~2.3.2).
   \end{itemize}

   We acknowledge that the identification
   of three spectroscopic members out of the GROND photometric sample
   of 832 sources was essential to determine the photo-$z$ of the cluster
   as well as to select its candidate (photometric) members.
   Indeed, the photo-$z$ distribution of the sources within $1^{\prime}$
   from the X-ray position of the cluster exhibits two substantial excesses
   at photometric redshifts of 1--1.1 and 1.2--1.3 with respect to
   the photo-$z$ distribution of the sources populating the remaining area
   imaged by GROND at all seven bands (Fig.~14a in Sect.~3.2.1).
   This dicotomy is caused, at least in part, by the following reasons.

   On one hand, the wavelength coverage yielded by the GROND dichroic filters
   for a $z = 1.1$ source is not optimal.
   This is shown in Fig.~17, where the transmission curves
   of the GROND bands are reproduced together with two synthetic SEDs
   (from Pierini et al. 2004a) redshifted to $z = 1.1$
   and normalized at the effective wavelength
   of the $\mathrm{z^{\prime}}$-band.
   One SED corresponds to an SSP model with age of 4 Gyr
   and solar metallicity (red); it represents a galaxy
   characterized by an old, passively evolving stellar population.
   The other corresponds to a composite stellar population model
   with constant star-formation rate (SFR), age of 0.1 Gyr,
   half-solar metallicity, attenuated by dust as for a typical
   nearby starburst (blue); it represents a very young, dusty starburst galaxy.
   The redshifted $4000~\mathrm{\AA}$ break, a fundamental feature
   in redshift determination, falls in between the $\mathrm{i^{\prime}}$-band
   and the $\mathrm{z^{\prime}}$-band.
   The redshifted [O\,II]($\mathrm{\lambda}=3727~\mathrm{\AA}$) emission line
   falls at the red edge of the $\mathrm{i^{\prime}}$-band,
   whereas the redshifted [O\,III]($\mathrm{\lambda}=5007~\mathrm{\AA}$)
   and $\mathrm{H \alpha}$($\mathrm{\lambda}=6563~\mathrm{\AA}$) emission lines
   fall, respectively, in regions of the $\mathrm{z^{\prime}}$
   and $\mathrm{J}$ filters with poor transmission efficiency.

   On the other hand, some ambiguity is favoured by the photometric uncertainty
   yielded by the present, unoptimized GROND observations,
   in particular at near-IR wavelengths (Fig.~7 in Sect.~2.3.2).
   It is known that the photometric redshift technique
   can be reliably extended from $z \sim 1$ to $z \sim 2$
   only if deep near- and mid-IR photometry is available
   (e.g., Bolzonella, Miralles, Pell\'o 2000; Ilbert et al. 2009).

   In spite of the unfortunate mapping of the rest-frame SED
   of a $z = 1.1$ galaxy yielded by the GROND channels
   and the limited photometric accuracy of the present near-IR data,
   it is remarkable that the photometric and spectroscopic estimates
   of the redshift of XMMU\,J0338.7$+$0030 coincide within an uncertainty
   of 0.09 (cf. e.g. Lerchster et al. 2011).
   Indeed, the availability of spectroscopic redshifts for only 0.6\%
   of the GROND sources (and in the narrow redshift range 0.85--1.1) prevented
   any meaningful training of the solutions of the photo-$z$ code
   {\em le Phare}.

   Another exquisite possibility offered
   by the simultaneous optical/near-IR imaging with GROND
   is the characterization of the galaxy population of a high-$z$ cluster
   on the basis of SEDs - not of a single colour--magnitude diagram -
   in one and the same study
   (cf. Mullis et al. 2005 and Lidman et al. 2008 for XMMU\,J2235.3$-$2557;
   Lidman et al. 2004 and Tanaka et al. 2009 for RDCS\,J1252.9$-$2927;
   Mei et al. 2006a and Tanaka et al. \cite{tanaka08} for RDCS\,J0910$+$5422).
   For instance, the distribution in the $i^{\prime} - z^{\prime}$
   vs $z^{\prime}$ colour--magnitude diagram of the 44 photo-$z$ selected
   members of XMMU\,J0338.7$+$0030 resembles that of the members
   of the cluster RDCS\,J0910$+$5422 at $z = 1.106$, as determined
   by Mei et al. (2006a), as shown in Fig.~15.
   At the same time, the fiducial members of XMMU\,J0338.7$+$0030
   that are classified as elliptical galaxies
   populate a red locus in the $i^{\prime} - Ks$ vs $Ks$ colour--magnitude
   diagram, which is broadly consistent with the red sequence
   of RDCS\,J0910$+$5422 determined by Tanaka et al. (\cite{tanaka08}),
   as shown in Fig.~18b.

   The rather neat progression from red, passively evolving,
   elliptical galaxies to blue, star-forming, late-type spirals and irregulars
   in Fig.~15b is not preserved in other colour--magnitude diagrams
   that are used to constrain distance and properties
   of the galaxy population of high-$z$ clusters.
   The identification of a red-sequence and the classification
   of a cluster member galaxy as passively evolving or star forming
   become much more ambiguous in those colour--magnitude diagrams
   that involve broad-band filters that do not map the redshifted
   $4000~\mathrm{\AA}$ break.
   This is demonstrated by the distribution of the fiducial members
   of XMMU\,J0338.7$+$0030 in the $i^{\prime} - Ks$ vs $Ks$
   and $J - Ks$ vs $Ks$ colour--magnitude diagrams
   (Figs.~18 and 19, respectively).
   As a reference, we reproduce the red-sequences
   of the spectroscopically confirmed clusters
   RDCS\,J0910$+$5422 at $z = 1.106$ (Tanaka et al. \cite{tanaka08}),
   RDCS\,J1252.9$-$2927 at $z = 1.24$ (Lidman et al. \cite{lidman04};
   Tanaka et al. \cite{tanaka09}) and XMMU\,J2235.3$-$2557 at $z = 1.39$ (Lidman et al. \cite{lidman08})\footnote{The observed red-sequences and colour thresholds for $i^{\prime} - Ks$ selected EROs (Pozzetti \& Mannucci \cite{pozzetti00}) and $J - Ks$ selected distant red galaxies (DRGs, Franx et al. \cite{franx03}) reproduced in the two figures were corrected for the different filter transmission functions, assuming a $4~\mathrm{Gyr}$-old SSP with solar metallicity (see Pierini et al. \cite{pierini04a}).}.

   A galaxy classification based on either spectro-photometry or morphology
   indicates that there are by far not as many early-type galaxies
   within a cluster-centric distance of $1^{\prime}$ in XMMU\,J0338.7$+$0030
   as in RDCS\,J0910$+$5422: 3 vs 31 (Sect.~3.2.2).
   A high percentage of passively evolving galaxies
   among the brightest members of a cluster should not necessarily be expected
   if the cluster is observed at an early stage of its halo assembly
   (e.g., Tanaka et al. 2008; see also Braglia et al. 2009
   for the most massive clusters at $z \sim 0.3$).
   This could be the case of XMMU\,J0338.7$+$0030, according to the projected
   distribution of its candidate member galaxies (Figs.~4 and 16).

   The spectroscopically confirmed, optically-selected proto-cluster
   Cl\,0332$-$2742 at $z = 1.6$ provides evidence of the build-up
   of the red sequence during its assembly (Kurk et al. 2009).
   Conversely, the X-ray luminous cluster XMMU\,J2235.3$-$2557 at $z = 1.393$
   (Strazzullo et al. 2010) exhibits a tight red-sequence of massive galaxies,
   with overall old stellar populations, generally early-type morphology,
   typically showing early-type spectral features
   and rest-frame far-UV emission consistent with very low SFRs.
   The massive cluster XMMU\,J1230.3$+$1339 at $z = 0.975$
   exhibits a richly populated red sequence,
   but also a few, spectro-photometrically classified, early-type galaxies
   that potentially host some residual star-formation activity
   (Lerchster et al. 2011).
   Consistently, the X-ray selected cluster RDCS\,J0910$+$5422,
   exhibits a red sequence although its early-type galaxy population
   appears to be still forming (Mei et al. 2006a; Tanaka et al. 2008).
   With a total mass of about $10^{14}~\mathrm{M}_{\sun}$
   (Tanaka et al. 2008), RDCS\,J0910$+$5422 is as massive
   as XMMU\,J0338.7$+$0030 and four times less massive
   than XMMU\,J1230.3$+$1339 at $z = 0.975$ (Fassbender et al. 2011a;
   Lerchster et al. 2011).

   If XMMU\,J0338.7$+$0030 lacks passively evolving galaxies,
   it hosts a high percentage of star-forming systems
   (late-type galaxies or starbursts),
   judging from the spectro-photometric classification
   of its candidate members (Fig.~15b).
   Nine of its 10 candidate members that are spectro-photometrically
   classified as starbursts are located well beyond a cluster-centric distance
   of $1^{\prime}$ (right panel of Fig.~4).
   Consistently, the spectroscopically confirmed member
   at a cluster-centric distance of $2.3^{\prime}$ (ID$=$12)
   is a bright galaxy with a quite strong [O\,II] line emission (Sect.~2.4.2).
   The other bright spectroscopic member with evidence
   of [O\,II] line emission (ID$=$15) is a photometric member as well
   but is only $50^{\prime \prime}$ away from the cluster centre
   and sits on the red sequence in Fig.~15
   as well as the two spectroscopic members classified as elliptical galaxies
   (ID$=$2 and ID$=$15) at cluster-centric distances of $28^{\prime \prime}$
   and $2.1^{\prime \prime}$, respectively.

   Interestingly, there is evidence of sustained star-formation activity
   at the bright end of the galaxy luminosity function in the X-ray luminous
   cluster XMMU\,J1007.4$+$1237 at $z = 1.555$ (Fassbender et al. 2011b)
   and in dense environments at $z \sim 1$ (e.g., Elbaz et al. 2007;
   Gerke et al. 2007; Cooper et al. 2008).
   Several [O\,II] line emitters are found
   in the luminous, red galaxy population of XMMU\,J0302.2$-$0001
   at $z = 1.185$ with a total mass of a few times $10^{14}~\mathrm{M}_{\sun}$
   (\u{S}uhada et al. 2011). 
   Spectroscopic cluster members with [O\,II] line emission
   can populate the locus of old, passively evolving galaxies
   or be even redder than the red sequence (see XMMU\,J2235.3$-$2557;
   Lidman et al. 2008).
   Consistently, a non negligible fraction (20\%)
   of the luminous infrared galaxies in 16 optically-selected galaxy clusters
   at $0.4 < z < 0.8$ sits on the corresponding red sequences
   (Finn et al. 2010), the remaining 80\% laying below.
   This is no surprise when the effect of dust attenuation
   on the observed SED of a galaxy is computed in a physical way,
   tailored to its star-formation mode\footnote{Assumptions on the relative distribution of dust and stars as well as on the extinction law seem to impact the estimates of SFR and stellar mass in a systematic way (e.g.,  K\"{u}pc\"{u} Yolda\c{s} et al. 2007).} (cf. Pierini et al. 2004a, 2004b, 2005).

   In addition, star-forming galaxies at $z \sim 1$--2.5,
   selected in the rest-frame ultraviolet (UV)
   and with stellar masses of $10^{10}$--$10^{11}~\mathrm{M}_{\sun}$,
   exhibit evidence of complex configurations of stars
   and dusty interstellar medium (e.g., Noll et al. 2007, 2009).
   This points to the existence of dusty gas outflows,
   likely driven by feedback from supernovae, at high redshifts
   (cf. Steidel et al. 2010), but gravitational interactions cannot be excluded\footnote{Hence, some complexity in the relation between bolometric infrared luminosity associated with re-emission by dust at mid-IR through sub-millimeter wavelengths and colour temperature of the warm and cold dust components can be expected. This could indeed have already been observed (cf. Hwang et al. 2010).}.
   Extended distributions of dust not associated with the main bodies
   of galaxies have been discovered in the nearby Stephan's Quintet
   (Natale et al. 2010) and M81 triplet of galaxies (Walter et al. 2011).
   Walter et al. (2011) highlight the importance of tidal stripping
   for the metal enrichment of the intergalactic medium at high redshifts
   in addition to outflows.
   Here we note that the additional cooling of the ICM on dust grains
   impacts on the X-ray scaling properties of groups and clusters of galaxies
   (e.g., da Silva et al. 2009; Natale et al. 2010).
   Furthermore, early preheating of the ICM is likely responsible for
   the observed evolution of the X-ray scaling relations of galaxy clusters
   out to $z \sim 1.5$ (Reichert et al. 2011).

   In conclusion, XMMU\,J0338.7$+$0030 appears as a bound, low-mass system
   of galaxies at $z = 1.1$ that is caught in an early assembly phase
   and where the shaping of the bright end of the galaxy luminosity function
   and nurturing effects on galaxy evolution
   (see Boselli \& Gavazzi 2006 for a review) can be sought.
   A full understanding of the properties of the galaxy population and ICM
   of this cluster requires a more in-depth study based on spectroscopy
   and multi-wavelength imaging, which offer better statistics
   and/or superior quality.


\section{Conclusions}

   This paper presents, in particular, results
   from the first pointed observations of a high-$z$ cluster with GROND,
   the seven-channel imager mounted at the MPI/ESO 2.2m telescope of La Silla,
   Chile.
   The target was the X-ray selected, weak, marginally extended source
   XMMU\,J0338.7$+$0030 in the XMM-{\em Newton} Distant Cluster Project
   (XDCP) survey.

   Follow-up imaging with OMEGA2000 at the 3.5m CAHA telescope
   enabled a comparison of the reddest $z - H$ colours
   of the galaxies within a cluster-centric distance of $45^{\prime \prime}$
   with predictions from simple stellar population evolutionary models,
   which suggested a redshift of $z = 1.45 \pm 0.15$ ($1 \mathrm{\sigma}$).
   Later available VLT/FORS2 spectroscopy identified XMMU\,J0338.7$+$0030
   as a galaxy cluster at $z = 1.097 \pm 0.002$ ($1 \mathrm{\sigma}$)
   with four spectroscopic members.
   From the determination of the flux in the soft 0.5--$2~\mathrm{keV}$ band,
   the bolometric luminosity of the cluster was inferred.
   The ensuing X-ray estimate of the total mass of XMMU\,J0338.7$+$0030,
   based on a luminosity scaling relation,
   is equal to $M_{200} \sim 10^{14}~\mathrm{M}_{\sun}$. 

   About six hours of simultaneous imaging in the GROND
   $\mathrm{g^{\prime}, r^{\prime}, i^{\prime}, z^{\prime}, J, H, Ks}$ bands
   enabled us to extract photometry for 832 sources detected
   down to $z^{\prime}_{\mathrm{AB}} \sim 26$ ($1 \mathrm{\sigma}$).
   Existing coverage of the same region by the shallower
   {\em Sloan} Digital Sky Survey and 2 Micron All Sky Survey
   allows a robust photometric calibration to be obtained
   for the optical and near-infrared (IR) channels of GROND, respectively.
   This is the basis for the application
   of the photometric redshift technique,
   as implemented in the publicly available code {\em le Phare}.

   With the GROND data, we confirmed the existence of an optical counterpart
   to XMMU\,J0338.7$+$0030 on the basis of five consistent and independent
   pieces of evidence.
   In particular, the Ks-band galaxy number counts for the cluster region
   imaged with GROND exhibit an excess by a factor of two
   at 20.5--$22.5~\mathrm{AB~mag}$ with respect to those
   from survey areas/deep fields.
   In addition, the photometric redshifts of the three spectroscopic members
   imaged with GROND ($1.12 \pm 0.09$, $1 \mathrm{\sigma}$)
   is consistent with the spectroscopic redshift of the cluster.
   Furthermore, the distribution of the 44 candidate members
   of XMMU\,J0338.7$+$0030 (with photometric redshifts within the range
   1.01--1.23) in the $i^{\prime} - z^{\prime}$ vs $z^{\prime}$
   colour--magnitude diagram exhibits a red locus and a bluer one
   that are consistent with the red sequence
   and the distribution of the star-forming galaxy population
   of the similarly massive cluster RDCS\,J0910$+$5422 at $z = 1.106$.

   XMMU\,J0338.7$+$0030 seems to host a galaxy population
   that can still undergo significant bursts of star-formation activity,
   as also found in other XDCP high-$z$ clusters.
   Part of these star-forming galaxies can exhibit colours that are at least
   as red as those of passively evolving galaxies, owing to dust attenuation,
   in agreement with results in the literature.
   A comparison of the projected distributions of the X-ray emitting plasma
   and the photometric cluster members suggests that XMMU\,J0338.7$+$0030
   is likely caught in an early phase of its assembly:
   its star-forming galaxies fall in and face an environment
   that rapidly changes.

   Finally, we acknowledge that the identification
   of three spectroscopic members out of the 832 GROND sources was essential
   to determine the photo-$z$ of the cluster as well as to select
   its candidate (photometric) members.
   However, it is remarkable that these two estimates of the redshift
   of XMMU\,J0338.7$+$0030 coincide within an uncertainty of 0.09,
   given that the so-called $4000~\mathrm{\AA}$ break
   (a fundamental feature for photo-$z$ determination) falls in between
   the GROND $\mathrm{i^{\prime}}$ and $\mathrm{z^{\prime}}$ channels
   for a $z = 1.1$ galaxy and the present unoptimized observations with GROND
   yielded a limited photometric accuracy in the near-$IR$ bands.
   We also stress that no training of the solutions of the photo-$z$ code
   was performed, because the available spectroscopic information
   is limited to only 0.6\% of the GROND sources.

   For these reasons, this unique imager is useful
   for establishing optical counterparts to X-ray selected clusters,
   even in absence of spectroscopic priors.
   This is particularly true for high throughput surveys for clusters
   like the eROSITA all-sky survey, which is expected to mostly deliver
   massive clusters at intermediate redshifts.
   These clusters should host a wealthy population of old, passively evolving
   galaxies, which define a neat red sequence in at least one
   of the colour--magnitude diagrams that can be built
   out of the GROND optical/near-IR photometry obtained
   in less than two hours of observations.

\begin{acknowledgements}

DP thanks the anonymous referee for her/his insightful comments, which led to a signficant improvement on the robustness and presentation quality of the results in the final version of the paper.

This research was supported by the DFG cluster of excellence ``Origin and Structure of the Universe'' (http://www.universe-cluster.de) through EXC project number 153, the DFG under grants Schw 536/24-1, Schw 536/24-2, BO 702/16-3, and the DLR under grants 50 OR 0405 and 50 QR 0802.

RS acknowledges support by the DFG in the program SPP 1177.

HQ thanks the FONDAP Centro de Astrof\'{\i}sica for partial support.

FZ acknowledges support from and participation in the International Max-Plank Research School on Astrophysics at the Ludwig-Maximilians University.

DP acknowledges useful feedback from O. Ilbert.

DP acknowledges the kind hospitality at the Max-Planck-Institut f\"ur extraterrestrische Physik (MPE).

This research has made use of observations collected at the European Organisation for Astronomical Research in the Southern Hemisphere (ESO), Chile (079.A$-$0634).

This research has made use of observations collected at the Centro Astron\'omico Hispano Alem\'an (CAHA) at Calar Alto, Spain operated jointly by the Max-Planck-Institut f\"ur Astronomie and the Istituto de Astrof\'{\i}sica de Andaluc\'{\i}a (CSIC).

This work has made use of the SDSS database. Funding for the SDSS and SDSS-II has been provided by the Alfred P. Sloan Foundation, the Participating Institutions, the National Science Foundation, the U.S. Department of Energy, the National Aeronautics and Space Administration, the Japanese Monbukagakusho, the Max Planck Society, and the Higher Education Funding Council for England. The SDSS Web Site is http://www.sdss.org/.

The SDSS is managed by the Astrophysical Research Consortium for the Participating Institutions. The Participating Institutions are the American Museum of Natural History, Astrophysical Institute Potsdam, University of Basel, University of Cambridge, Case Western Reserve University, University of Chicago, Drexel University, Fermilab, the Institute for Advanced Study, the Japan Participation Group, Johns Hopkins University, the Joint Institute for Nuclear Astrophysics, the Kavli Institute for Particle Astrophysics and Cosmology, the Korean Scientist Group, the Chinese Academy of Sciences (LAMOST), Los Alamos National Laboratory, the Max-Planck-Institute for Astronomy (MPIA), the Max-Planck-Institute for Astrophysics (MPA), New Mexico State University, Ohio State University, University of Pittsburgh, University of Portsmouth, Princeton University, the United States Naval Observatory, and the University of Washington.

This publication makes use of data products from the Two Micron All Sky Survey, which is a joint project of the University of Massachusetts and the Infrared Processing and Analysis Center/California Institute of Technology, funded by the National Aeronautics and Space Administration and the National Science Foundation.

This research has made use of the NASA/IPAC Extragalactic Data base (NED) which is operated by the Jet Propulsion Laboratory, California Institute of Technology, under contract with the National Aeronautics and Space Administration.

\end{acknowledgements}

\begin{appendix}

\section{Photometric redshifts: accuracy and reliability} 

   \begin{figure}
   \centering
   \vskip -0.75truecm
      \includegraphics[width=9cm]{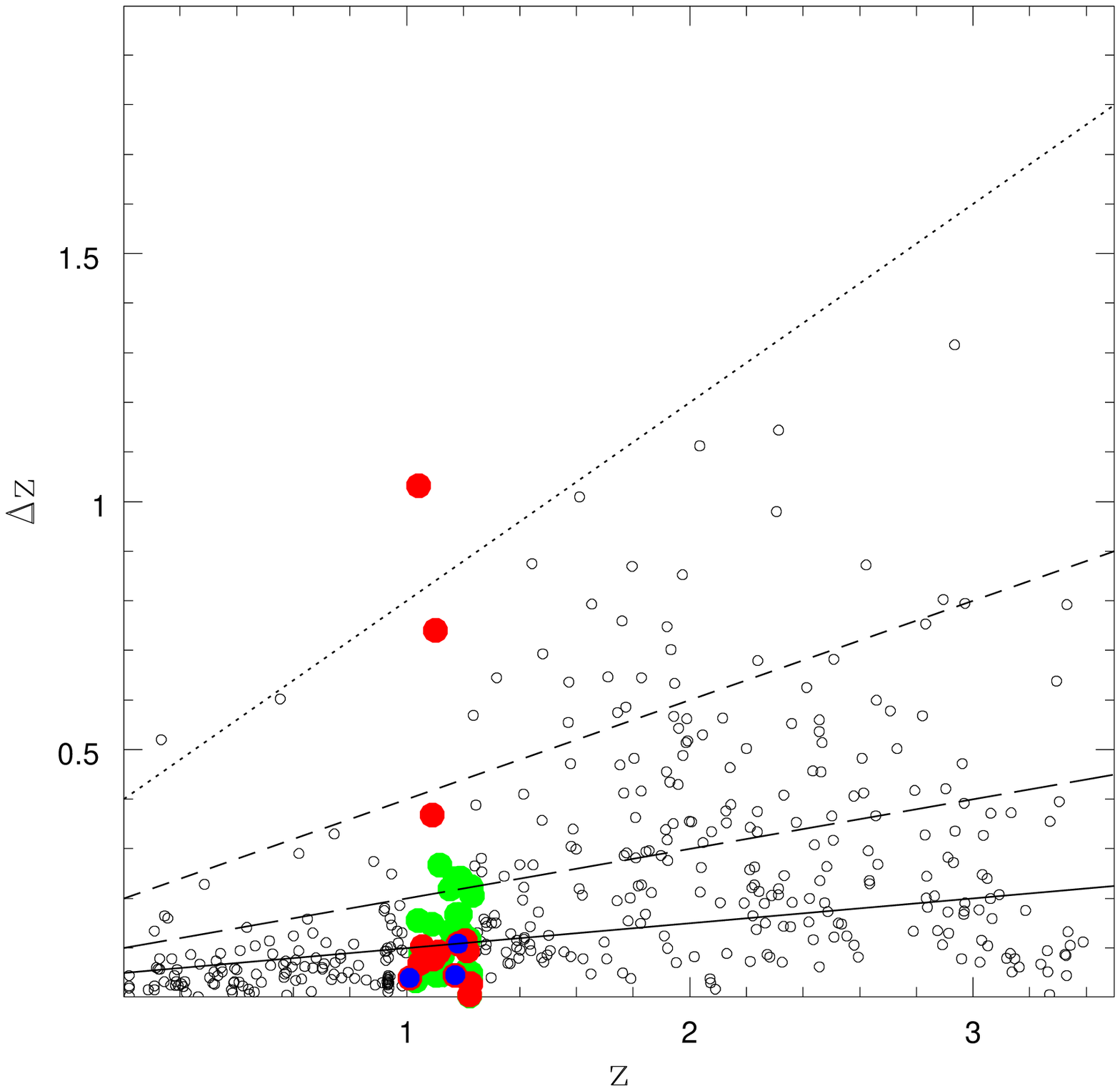}
      \vskip -0.15truecm
      \caption{Distribution of the 68\% range of the photo-$z$ solutions
               ($\Delta z$) as a function of the associated best-fit photo-$z$
               for the 436 GROND sources identified as galaxies
               (empty circles).
               Candidate members of XMMU\,J0338.7$+$0030
               are reproduced with red (green) filled circles
               if they are within (outside) the bona fide cluster region;
               the three spectroscopic members among them
               are marked with blue filled circles (see Sect.~3.3.2).
               The dotted, short-dashed, long-dashed and solid lines
               represent values of $\Delta z$ equal to 0.4, 0.2, 0.1
               and $0.05 \times (1 + z)$, respectively.
              }
      \label{app1fig1}
   \end{figure}
%

   \begin{figure}
   \centering
   \vskip -0.75truecm
      \includegraphics[width=9cm]{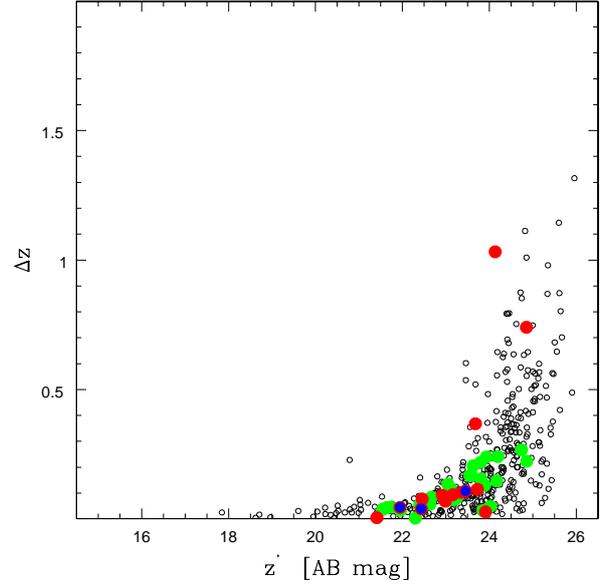}
      \vskip -0.15truecm
      \caption{Distribution of $\Delta z$ as a function of
               the $\mathrm{z^{\prime}}$-band magnitude
               for the same GROND sources displayed in Fig.~A.1.
               Symbols have the same meaning as well.
              }
      \label{app1fig2}
   \end{figure}
%

   \begin{figure}
   \centering
   \vskip -0.5truecm
      \includegraphics[width=9cm]{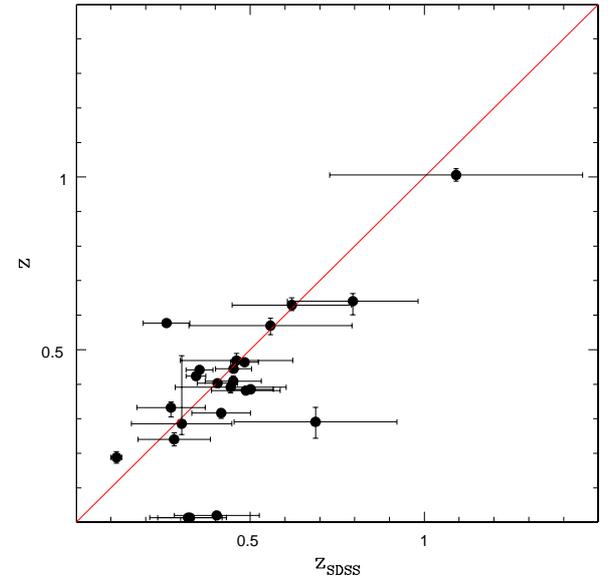}
      \vskip -0.15truecm
      \caption{Comparison of photometric redshifts for the 24 galaxies
               that are in common between our GROND photometric catalogue
               and the SDSS DR7 one (Sect.~2.4.1).
               For the SDSS photo-$z$'s, uncertainties correspond
               to $\pm 1 \mathrm{\sigma}$ values.
               The red line represents equality.
              }
      \label{app1fig3}
   \end{figure}
%

   \begin{figure}
   \centering
   \vskip -0.75truecm
      \includegraphics[width=9cm]{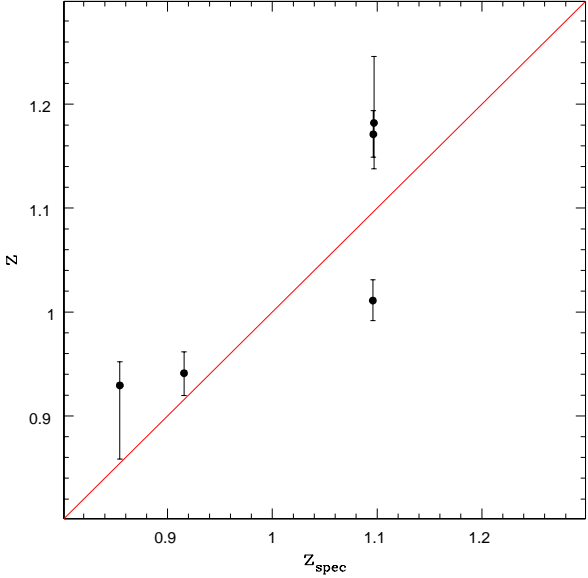}
      \vskip -0.15truecm
      \caption{Comparison of photometric and spectroscopic redshifts
               for the five sources in the GROND photometric catalogue
               observed with FORS2 (Sect.~2.4.1).
               For each spectroscopic value, we show
               the 68\% range of the photometric estimates.
               The red line represents equality.
              }
         \label{app1fig4}
   \end{figure}

   Here we discuss the accuracy of the photometric estimates
   of the distances to the individual sources
   in the $3.9 \times 4.3~\mathrm{arcmin}^2$ region of XMMU\,J0338.7$+$0030
   imaged with GROND that is achieved through the GROND multi-band photometry
   and the photo-$z$ code {\em le Phare} (see Sect.~3.2).
   At variance with works on survey areas/deep fields, we do so
   on the basis of photometric redshifts, mostly, because spectroscopic ones
   are available for only 0.6\% of the GROND photometric sample (Sect.~2.4.1).

   In particular, we reproduce the behaviour of the 68\% range
   of the photo-$z$ solutions yielded by {\em le Phare},
   $\Delta z$\footnote{Given its definition, $\Delta z$ corresponds to twice the value expected for a Gaussian photo-$z$ distribution centred at the best-fit value.}, as a function of the best-fit photo-$z$ or $\mathrm{z^{\prime}}$-band
   magnitude for the 436 sources out of 832 that are classified as galaxies
   by the best-fit solution (Figs.~A.1 and A.2, respectively).
   The CFHTLS galaxy templates without emission lines are used.
   The 42 galaxies selected as fiducial cluster members
   (i.e., with $1.01 \le z \le 1.23$, see Sect.~3.3.1)
   are colour-coded as in Fig.~15 (Sect.~3.2.2).

   Overall, the distribution of these photometric redshifts
   is well described by a straight line of equation
   $\Delta z \sim 0.05 \times (1 + z)$ up to $z \sim 1.5$;
   only 2 photo-$z$ estimates (i.e., 0.4\%) behave as outliers,
   in the sense that they exhibit a value of $\Delta z > 0.4 \times (1 + z)$.
   Unsurprisingly, photometric redshifts lower than 1 are assigned
   with an accuracy that is better than $\Delta z \sim 0.05 \times (1 + z)$,
   whereas photometric redshifts between 1.5 and 3 exhibit a mean accuracy
   equal to $\Delta z \sim 0.1 \times (1 + z)$ but with a large scatter.
   In addition, an accuracy of 0.2 or better characterizes
   the photometric redshifts attributed to galaxies that are brighter
   than $z^{\prime} \sim 23.5$, which corresponds to a $10 \mathrm{\sigma}$
   detection (see Sect.~2.3.2).

   The candidate member galaxies of XMMU\,J0338.7$+$0030
   tend to exhibit the lowest values of $\Delta z$ among the galaxies
   with the same $\mathrm{z^{\prime}}$-band magnitude (Fig.~A.2).
   However, three fiducial members within a cluster-centric distance
   of $1^{\prime}$ exhibit values of $\Delta z$ between 0.3 and 1.1:
   one is an outlier (Fig.~A.1).

   As a consistency check of our set of photometric redshifts,
   we compared photo-$z$'s for 24 GROND sources
   with counterparts in the SDSS DR7 (see Sect.~2.3.2).
   The comparison SDSS photometric redshifts were calculated
   using a Neural Network method (Oyaizu et al. 2008).
   The agreement between these two independent sets
   of photo-$z$ estimates is good up to $z \sim 1$.
   As expected, our values have significantly lower uncertainties
   since they are based on medium--deep photometry
   in the $\mathrm{g^{\prime}, r^{\prime}, i^{\prime}, z^{\prime}, J, H, Ks}$
   bands (e.g., Bolzonella, Miralles, Pell\'o 2000).
   However, the $\mathrm{u}$-band photometry that is available
   from the SDSS shallower imaging in the $\mathrm{u, g, r, i, z}$ bands
   (Abazajian et al. 2009 and references therein)
   better helps constraining photo-$z$'s of galaxies at $z \le 0.4$
   (Bolzonella, Miralles, Pell\'o 2000).

   Finally, we reproduce photometric redshifts vs spectroscopic ones
   for the five extragalactic sources in the GROND photometric catalogue
   that were observed with VLT-FORS2 (see Sect.~2.4.1) in Fig.~A.4.
   In spite of a slight tendency towards higher values,
   our photometric redshifts, computed without any training of the solutions
   of the photo-$z$ code {\em le Phare}, agree with the spectroscopic ones
   within $\pm 0.1$.

\section{On the z$^{\prime}$ - H vs H colour--magnitude diagram of XMMU\,J0338.7$+$0030} 

   \begin{figure*}
   \centering
   \vskip -4.85truecm
      \includegraphics[width=15cm]{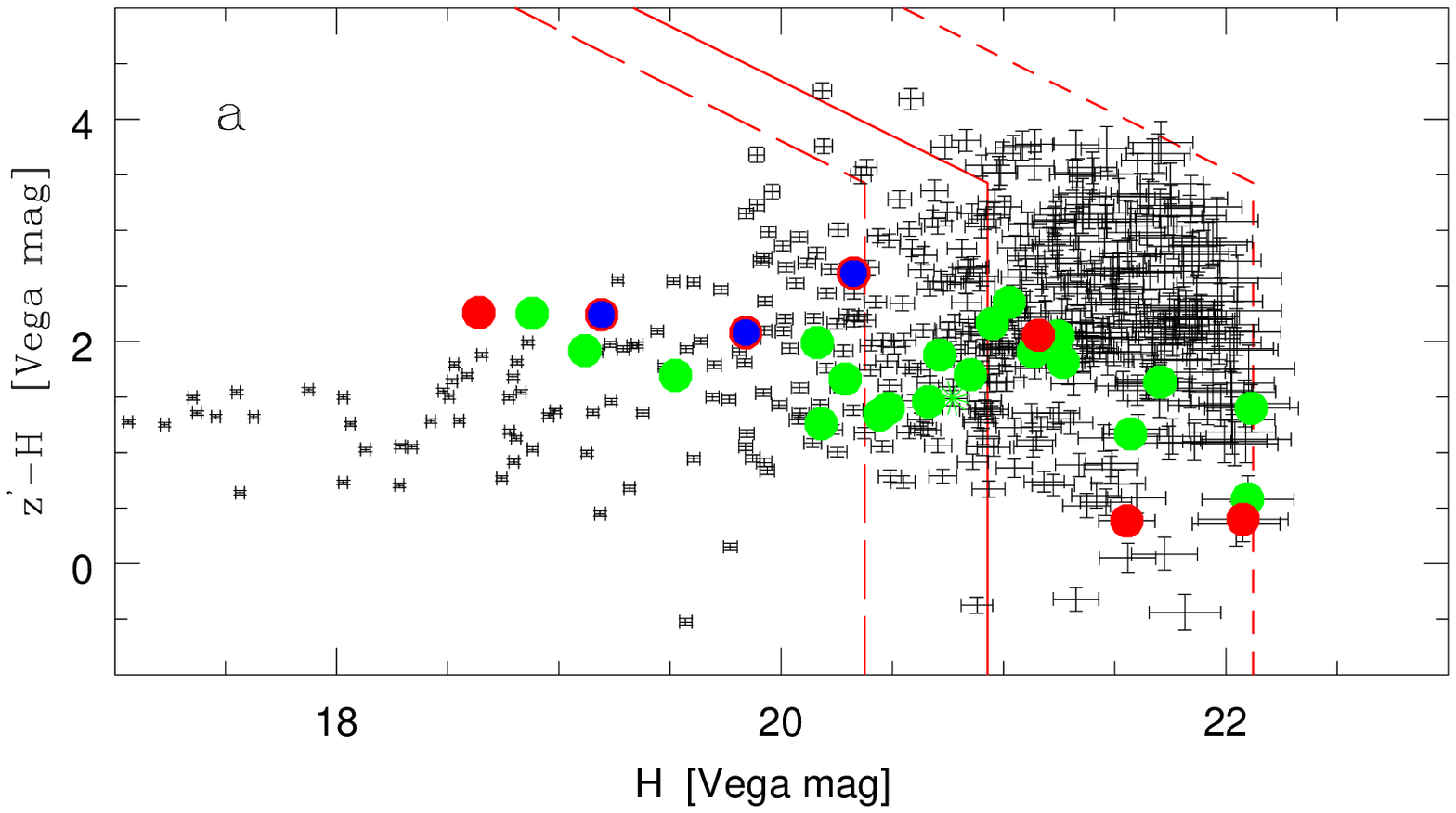}
      \vskip -7.5truecm
      \includegraphics[width=15cm]{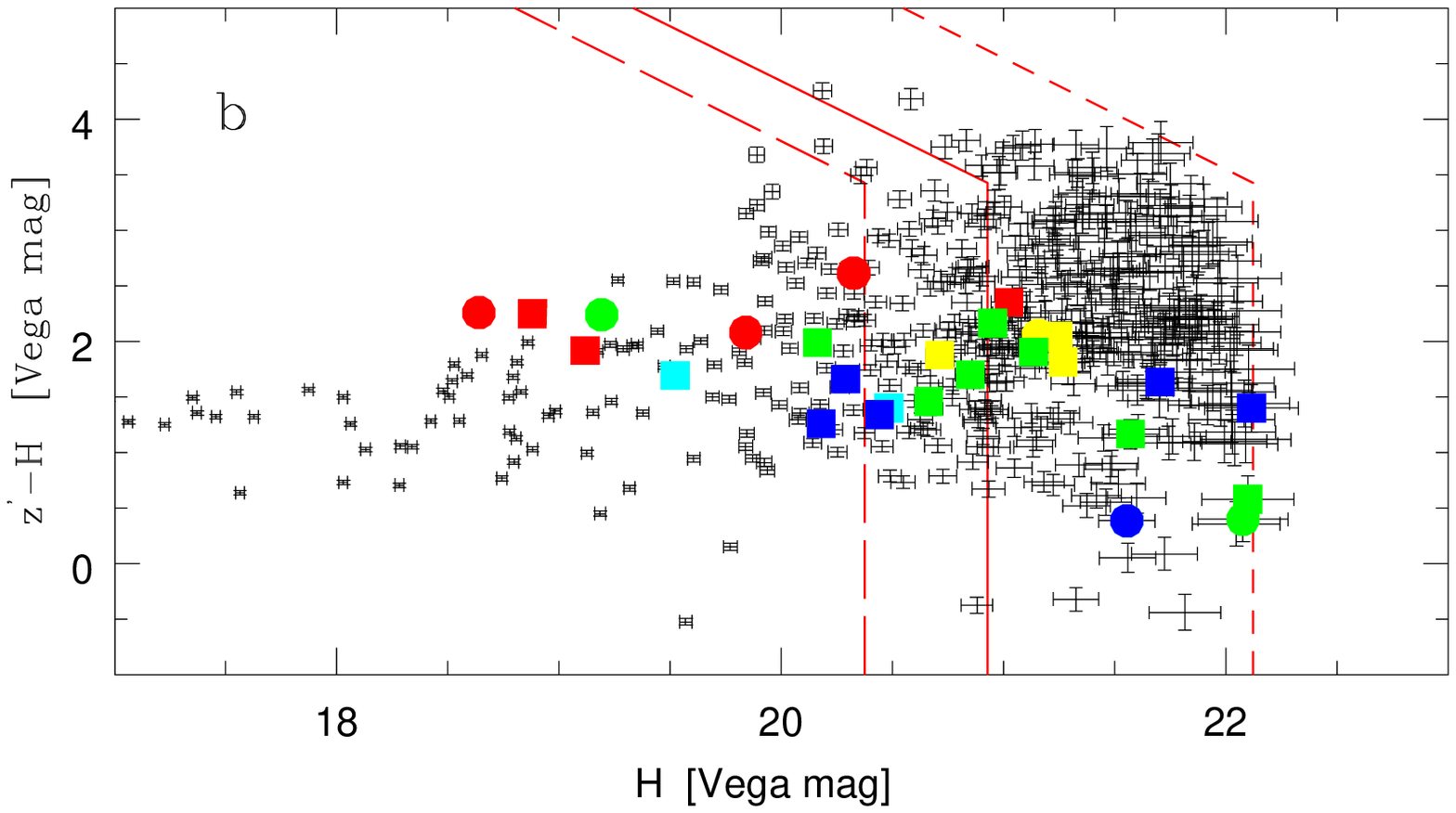}
      \vskip -2.8truecm
      \caption{$z^{\prime} - H$ vs $H$ colour--magnitude diagram
               for all GROND sources in the XMMU\,J0338.7$+$0030 region
               with flux detections
               in the $\mathrm{z^{\prime}, H}$ bands
               at a significance level $\ge 1 \mathrm{\sigma}$,
               whatever the classification of the source by {\em Le Phare}
               (i.e., stars are included).
               Photometric errors ($1 \mathrm{\sigma}$) are also shown.
               As in Fig.~15, candidate members of XMMU\,J0338.7$+$0030
               are reproduced with red (green) filled circles if they are
               within (outside) the bona fide cluster region (panel a).
               The three spectroscopic cluster members
               in the GROND photometric catalogue are marked
               with blue filled circles there.
               Green asterisks represent the two QSO at a cluster-centric
               distance greater than $1^{\prime}$.
               Alternatively (panel b), photometric cluster members classified
               as galaxies are colour-coded according to
               their spectro-photometric type, which is E (red), Sbc(yellow),
               Scd (green), Im (cyan) or SB (blue).
               In each panel, the red short-dashed, solid and long-dashed lines
               represent the $1 \mathrm{\sigma}$, $3 \mathrm{\sigma}$,
               and $5 \mathrm{\sigma}$ flux thresholds
               in the $\mathrm{z^{\prime}, H}$ bands (cf. Sect.~2.3.2).
               Photometric members classified as elliptical galaxies
               define a narrow red locus in this colour--magnitude diagram.
              }
         \label{app2fig1}
   \end{figure*}

   Here we investigate the origin of the discrepancy
   between the spectroscopic redshift of XMMU\,J0338.7$+$0030
   ($z = 1.097 \pm 0.002$, Sect.~2.4.2)
   and the value that was suggested from the analysis
   of the $z - H$ vs $H$ colour--magnitude diagram obtained
   from the CAHA imaging ($z \sim 1.45 \pm 0.15$, Sect.~2.2.2).
   There total magnitudes correspond to Kron aperture magnitudes
   i.e., computed as {\em SExtractor} MAG\_AUTO (Sect.~2.1.2).

   Figure B.1 shows the $z^{\prime} - H$ vs $H$ colour--magnitude diagram
   of the same cluster obtained from the GROND simultaneous multi-band imaging,
   where total magnitudes are computed following method $B$ (see Sect.~2.3.2).
   Its comparison with the analogous diagram in Fig.~3 (Sect.~2.2.2)
   reveals that the spectroscopic members with ID$=$2 and ID$=$16
   exhibit $z^{\prime} - H$ colours that are
   about $0.8~\mathrm{Vega~mag}$ and $0.6~\mathrm{Vega~mag}$ bluer
   than their $z - H$ colours listed in Table 1 (Sect.~2.2.2).
   Therefore, they lie closer to the spectroscopic member with ID$=$15
   (with $z^{\prime} - H = 2.24~\mathrm{Vega~mag}$, in full agreement
   with its $z - H$ colour listed in Table 1) in Fig.~B.1 than in Fig.~3.
   In addition, the existence of a red sequence is suggested in Fig.~B.1.
   If we compute total magnitudes following method $A$,
   the spectroscopic members with ID$=$2 and ID$=$16
   exhibit $z^{\prime} - H$ colours
   that are about $0.4~\mathrm{Vega~mag}$ bluer than their $z - H$ colours
   in both cases, whereas the spectroscopic member with ID$=$15
   exhibits a $z^{\prime} - H$ colour that is about $0.3~\mathrm{Vega~mag}$
   redder than its $z - H$ colour this time.
   The $\mathrm{H}$-band magnitudes of these three galaxies
   can differ by up to $\pm 0.5~\mathrm{Vega~mag}$
   with respect to those listed in Table 1, as a function of object
   and method ($A$ or $B$).

   The method adopted to compute total magnitudes certainly impacts
   the estimated brightness of sources like these three galaxies,
   which are at the limit bewteen being considered partly resolved
   or point-like for resolutions of 1.1--$1.3^{\prime \prime}$.
   However, it is likely the stacking of individual frames
   without PSF-matching in the case of the OMEGA2000 images
   that explains the previous differences in colours.

   We note that an SSP model with Salpeter IMF and solar metallicity
   exhibits $z - H \sim 2.4~\mathrm{Vega~mag}$ when observed at $z = 1.1$,
   for a formation redhift between 3 and 10 (fig.~3 in Fassbender et al.,
   2011c).
   This colour is about $0.8~\mathrm{Vega~mag}$ bluer
   than the $z - H$ colour of the same model observed at $z = 1.45$ (Fig.~3).
   It also agrees well with the distribution
   of the spectro-photometrically classified elliptical galaxies in Fig.~B.1,
   modulo (small) transformations between different $\mathrm{z, H}$ filters.
   At the same time, it is not inconsistent with the distribution
   of the spectroscopic members with absent/weak [OII] line emission
   (ID$=$2 and ID$=$15) in Fig.~3, taking into account
   the photometric uncertainties.

\end{appendix}

\end{document}